   \documentclass{elsart}

   \usepackage{graphicx}
    \usepackage{latexsym}
    \usepackage{tabularx}
   \usepackage{amsmath}
 \usepackage{amssymb}
 
 \journal{Physics Reports}

\begin{document}

\begin{frontmatter}



\title{Gravity and the Thermodynamics of Horizons}


\author{T.  Padmanabhan}
\ead{nabhan@iucaa.ernet.in}
\ead[url]{http://www.iucaa.ernet.in/~paddy}
\address{IUCAA, Post Bag 4, Ganeshkhind, Pune 411 007, India.}

     
\begin{abstract}

Spacetimes with horizons show a resemblance to thermodynamic systems and it is possible to associate  the notions of temperature and entropy with them. Several aspects of this connection are reviewed in a manner appropriate for  broad readership. The approach uses two essential principles: (a) the physical theories  must be formulated for each observer entirely in terms of variables any given observer can access and (b) consistent formulation of quantum field theory requires analytic continuation to the complex plane.
These two principles, when used together in spacetimes with horizons, are powerful enough to provide several results in a unified manner.  Since  spacetimes with horizons have a generic behaviour under  analytic continuation, standard results of quantum field theory in curved spacetimes with horizons can be obtained directly (Sections
III to VII). The
requirements (a) and (b) also  put  strong constraints on the action principle describing the gravity and, in fact, 
one can obtain the Einstein-Hilbert action from the thermodynamic considerations (Section VIII).
The review emphasises the thermodynamic aspects of horizons, which could be obtained from general
principles and is expected to remain valid,  independent of the microscopic
description (`statistical mechanics') of horizons.

\bigskip

\emph{We combine probabilities by multiplying, but we combine the actions ... by adding;  ...... since the
 logarithm of the probability is necessarily negative, we may
identify action  provisionally with minus the logarithm of the statistical probability of the state...}

\hfill Eddington (1920)\ \cite{Eddington:1920}

\smallskip

\noindent\emph{The mathematicians can go beyond this Schwarzschild radius and get inside, but
I would maintain that this inside region is not physical space, .... and should not be taken
  into account in any physical theory.}
  
\hfill Dirac (1962)\ \cite{Dirac:1962}

\end{abstract}

\end{frontmatter}

\tableofcontents

\section{Introduction}

   The simplest solution  to Einstein's equations in general relativity --- the Schwarzschild  solution ---
  exhibits a singular behaviour 
when expressed in the most natural coordinate system which makes the symmetries of 
  the solution obvious. One of the metric coefficients ($g_{tt}$)
  vanishes on a surface $\mathcal{H}$ of finite area while another ($g_{rr}$) diverges on the same surface.
  After some  initial confusion, it was realized that these singularities are  due
  to bad choice of coordinates. But the surface $\mathcal{H}$ brought in new physical features
  which have kept physicists active in the field for  decades.  
  
  Detailed investigations 
 in the 1970s showed that  the Schwarzschild solution and its generalisations  (with horizons)
 have an uncanny relationship with 
    laws of thermodynamics. [A description of
 classical aspects of black hole thermodynamics can be found in 
 \cite{Bardeen:1973gs,BHLS} and  \cite{membrane}].  The work of 
 Bekenstein moved these ideas forward  \cite{Bekenstein:1972tm,Bekenstein:1973ur,Bekenstein:1974ax}
 and one was initially led to a system with entropy but no temperature.
 This paradox was resolved   when
  the black hole evaporation was discovered \cite{Hawking:1975sw} and it was very
  soon realized that there is an intimate connection between horizons and temperature 
 \cite{Fulling:1973md,Davies:1975th,Gerlachbh}.

  Later work over three decades has re-derived these results and extended them in many different
  directions but --- unfortunately --- without any further  insight. It is probably fair to say that the
  ``deep'' relation between thermodynamics, quantum theory and general relativity, which was hoped for,
  is still elusive in the conventional approaches.
  
  This review focuses on certain specific aspects of thermodynamics of horizons and attempts to unravel
a deeper relationship between thermodynamics of horizons and
   gravity. Most of the material is aimed at a broader readership than  the experts in the field.
  In order to keep the review  self contained and of reasonable length, it is necessary to concentrate on
  some simple models (mentioning generalisations, when appropriate, only briefly) and deal directly
  with semi classical and quantum mechanical aspects. (Hence many of the beautiful results of 
  classical black hole thermodynamics  will not be discussed here. Approaches 
  based on string theory and loop gravity will be only briefly touched upon.) The broad aim 
  of the review will be to analyse 
 the following  important conceptual issues: 
  \begin{itemize}
  \item 
  What is the key physics
  ({\it viz.} the minimal set of assumptions)
   which leads to the association of a temperature with
  a horizon ? 
  Can one associate a temperature with \emph{any} horizon~? 
  \item Do all horizons, which hide information, possess an entropy ? If so, how can one understand the entropy
  and temperature of horizons in a broader context than that of, say, black holes ? 
  What are the microscopic degrees of freedom associated with this entropy ?
  \item Do all observers attribute a temperature and entropy to the horizon in spite of the fact that
  the amount of information accessible to  different observers is different ? If the answer is ``no'',
  how does one reconcile dynamical effects related to, say, black hole evaporation, with 
  general covariance ?
  \item What is the connection between the above results and gravity, since horizons of certain kind
  can exist even in flat spacetime in the absence of gravity ?
  \end{itemize}

  All these issues are subtle and controversial to different degrees. Current thinking favours --- correctly
  --- the view that a temperature can be associated with 
  any horizon  and the initial sections of the review will concentrate on
  this question. The second set of issues raised above are not really settled in the literature
  and fair  diversity of views prevails. We shall try to sort this out and clarify matters 
   though there are still several open issues. The answer to the question 
   raised in the third item above is indeed ``no''
  and one requires serious rethinking about the concept of general covariance in quantum theory.
  We will describe, in the latter half of the review, 
  a possible  reinterpretation of  the formalism so that each observer will have
  a consistent description. This analysis also leads to a deeper connection between gravity
   and spacetime thermodynamics, thereby shedding light on the last issue.

\begin{figure}[htbp] 
 \begin{center}
\includegraphics[scale=0.6]{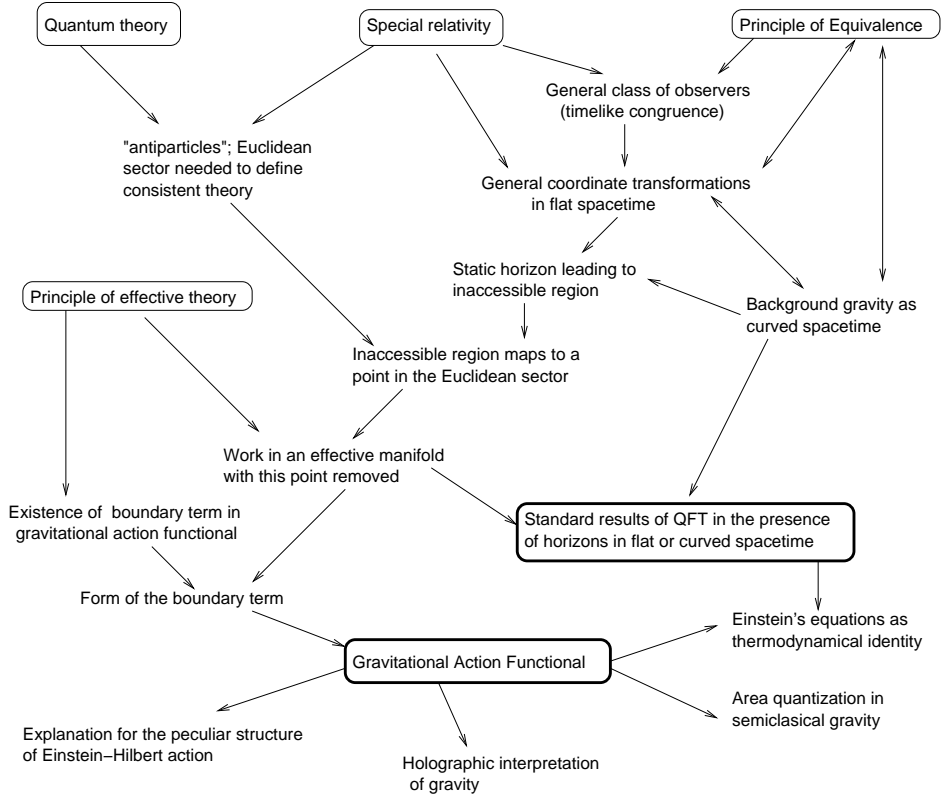}
 \caption{Summary of the logical structure of the approach adopted in this review}
 \label{flowchart}
\end{center}
 \end{figure}

The logical structure of our approach
(summarized in the last Section and Fig. \ref{flowchart} in page \pageref{flowchart}) will be as follows:
 Families of observers exist in any spacetime, who --- classically ---  have  access to only limited portions of the spacetime because of the existence of horizons. This leads to two effects when the horizon is (at least approximately) static:
\begin{itemize}
\item
The Euclidean version of the quantum field theory needs to be formulated in an \emph{effective} spacetime manifold 
obtained by removing the region blocked by the horizon. 
When the horizon is static, this {\it effective} 
manifold will have a nontrivial topology and leads to the association of a temperature with the horizon (Sections III to VI). This arises because  the quantum theory contains information which  classical
theory does not have, due to non-zero  correlation functions on a spacelike hypersurface across the horizon.
\item
The gravitational action functional, when formulated in terms of  the variables the family of observers can access,  \emph{will} have a boundary term  proportional to the horizon area. This is equivalent to associating a constant entropy per unit area of any horizon. Further, it is possible to obtain the Einstein-Hilbert action using the structure of the boundary term. Among other things, this
\emph{clarifies a peculiar relation between the boundary and surface terms of the Einstein-Hilbert action} (Section VIII). This idea lends itself to further generalisations and leads  to specific results in the semiclassical limit of quantum gravity.
\end{itemize}

Throughout the discussion, we emphasize the `thermodynamical' aspects of horizons rather than 
the `statistical mechanics'  based on microscopic models, like string theory or loop gravity.
While there has been considerable amount of work in recent years in the latter approaches (briefly 
discussed in Section \ref{qgbh}), most of the results obtained by these approaches are necessarily
model dependent. On the other hand, since any viable microscopic model for quantum gravity
reduces to Einstein gravity in the long wavelength limit, it is possible to obtain several
general results in the semi-classical limit of the theory which are independent of the microscopic
details. This is analogous to the fact that the thermodynamical description of a gas, say, is 
broadly independent of the microscopic Hamiltonian which describes the behaviour of molecules
in the gas. While such a microscopic description is definitely worth pursuing, one also needs
to appreciate how much progress one can make in a reasonably model independent manner using
essentially the structure of classical gravity. As we shall see, one can make significant progress in
understanding the thermodynamics of  horizon by this approach which should be thought of as 
complementing the more microscopic descriptions like the ones based on string theory.

We follow the sign conventions of  \cite{mtw} with the signature $(- + + +)$
and use units with $G=\hbar=c=1$. But, unlike \cite{mtw}, we let the Latin indices cover 0,1,2,3 while the 
Greek indices cover 1,2,3. The 
background material relevant to this review can be found
 in several text books \cite{Birrel:bkqft,Fulling:bkqft,Wald:bkqft} and review articles 
 \cite{Dewitt:1975ys,Takagi:1986kn,Sriramkumar:1999nw,Bousso:2002ju,Brout:1995rd,Wald:1999vt,Sciama:1981hr}.  

\section{Horizon for a family of observers}\label{horizonsr}

Classical and quantum theories based on non-relativistic physics use the notion of absolute time 
and allow for information to be transmitted with arbitrarily large velocity.  
An event $\mathcal{P}(T_0,X_0^\alpha)$ can, in principle, influence {\it all} 
events at $T\geq T_0$ and be 
influenced by all events at $T\leq T_0$. There is no horizon limiting one's 
region of influence in non-relativistic
theories. 

The situation changes in special relativity, which introduces a maximal speed $c$ 
(equal to unity in our choice of units) for the propagation
 of signals.  An event $\mathcal{P} (T_0,{\bf X}_0)$ can 
 now acquire information
only from the events $\mathcal{P}(T,{\bf X})$ in the ``backward" light cone $|{\bf X}_0-{\bf X}|\leq (T_0-T)$ and can send information only 
to events in the ``forward" light cone $|{\bf X}-{\bf X}_0|\leq (T-T_0)$. The light cones 
$\mathcal{C} (\mathcal{P})$   at $\mathcal{P}$, defined by the equation
$\mathcal{C}(X^a) \equiv |{\bf X}-{\bf X}_0|^2- (T-T_0)^2 =0$, divide
 the spacetime into two regions which are either causally 
connected or causally disconnected
  to $\mathcal{P}$. This light cone structure is invariant under Lorentz transformations.  
  The normal $n_a=\partial_a \mathcal{C} (T-T_0, \textbf{X} - \textbf{X}_0)$ 
   to the light cone  $\mathcal{C}(\mathcal{P})$
   is a null vector ($n^a n_a =0$) and the light cone is a null surface. 
   
   Consider now a timelike curve $X^a(t)$ in the spacetime, parametrised by the proper time $t$ of the clock
   moving along that curve. We can construct past light cone $\mathcal{C}(t)$
   for each event $\mathcal{P}[X^a(t)]$ on this trajectory. The union $U$ of all these past light cones 
   $\{\mathcal{C}(t),-\infty\leq t \leq \infty\}$ determines whether an observer on the trajectory $X^a(t)$
   can receive information from all events in the spacetime or not. If $U$ has a nontrivial boundary, there will be regions in the spacetime from which this observer cannot receive signals. (We shall always  use the term ``observer''   as synonymous
  to a time-like curve in the spacetime,  without any other additional, implied, connotations.) 
  In fact, one can extend this notion to a family of timelike curves which fill a region of spacetime.
  We shall call such a family of curves with reasonable notions of smoothness a  ``congruence"; it is 
  possible to define this concept with greater level of abstraction (see e.g. \cite{Hawking:1973el}) which is not required for our 
  purpose. 
 Given a congruence of time-like curves (``family of observers"),
the boundary of the union of their causal pasts (which is essentially  the boundary of the union of backward
light cones) will define a {\it horizon} for this set of observers. We will assume that each of the timelike curves has
 been extended to the maximum possible value for the proper time parametrising the curve. If the curves do not hit any spacetime singularity, then this requires extending the proper time to infinite values. 
This horizon is
  dependent on the family of observers that is chosen, but is coordinate independent.
We shall call the horizon defined by the above procedure as \emph{causal} horizon
  in order to distinguish it from horizons defined through other criteria, some of which we will
  discuss in Section \ref{stexample}.
  
An important example (in flat spacetime) of a set of observers with horizon, which we shall
  repeatedly come across as a prototype,  is a class of trajectories $X^i (t)=(T(t),X(t),0,0)$:
  \begin{equation}
 \kappa T = N\sinh (\kappa t), \quad\kappa X = N \cosh (\kappa t), 
  \label{trajectory}
  \end{equation}
  where $N $ and $\kappa$  are constants.
  The quantity 
    $(Nt) $ is the proper time of the clock carried by the observer
  with the trajectory $N=$ constant.
  Physically, for finite $t$, 
  these trajectories (for different $N$) represent observers moving with (different) 
  uniform acceleration  $(\kappa/N)$  along the 
  X-axis. The velocity $(dX/dT)=\tanh (\kappa t)$ approaches the speed of light as $t\to\pm \infty$.

  For all $ N>0, \kappa>0 $, these trajectories are hyperbolas confined to the `right wedge'  of the 
  spacetime  $(\mathcal{R})$ defined by $ X>0, |T|<X$  and these observers 
  cannot access any information in the region $ T > X $.
  Hence, for this class of observers, the null light cone surface,
   $( T-X )=0$, acts as a  horizon.
  An inertial observer with the trajectory
  $(T=t, X=x, 0, 0)$ for all $t$  will be able to access information from the region $T>X$
  at sufficiently late times. 
  The accelerated
  observer, on the other hand, will not be able to access information from half the spacetime
  even when $t \to \infty$. 
 
  Similarly,  Eq.~(\ref{trajectory}) with $N <0$ represents a class of observers
  accelerating along negative x-axis and confined to the `left wedge' $(\mathcal{L})$
  defined by $ X<0, |T|<|X|$
  who will not have access to the region  $ (T+X)>0$.
 This example shows that the horizon structure is ``observer dependent"
  and arises because of the nature of timelike congruence which is chosen to define it.  
 
 These ideas generalise in a straight forward manner to curved spacetime.  As a simple example,
 consider a class of spacetimes with the metric
 \begin{equation}
 ds^2=\Omega^2(X^a)(-dT^2+dX^2)+dL_\perp^2
 \end{equation}
 where $\Omega(X^a)$ is a nonzero, finite, function everywhere (except possibly on events at which the spacetime has curvature singularities) and $dL_\perp^2$ vanishes on the $T-X$ plane. For light rays propagating in the $T-X$ plane, with $ds^2=0$, the trajectories are lines at $45^o$, just as in flat space time. The 
 congruence  in Eq.~(\ref{trajectory}) will again have a horizon given by the surface $(T-X)=0$ in this spacetime.   Another class of observers with the trajectories
  $(T=t, X=x, 0, 0)$ for all $t$  will be able to access information from the region $T>X$
  at sufficiently late times (provided the trajectory can be extended without hitting a spacetime singularity).
  Once again, it is clear that the horizon is linked to the choice of a congruence of timelike curves.

Given any family of observers in a spacetime, it is most convenient to interpret the results of observations performed by these observer in a frame in which these observers are at rest. So the natural coordinate system $(t,{\bf x})$ 
attached to any timelike congruence is the one in which each trajectory of the congruence corresponds to ${\bf x}=$
constant. (This condition, of course, does not uniquely fix the coordinate system but is sufficient for 
our purposes.) For the accelerated observers
  introduced above, such a
    coordinate system  is already provided by  Eq.~(\ref{trajectory}) itself with
   $(t, N, Y, Z)$ now  being interpreted as  a new coordinate system, related to the inertial coordinate system
   $ (T, X, Y, Z)$ with all the coordinates taking the range
    $(-\infty,  \infty)$.
   The transformations in Eq.~(\ref{trajectory})   do not
   leave the form of the line interval $ds^2 = -dT^2 + |d\textbf{X}|^2$ invariant;
    the line interval in the new coordinates is given by 
   \begin{equation}
   ds^2\equiv g_{ab}(x) dx^a dx^b = -N^2 dt^2 + dN^2/\kappa^2 + dL_\perp^2 
   \label{rindlermetric}
   \end{equation}
   The light cones $T^2 =|X| ^2$ in the $(Y,Z)=$ constant sector,
now corresponds to the surface $N=0$ in this new coordinate
   system (usually called the Rindler frame). Thus the Rindler frame is a static coordinate system
   with  the $g_{00}=0$  surface --- which is just the light cone through the origin 
   of the inertial frame ---
 dividing
   the frame into two causally disconnected regions.  
   Since the transformations in Eq.~(\ref{trajectory}) covers only the right and left wedges,
   the metric in Eq.~(\ref{rindlermetric}) is valid only in these two regions. Both the branches
   of the light cone $X=+T$ and $X=-T$ collapse to the line $N=0$. The top wedge, $\mathcal{F}(|X|<T, T>0$)
    and the bottom wedge $\mathcal{P}(|X|<T, T<0$)  of the Minkowski space {\it disappear} in this representation.
    (We shall see below how similar coordinates can be introduced in  $\mathcal{F},\mathcal{P}$ as well; see
Eq.~(\ref{exptwo})).

   The metric in  Eq.~(\ref{rindlermetric}) is static even though the 
   transformations in Eq.~(\ref{trajectory}) appear to depend on time in a nontrivial manner.
   This static nature 
   can be understood as follows: The Minkowski spacetime 
 possesses invariance under translations, rotations and Lorentz boosts
which are characterised by the existence of a set of ten
Killing vector fields. Consider any linear combination $V^i$
of these Killing vector fields which is timelike in a sub-region $\mathcal{S}$ of Minkowski spacetime.
 The integral curves to this vector field $V^i$ will
define   timelike curves in $\mathcal{S}$. If one treats these 
curves as the trajectories of a family of hypothetical observers, then one can
set up an appropriate  coordinate system
for this observer. Since the four velocity of the observer is along the
Killing vector field, it is obvious that the metric components
in this coordinate system will not depend on the time coordinate.  
   A sufficiently general Killing vector field which incorporates the effects of translations, rotations and 
   boosts can be written as  $V^i = (1+\kappa X, \kappa T - \lambda Y, \lambda X - \rho Z, \rho Y)$
   where $\kappa, \lambda$ and $\rho$ are constants.
   When $\lambda = \rho =0$, the field  $V^i$  generates the effects of Lorentz boost
   along the $X-$axis and the trajectories in Eq.~(\ref{trajectory}) are the
    integral curves of this Killing vector field.
The static nature of Eq.~(\ref{rindlermetric}) reflects the invariance under Lorentz boosts
along the $X-$axis. One simple way of 
proving this is to note that Lorentz boosts along $X-$axis  ``corresponds to" a rotation 
in the $X-T$ plane by an imaginary angle; or, equivalently, Lorentz boost will ``correspond to''
rotation in terms of the imaginary time coordinates $T_E=iT, t_E = it$. In
Eq.~(\ref{trajectory})  $t\to t+ \epsilon$ does represent  a rotation in the $X-T_E$ plane
on a circle of radius $N$. 
Clearly,  Eq.~(\ref{trajectory}) is just  one among several
possible trajectories for observers such that the resulting metric [like the one in 
 Eq.~(\ref{rindlermetric})] will be static. 
   (For example, the Killing vector field with $\rho =0$ corresponds to a rotating observer
   while $\lambda =\kappa, \rho=0$ leads to a cusped trajectory.) 
 Many of these are analysed in literature (see, for example,
 \cite{Letaw:1981ik,Letaw:1981yv,Padmanabhan:1982apsci,Sriramkumar:1999nw}) but none of them 
 lead to results as 
 significant  as Eq.~(\ref{rindlermetric}).
This is because  Eq.~(\ref{rindlermetric}) is a good approximation
 to a very wide class of metrics near the horizon. We shall now discuss this feature.

    Motivated by Eq.~(\ref{rindlermetric}),
  let us   consider a more a general class of metrics
     which are: (i)
  static  in the given coordinate system, $g_{0\alpha} =0, g_{ab} (t,{\bf x}) = g_{ab}({\bf x})$;
   (ii) $g_{00}({\bf x}) \equiv -N^2({\bf x})$ vanishes on some 2-surface $\mathcal{H}$ defined
   by the equation $N^2 =0$, (iii) $\partial_\alpha N$ is finite and non zero on $\mathcal{H}$
   and (iv) all other metric components and curvature
   remain finite and regular on $\mathcal{H}$.  
 The line element will now be:
 \begin{equation}
   ds^2=-N^2 (x^\alpha)  dt^2 +  \gamma_{\alpha\beta} (x^\alpha) dx^\alpha dx^\beta
   \label{startmetric}
   \end{equation}
The comoving observers in this frame have trajectories ${\bf x}=$
    constant, four-velocity $u_a=-N\delta^0_a$ and four acceleration $a^i=
    u^j\nabla_ju^i=(0,{\bf a})$ which has the purely
    spatial components $a_\alpha=(\partial_\alpha N)/N$.
 The unit normal $n_\alpha$ to the $N=$ constant surface is given by
    $n_\alpha =\partial_\alpha N (g^{\mu\nu}\partial_\mu N  \partial_\nu N)^{-1/2} =
    a_\alpha (a_\beta a^\beta)^{-1/2}$. A simple computation now shows that 
    the normal component of the acceleration $a^i n_i = a^\alpha n_\alpha$, `redshifted'
    by a factor $N$, has the value
\begin{equation}
    N(n_\alpha a^\alpha) = ( g^{\alpha\beta} \partial_\alpha N \partial_\beta N)^{1/2}\equiv Na({\bf x})
\label{defkappa}
\end{equation}
where the last equation defines the function $a$. From our assumptions, it follows that on the horizon $N=0$,
this quantity has a finite limit $Na\to \kappa$; the $\kappa$ is called the surface gravity of the horizon.

These static spacetimes, however,  have a more natural coordinate system defined in terms of the level surfaces of $N$. That is, we transform from the original space coordinates $x^\mu$ in Eq.(\ref{startmetric}) to the set $(N,y^A), A=2,3$ by treating $N$ as one of the  spatial coordinates. The $y^A$ denotes the two
transverse coordinates on the $N=$ constant surface. (Upper case Latin letters go over the coordinates 2,3 on the $t=$ constant, $N=$ constant surface). This
can be always done locally, but possibly not globally, because $N$ could be multiple valued etc. We, however,
 need this description only locally. The components of acceleration in the $(N,y^A)$ coordinates are
\begin{equation}
a^N=a^\mu\partial_\mu N=Na^2,a^B=a^\mu\frac{\partial y^B}{\partial x^\mu},a_B=0,a_N=\frac{1}{N}
\end{equation}
Using these we can express the metric in the new coordinates as 
\begin{equation}
g^{NN}=\gamma^{\mu\nu}\partial_\mu N\partial_\nu N=N^2a^2;\;\; g^{NA}=Na^A
\end{equation}
etc. The line element now becomes:
\begin{equation}
ds^2=-N^2dt^2+ \frac{dN^2}{(Na)^{2}}+
\sigma_{AB}(dy^A-\frac{a^A dN}{Na^2})(dy^B-\frac{a^BdN}{Na^2})
\label{iso}
\end{equation}
The original 7 degrees of freedom in $(N,\gamma_{\mu\nu})$ are now reduced to 6 degrees of freedom in $(a,a^A,
\sigma_{AB})$, because of our choice for $g_{00}$. This reduction  is similar to what happens in 
the synchronous coordinate system which  makes
$N=1$, but the synchronous frame loses the static nature \cite{lltwo}. In contrast, Eq.(\ref{iso}) describes the spacetime in terms of
the magnitude  of acceleration $a$, the transverse components $a^A$ and the  metric $\sigma_{AB}$ on 
the two surface and maintains the $t-$independence. 
The $N$ is now merely a coordinate and the spacetime geometry is  described in terms of 
$(a,a^A,\sigma_{AB})$
all of which are, in general, functions of $(N,y^A)$. In well known, spherically symmetric spacetimes with horizon, we will have $a=a(N),a^A=0$ if we choose $y^A=(\theta,\phi)$. Important features of dynamics are usually encoded in 
the function $a(N,y^A)$.

 Near the $N\to 0$ surface, $Na\to \kappa$, the surface gravity, and the metric reduces to the Rindler form in
 Eq.(\ref{rindlermetric}):
   \begin{equation}
   ds^2=-N^2dt^2+ \frac{dN^2}{(Na)^{2}}+dL_\perp^2
   \simeq-N^2 dt^2 + \frac{dN^2}{\kappa^2} +dL_\perp^2
   \label{dsfirst}
   \end{equation}
   where   the second equality is applicable 
   close to  $\mathcal{H}$. Thus the metric in Eq.~(\ref{rindlermetric})
    is a good approximation to a large class of static metrics with $g_{00}$ vanishing on a surface.   
(It is, of course, possible for $N$ to vanish on more than one surface so that the spacetime
 has multiple horizons;
this is a more complicated situation and requires a different treatment, which we will discuss in 
Section~\ref{multihorizon}).
   
   There is an interesting extension of the metric in Eq.~(\ref{rindlermetric}) or Eq.~(\ref{dsfirst})
   which is worth mentioning. Changing to the variable from $N$ to $l$ with 
\begin{equation}
dl=\frac{dN}{a}=\frac{NdN}{Na};\quad\quad l\approx\frac{1}{2\kappa}N^2
\end{equation}
where the second relation is applicable near the horizon with $Na\approx \kappa$,
   we can cast the line element in the form
   \begin{equation}
   ds^2 =- f(l)  dt^2 + \frac{dl^2}{f(l)}  + dL_\perp^2 \approx - 2 \kappa l \ dt^2 + \frac{dl^2}{2\kappa l}  + dL_\perp^2
   \label{standardhorizon}
   \end{equation}
   where the second equation is applicable near the horizon with
    $l\approx(1/2\kappa)N^2$.
    More generally, the function $f(l)$ is obtained by expressing $N$ in terms of $l$.
   Many examples of horizons in curved spacetime we come across  have this structure with
   $g_{00} = -g^{11}$ and hence this is a convenient form to use. 
   
\begin{figure}[htbp] 
\begin{center}
\includegraphics[scale=0.5]{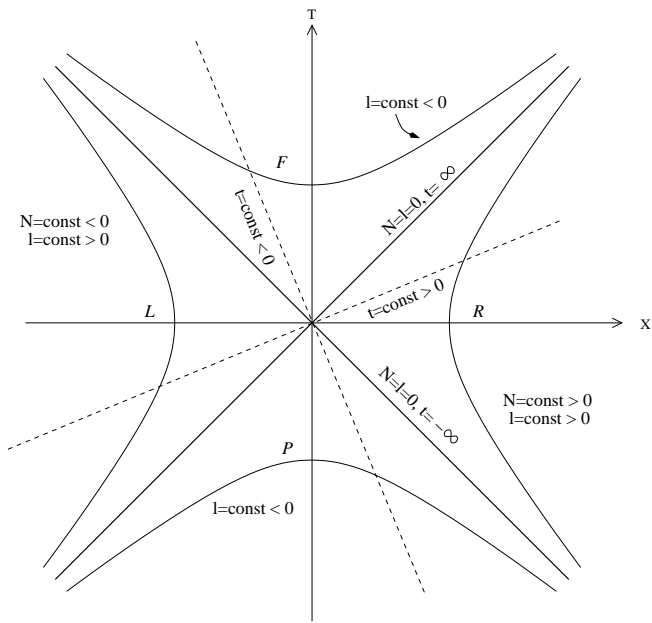}
  \caption{The global manifold with different coordinate systems in the four quadrants. See 
  text for discussion.}
\label{hyperfig}
\end{center}
\end{figure}

There is a further advantage in using the variable $l $. The original transformations
   from $(T,X)$ to $(t,N)$ given by Eq.~(\ref{trajectory}) maps the right and left wedges
   ($\mathcal{R}, \mathcal{L}$)  into ($N>0,  N<0)$ regions. Half of Minkowski spacetime contained in the 
   future light cone ($\mathcal{F}$) through the origin ($|X|<T,T>0$) and 
    past light cone ($\mathcal{P}$) through the origin ($|X|<T,T<0$)
    is  \emph{not} covered by the $(t,N)$ coordinate system  of Eq.~(\ref{trajectory}) at all. But, 
   if we now extend
   $l $ to negative values then it is possible to use this $(t,l)$ coordinate system to cover
   all the four quadrants of the Minkowski spacetime.  
   The complete set of transformations we need are: 
   \begin{equation}
   \kappa T=\sqrt{2\kappa l} \sinh (\kappa t); \quad \kappa X=\pm \sqrt{2\kappa l} \cosh (\kappa t)
   \label{expone}
   \end{equation}
   for $|X|>|T|$ with the positive sign in $\mathcal{R}$ and negative sign in 
   $\mathcal{L}$ and 
 \begin{equation}
   \kappa T=\pm \sqrt{-2\kappa l} \cosh (\kappa t); \quad \kappa X= \sqrt{-2\kappa l} \sinh (\kappa t)
   \label{exptwo}
   \end{equation}
   for $|X|<|T|$ with the positive sign in $\mathcal{F}$ and negative sign in 
   $\mathcal{P}$.
   Clearly, $l<0$ is used in $\mathcal{F}$ and  $\mathcal{P}$.
  Note that $t$ is timelike and $l $ is spacelike in Eq.~(\ref{standardhorizon}) only for $l>0$
  with their roles reversed for $l<0$. A given value of $(t,l)$ corresponds to a pair
  of points in $\mathcal{R}$ and $\mathcal{L}$ for $l>0$ and to pair of points in
  $\mathcal{F}$ and $\mathcal{P}$ for $l<0$. 
    Figure \ref{hyperfig} shows the geometrical features of the coordinate systems.
  
  The following crucial difference between the $(t,N)$ coordinates and $(t,l)$ coordinates must be stressed:
  In the $(t,N)$ coordinates, $t$ is everywhere timelike (see the second equation of Eq.~(\ref{dsfirst})) and the two regions $N>0$ and $N<0$ are completely disconnected. In the $(t,l)$ coordinates, $t$ is timelike where $l>0$ and spacelike where $l<0$ (see Eq.~(\ref{standardhorizon})) and the surface $l=0$
  acts as a ``one-way membrane"; signals can go from $l>0$ to $l<0$ but not the other way around. When we talk of $l=0$ surface as a horizon, we often have the interpretation based on this feature.

  In Eq.~(\ref{startmetric}), (\ref{standardhorizon}) etc., we have defined $N$ and $l $ such that
  the horizon is at $N=l=0$. This, of course, is not needed and our results continue to hold 
  when $f=0$ at some finite $l=l_H$. In spherically symmetric spacetimes it is often convenient
  to take $0\le l < \infty$ and have the horizon at some finite value $l=l_H$.

   Metrics of the kind in Eq.~(\ref{startmetric}) could
    describe either genuinely curved spacetimes or flat spacetime in some non inertial coordinate
    system. The local physics of the horizons really does not depend on whether the spacetime is 
    curved or flat and
    we shall present several arguments  in favour of the ``democratic'' treatment
    of horizons. In that spirit, we do not worry whether Eq.~(\ref{startmetric}) represents flat or 
    curved spacetime. 

We have assumed that the spacetime in Eq.~(\ref{iso}) is static. It is possible to generalise some 
 of our results
to {\it stationary} spacetimes, which have $g_{0\mu}\neq 0$ but with all metric coefficients remaining time independent. 
A uniformly rotating frame as well as curved spacetimes like Kerr metric belong to this
class and pose some amount of mathematical  difficulties. These difficulties
can be overcome, but only by complicating the formalism and obscuring the simple physical insights.
It is
more difficult to extend the results to general, time dependent, horizons
(for a discussion of issues involved in providing a general definition of horizon, see e.g.,  \cite{Ashtekar:2003hk,Date:2001xj}). If one considers the static  horizons as analogous to
equilibrium thermodynamics then the analogue of time dependent horizons will be non equilibrium thermodynamics. The usual
approach in thermodynamics is to begin with the study of equilibrium thermodynamics in order to define different thermodynamical variables etc. and then proceed to time dependent non equilibrium processes. These extreme limits are connected by quasi-static systems,
which can again be handled by a straight forward generalisation of the static case. We shall adopt a similar philosophy in our study of horizons and develop the notion of thermodynamical variables like temperature, entropy etc. for the horizons using static spacetimes of the form
in Eq.~(\ref{iso}) thereby precluding from consideration, stationary metrics  like that of  rotating frame or
Kerr spacetime.
 While stationary and time dependent metrics will be more complicated to analyse, we do not expect any new serious conceptual features to arise due to time dependence. What is more, the static horizons themselves have a rich amount of physics which needs to be understood. 
    
     The  coordinate systems having metrics of the form Eq.~(\ref{dsfirst})
    have several interesting, generic, features which we shall now briefly describe.

    \subsection{Horizon and infinite redshift}
    
   In the metrics of the form in  Eq.~(\ref{dsfirst}), the $N=0$ surface acts as a horizon
   and the coordinates $(t,N)$ and $(t,l)$  are badly behaved near this surface.
   This is most easily seen by considering the light rays traveling along the $N-$direction
   in Eq.~(\ref{dsfirst}) 
   with $y^A =$ constant. These light rays are determined by the equation $(dt/dN) = \pm (1/N^2a) $
   and as $N \to 0$, we get $(dt/dN) \approx \pm (1/N\kappa) $. The slopes of the light cones diverge making the $N=0$ surface act as a
   one way membrane
   in the $(t,l)$ coordinates and as a barrier 
     dividing the spacetime into two causally disconnected regions in the $(t,N)$ coordinates.
     This difference arises because the light cone $T=X$, for example, separates
     $\mathcal{R}$ from $\mathcal{F}$ and both regions are covered by the $(t,l)$ coordinates; in contrast,
     the region $\mathcal{F}$ (and $\mathcal{P}$) are not covered in the $(t,N)$ coordinates.

   This result is confirmed by the nature of the trajectories of material particles with constant energy, near  $N=0$.
    The Hamilton-Jacobi  (HJ) 
   equation for the action $A$ describing a particle of mass $m$ is $\partial_a A\partial^a A=-m^2$.  
   In a spacetime with the metric in Eq.(\ref{iso}) the standard substitution $A=-Et +f(x^\alpha)$, reduces it to:
   \begin{equation}
   N^4a^2\left(\frac{\partial f}{\partial N}\right)^2=E^2-N^2[m^2+(\partial_\perp f)^2] 
\label{hjmodes}
   \end{equation}
where $(\partial_\perp f)^2$ is the contribution from transverse derivatives.  Near $N=0$, the solution  is universal, independent of $m$ and the transverse degrees of freedom:
   \begin{equation}
   A\approx-Et \pm E\int\frac{dN}{N^2a}\approx-E(t\pm\xi)
   \label{hjapprox}
   \end{equation}
   where 
   \begin{equation}
   \xi \equiv \int \frac{dN}{N^2a} = \int \frac{dl}{f(l)}
   \label{NEW}
   \end{equation}
    is called the {\it tortoise coordinate} and behaves as $\xi\simeq (1/\kappa)\ln N$ near the horizon.
   The trajectories are $N\cong ({\rm constant}) \exp(\pm \kappa t)$ clearly
   showing that the horizon (at $N=0$) cannot be reached in finite time $t$ from either side.
   
    Let us next consider the redshift of a photon emitted at $(t_e, N_e, y^A)$, 
   where $N_e$ is close to the horizon surface $\mathcal{H}$,  and is observed
   at $(t, N, y^A)$.  The frequencies at emission $\omega(t_e)$ and detection $\omega(t)$  are related by
   $[\omega(t)/\omega(t_e)]=[N_e/N] $. 
   The trajectory of the out-going photon is given by 
   \begin{equation}
   t-t_e =  \int_{N_e}^N \frac{dN}{N^2a} = -\frac{1}{\kappa} \ln N_e + \textrm{constant}
   \label{photonpath}
   \end{equation}
  where we have approximated the integral by the dominant 
   contribution near $N_e =0$. This gives $N_e  \propto \exp(- \kappa t)$,
   leading to the exponentially redshifted frequency 
   $\omega(t)\propto N_e\propto \exp(-\kappa t)$.

  \subsection{ Inertial coordinate system near the horizon}  
  
 The bad behaviour of the metric
  near $N=0$ is connected with the fact that the observers at constant-${\bf x}$  perceive a horizon at $N=0$. 
Given a congruence of timelike curves, with a non-trivial boundary for their union of past light cones,
there will be trajectories in this congruence which are arbitrarily close to the boundary. Since each trajectory
is labelled by a ${\bf x}=$ constant curve in the comoving coordinate system, it follows that the metric in this coordinate system will behave badly at the boundary.

   The action functional in Eq.~(\ref{hjapprox}) corresponds to a particle with constant
    energy in the $(t,{\bf x})$
   coordinate system, since we have 
separated the HJ equation with $(\partial A/\partial t)=-E=$ constant. 
Since this coordinate system is badly behaved at the horizon, the  trajectory
takes infinite coordinate time to reach the horizon from either direction. In a different coordinate
system which is regular at the horizon, the trajectories can cross the horizon at finite time.
This is clear from the fact that 
one can introduce a local
inertial frame
even near the horizon; the observers at rest in this frame (freely falling observers) will have regular
 trajectories which will
cross the horizon.
 If we use a coordinate system in which
 freely falling  observers are at rest and use their clocks to measure time, there will be no pathology at
  the horizon.
  In case of flat spacetime, the freely falling trajectories are obtained 
 by choosing the action functional which behaves as $A=-E'T+F({\bf X})$.
 The corresponding ``good" coordinate system is, of course,  the global 
 inertial frame.

  In the  general case, the required transformation is 
  \begin{equation}
  \kappa X=e^{ \kappa\xi} \cosh  \kappa t; \  \kappa T=e^{ \kappa\xi} \sinh  \kappa t
  \label{keytransform}
  \end{equation}
  where $\xi$ is defined by Eq.~(\ref{NEW}). This result can be obtained as follows:
  We first transform the line element in Eq.~(\ref{standardhorizon}) to
  the tortoise coordinate $\xi $:
  \begin{equation}
  ds^2 = N^2(\xi) (-dt^2 + d\xi^2)  + dL_\perp^2 ; 
  \end{equation}
  Introducing the null coordinates $u = (t-\xi), v=(t+\xi)$, we see that 
  near the horizon, $N \approx 
  \exp[ \kappa\xi] = \exp[( \kappa/2)(v-u)]$ which is singular as $\xi \to -\infty$. This suggests the transformations to two new null coordinates
  $(U,V)$ with  $ \kappa V=\exp[ \kappa v],  \kappa U=-\exp[- \kappa u]$ 
  which are regular at horizon.
  The corresponding $T$ and $X$ given by
  $ U=(T-X), V=(T+X)$. Putting it all together, we get the result in Eq.~(\ref{keytransform}).
  The metric in terms of $(T,X)$ coordinates has the form 
  \begin{equation}
  ds^2 =\frac{N^2}{ \kappa^2(X^2 - T^2)} (-dT^2 +dX^2) + dL_\perp^2
  \end{equation}
  where $N$ needs to be expressed in terms of $(T,X)$ using the coordinate transformations.
  In general, this metric will be quite complicated and will \emph{not} even be static.
  The  horizon at $N=0$
  corresponds to the light cones $T^2-X^2=0$ in these coordinates and 
  $[N^2/ \kappa^2(T^2-X^2)]$ is finite on the horizon by construction. Thus the $(T,X)$ coordinates are the 
  locally inertial coordinates near  $\mathcal{H}$.

  The transformations in Eq.~(\ref{keytransform})  show that $(X^2 - T^2)$ is purely a function
  of $N$ (or $l$) while $(X/T)$ is a function of $t$. Thus $t=$ constant curves are radial lines
  through the origin with the $X=0$ plane coinciding with $N=0$ plane. Curves of 
  $N=$ constant are hyperbolas (see figure \ref{hyperfig}).

  By very construction, the line element in the $(T,X)$ coordinates is well behaved near  the horizon,
  while the line element is pathological in the $(t,N)$ or $(t,l)$  coordinates because the transformations
  in Eq.~(\ref{keytransform}) are singular  at $N=l=0$. In the examples which we study
  the spacetime manifold will be well behaved near the horizon and this fact will be 
  correctly captured in the $(T,X)$ coordinates. The singular transformation from $(T,X)$ coordinates
  to $(t,l)$ coordinates is the cause for the bad behaviour of metric near $l=0$
  in these coordinates. But the family of observers, with respect to whom the horizon is defined to exist, will
find it natural to use the $(t,N)$ coordinate system and the ``bad'' behaviour of the metric tensor implies some 
{\it non-trivial physical phenomena} for these observers. Since any family of observers has a right to describe physics in the coordinate frame
in which they are at rest, we need to take these coordinates seriously.
(We will also see that  $(t,l)$ coordinates often have
  other interesting features which are not shared by the $(T,X)$ coordinates.
  For example, the metric can be static in $(t,l)$ coordinates but time dependent
  in $(T,X)$ coordinates.)
  
   The transformation in Eq.~(\ref{keytransform}) requires the knowledge of 
  the surface gravity $ \kappa $ on the horizon. If $N$ vanishes
  at more than one surface --- so that the spacetime has multiple horizons ---
  then we need different transformations of the kind in Eq.~(\ref{keytransform}) near
  each horizon with, in general, different values for $ \kappa$. We shall comment on this
  feature in Section \ref{multihorizon}.
  
  \subsection {Classical wave with exponential redshift}\label{infinitez}
  
  The fact that the time coordinates used by the freely falling and accelerated observers are related by a nonlinear
  transformation Eq.~(\ref{keytransform}) leads to  an interesting consequence. Consider a monochromatic 
  out-going wave  along
  the X-axis, given by $\phi(T,X)=\exp[-i\Omega(T- X)]$ with $\Omega>0$. 
  Any other observer who is inertial with respect to the $X=$ constant observer will see this as a 
  {\it monochromatic} wave, though  with a different frequency. But an accelerated observer,
  at $N=N_0 =$ constant  
   using the proper
  time co-ordinate $\tau \equiv N_0 t$ will see the same mode as varying in time as 
  \begin{equation}
  \phi = \phi(T(t),X(t))=\exp[i\Omega q e^{- \kappa t}]
  = \exp[i\Omega q \exp - (\kappa/N_0)\tau]
\label{expodamp}
  \end{equation}
  where we have used Eq.~(\ref{keytransform}) and $q\equiv \kappa^{-1} \exp( \kappa\xi) $.
  This is clearly not monochromatic and has a frequency which is being exponentially redshifted in time. The power spectrum   of this wave 
  is given by $P(\nu) = |f(\nu)|^2$ where $f(\nu)$ is the Fourier transform of $\phi(\tau)$ with
  respect to $\tau$:
  \begin{equation}
  \phi(\tau) = \int_{-\infty}^\infty \frac{d\nu}{2\pi} f(\nu) e^{-i\nu \tau}
  \label{ftf}
  \end{equation}
   Because of the exponential redshift, this power spectrum will \emph{not} vanish
  for $\nu<0$. Evaluating this Fourier transform (by changing to the variable 
  $\Omega q \exp[-( \kappa/N_0) \tau ] =z$
  and analytically continuing to Im $z$) one gets:
  \begin{equation}
 f(\nu) =(N_0/ \kappa) (\Omega q)^{i\nu N_0/\kappa } \Gamma(-i\nu N_0/ \kappa)
  e^{\pi \nu N_0/2 \kappa} 
 \label{powernu}
 \end{equation}
  This leads to the 
   the remarkable result
  that the power, per logarithmic band in frequency, at negative frequencies  
  is a Planckian at temperature $T= (\kappa/2\pi N_0)$: 
  \begin{equation}
 \nu\vert f (-\nu)\vert^2 = \frac{\beta}{e^{\beta \nu} - 1 } ; \quad \beta = \frac{2\pi N_0}{\kappa }
  \label{planck}
  \end{equation}
  and, more importantly,
\begin{equation} 
|f(-\nu)|^2/|f(\nu)|^2=\exp(-\beta\nu).
\label{nuandminusnu}
\end{equation}
  Though $f(\nu)$ in Eq.~(\ref{powernu})
   depends on $\Omega$, the power spectrum $|f(\nu)|^2$ is independent of 
  $\Omega$; monochromatic plane waves of any frequency (as measured by the freely falling observers at $X=$ constant) will appear to have Planckian
  power spectrum in terms of the (negative) frequency $\nu$, defined with respect to the proper time of 
   the accelerated observer located at $N=N_0=$ constant.  The scaling of the temperature $\beta^{-1}\propto N_0^{-1}\propto |g_{00}|^{-1/2}$ is precisely what is expected in general relativity for temperature.
   
   We saw earlier (see Eq.~\ref{photonpath}) that  waves propagating 
   from a region near the horizon will undergo
   exponential redshift.
  An observer detecting this  exponentially redshifted radiation 
   at late times $(t\to \infty)$, originating from  a region close 
   to $\mathcal{H}$ will attribute to this radiation a Planckian power spectrum given by  Eq.~(\ref{planck}). 
   This 
  result lies at the foundation of associating temperature with horizons.
  [The importance of exponential redshift is emphasised by several people including 
\cite{Grove:1986fz,Hu:1996br,Koks:1997rk,Hu:1996vu,Raval:1997vt,Visser:2001kq}.]

   The Planck spectrum in  Eq.~(\ref{planck})
  is in terms of the  frequency and $\beta$ has the  (correct) dimension of  time;   no $\hbar$
appears in the result. If we now switch  the variable to energy,
  invoking the basic tenets of quantum mechanics, and write $\beta \nu = (\beta/\hbar) (\hbar \nu)
  = (\beta/\hbar) E$, then one can identify a temperature $k_BT =( \kappa\hbar /2\pi c)$ which
  scales with $\hbar$. This ``quantum mechanical'' origin of temperature 
  is superficial because it arises merely because of a  change of  units from $\nu $ to $E$.
  An astronomer measuring frequency rather than photon energy  will see the spectrum in 
  Eq.~(\ref{planck}) as Planckian without any quantum mechanical input.

   It is fairly straightforward to construct  different time evolutions
  for a wave $\phi(t)$ such that the corresponding power spectrum $|f(\nu)|^2$ has the
  Planckian form. While the trajectory in Eq.~(\ref{trajectory}) was never constructed for this purpose
  and  leads to this result in a natural fashion, it is difficult to understand the physical origin of temperature
  or the Bose distribution for photons in this approach purely classically, especially since we started with a complex wave form. (The results for a real cosine wave is more intriguing; see \cite{tplsksrinia,tplsksrinib}). 
 The true importance
  of the above result  lies in the fact that, the mathematical operation involved in
  obtaining Eq.~(\ref{planck}), acquires physical meaning in terms of positive and negative
  frequency modes in quantum field theory which we shall  discuss later. This is suggested
by Eq.~(\ref{nuandminusnu}) itself. In the quantum theory of radiation, the amplitudes of the wave, with frequencies differing in sign,
cause absorption and emission of radiation by a system with two energy levels differing by $\delta E=h\nu$. Hence any system, which comes into steady state with this radiation in the accelerated frame, will have the ratio of populations in the two levels to be $\exp(-\beta E)$, giving an operational meaning to this temperature.

  \subsection{Field theory near the horizon: Dimensional reduction}\label{ftdim}
  
  The fact that $N\to 0$ on the horizon leads to interesting conclusions regarding
  the behaviour of  any classical (or quantum)
  field  near the horizon. Consider, for example, an interacting scalar field in a background spacetime described by the metric in Eq.(\ref{iso}),
  with the action: 
  \begin{eqnarray}
  A &=& -\int d^4x \sqrt{-g} \left( \frac{1}{2} \partial_a \phi \partial^a \phi +V \right)\\
 &=& \int dt dN d^2y\, \frac{\sqrt{\sigma}}{N^2a}\nonumber 
 \times \  \left[ \frac {\dot \phi^2}{2} -N^4a^2
 \left(  \frac{\partial \phi}{\partial N}\right)^2- N^2\left[\frac
  {(\partial_\perp \phi)^2}{2} + V\right] \right]
  \end{eqnarray}
where $(\partial_\perp \phi)^2$ denotes the contribution from the derivatives in the transverse directions
including cross terms of the type $(\partial_N \phi \partial_\perp  \phi)$.
  Near $N=0$, with $Na\to \kappa$, the action reduces to the form 
  \begin{equation}
  A\approx   \int \sqrt{\sigma} d^2x_\perp \int dt \int d\xi \,  \left\{ \frac{1}{2} \left[ \dot \phi^2 - 
  \left(  \frac{\partial \phi}{\partial \xi}\right)^2 \right]  \right\}
  \end{equation}
  where we have changed variable to $\xi$ defined in Eq.~(\ref{NEW}) [which behaves as
$\xi\approx (1/\kappa)\ln N$] and ignored terms which vanish as 
  $N\to 0$. Remarkably enough this action represents a two dimensional free field theory
  in the $(t,\xi)$ coordinates  which has the enhanced symmetry of invariance under
  the conformal transformations $g_{ab} \to f^2(t,\xi) g_{ab}$ [see e.g., Section 3 of \cite{Padmanabhan:2002ha}]. 
The solutions to the
  field equations near $\mathcal{H}$  are plane waves in the $(t,\xi)$ coordinates:
   \begin{equation}
   \phi_\pm = \exp[-i\omega (t \pm  \xi)] 
   = N^{\pm i\omega/\kappa}e^{-i\omega t}
   \label{twotwo}
   \end{equation}
   These modes are the same as $\phi=\exp iA$ where $A$ is the solution Eq.~(\ref{hjapprox})
    to the Hamilton-Jacobi equation; this is because the divergence of $(1/N)$ factor near the horizon 
   makes the WKB approximation almost exact near the horizon. The mathematics involved in this phenomenon
   is fundamentally the same as the one which leads to the ``no-hair-theorems" 
  (see, eg.,  \cite{Bekenstein:1998aw}) for the black hole.
  
There are several symmetry properties for these solutions which are worth mentioning:

   (a) The Rindler metric and the solution near $\mathcal{H}$ is  invariant under the rescaling
   $N\to \lambda N$,  in the sense that this transformation merely adds a phase  to 
   $\phi$. This scale invariance can also be demonstrated by studying the spatial part of the 
   wave equation \cite{Srinivasan:1998ty} near $\mathcal{H}$,  where the equation reduces to a
    Schrodinger equation
  for the zero energy eigenstate in
  the potential $V(N) = - \omega^2/N^2$ .
   This Schrodinger equation has the natural scale invariance with respect to $N\to \lambda N$
   which is reflected in our problem. 

(b) The relevant metric $ds^2=-N^2 dt^2+(dN/\kappa)^2$ in the $t-N$
plane is also invariant, up to a conformal factor, to the metric obtained by $N\to \rho=1/N$:
\begin{equation}
ds^2=-N^2 dt^2+\frac{dN^2}{\kappa^2}=\frac{1}{\rho^4}(-\rho^2 dt^2+\frac{d\rho^2}{\kappa^2})
\end{equation}
Since the two dimensional field theory is conformally invariant, if $\phi(t,N)$ is a solution, then
$\phi(t,1/N)$ is also a solution. This is clearly true for the solution in Eq.~(\ref{twotwo}). Since $N$ is a coordinate in our description, this connects up the infrared behaviour of the field theory with the ultraviolet behaviour.   

(c)  More directly, we note that the symmetries of the theory enhance significantly near the
$N=0$ hypersurface.
Conformal invariance,  similar to the one found above, occurs in the  gravitational sector as well.
Defining  $q=-\xi$ by $dq=-dN/N(Na)$, we see that $N\approx\exp(-\kappa q)$ near the horizon, where $Na\approx \kappa.$ The space part of the metric in Eq.(\ref{iso}) becomes, near the horizon
$dl^2=N^2(dq^2+e^{2\kappa q}dL_\perp^2)$
which is conformal to the metric of the anti-De Sitter (AdS) space. The horizon
becomes the $q\to\infty$ surface of the AdS space. These results hold in any dimension.

(d) Finally, one can construct the metric in the bulk by a Taylor
series expansion, from the form of the metric near the horizon,
along the lines of exercise 1 (page 290) of \cite{lltwo}.
These ideas work only because, algebraically, $N\to 0$ makes certain terms
in the diffeomorphisms vanish and increases the symmetry.
There is a strong indication that most of the results related to horizons
 will arise from the enhanced symmetry of the theory near
the $N=0$ surface (see e.g. \cite{Carlip:1999cy,Park:1999tj,Park:2001zn} and references cited therein).

\subsection {Examples of spacetimes with horizons}\label{stexample}   
   
While it is possible to have different kinds of solutions to Einstein's equations
with horizons, some of the solutions have attracted significantly more attention
than others. Table \ref{table:metricprop} summarises the features related to three of these solutions.
In each of these cases, the metric can be expressed in the form
Eq.~(\ref{standardhorizon}) with different forms of $f(l)$ given in the table.
All these cases  have only one horizon at
some surface $l=l_H$ and the surface gravity $\kappa$ is well defined. 
(We have relaxed the condition that the horizon occurs at $l=0$; hence $\kappa$ is 
defined as $(1/2) f'$ evaluated at the location of the horizon, $l=l_H$.)
The coordinates $(T,X)$ are well behaved near the 
horizon while the original coordinate system $(t,l)$ is singular at the horizon.
  Figure \ref{hyperfig} describes all the three cases of horizons 
  which we are interested in, with suitable definition for the coordinates.

 \begin{table*}

\begin{tabular*}{\linewidth}{>{$}l<{$}>{$}c<{$}>{$}c<{$}>{$}c<{$}}
   \hline
  \textrm{Metric} & \textrm{Rindler} & \textrm{Schwarzschild} & \textrm{De Sitter}\\
    \hline\hline
    \noalign{\medskip} 
  f(l) & 2\kappa l & \left[1 - \frac{2M}{l}\right] & ( 1 - H^2 l^2)\\
   \noalign{\medskip}
  \kappa=\frac{1}{2} f'(l_H) &  \kappa & \displaystyle{ \frac{1}{4 M }} &  - H \\
     \noalign{\medskip} 
     \xi & \frac{1}{2\kappa} \ln\kappa l &  l+ 2M \ln \left[ \frac{l}{2M} - 1\right] &  \frac{1}{2H}
     \ln \left( \frac{1-Hl}{1+Hl} \right) \\
      \noalign{\medskip}
     \kappa X& \sqrt{2\kappa l} \cosh\kappa t & \ \  e^{\frac{l}{4M}} \left[ \frac{l}{2M} - 1\right]^{1/2} 
               \cosh \left[\frac{t}{4M}\right] & \ \left( \frac{1-Hl}{1+Hl} \right)^{1/2} \cosh Ht \\	       
     \noalign{\medskip}         
   \kappa T & \sqrt{2\kappa l} \sinh\kappa t & \ \   e^{\frac{l}{4M}} \left[ \frac{l}{2M} - 1\right]^{1/2} 
              \sinh\left[\frac{t}{4M}\right] &\  \left( \frac{1-Hl}{1+Hl} \right)^{1/2} \sinh Ht\\	      
  \noalign{\bigskip}
   \hline
   \noalign{\bigskip}
   \end{tabular*}
   \caption{Properties of Rindler, Schwarzschild and De Sitter metrics}
    \label{table:metricprop}
   \end{table*}

In all the cases
the horizon at $l=l_H$ corresponds to the light cones through the origin 
$(T^2 - X^2)=0$ in the freely falling coordinate system.  it is conventional to call the $T=X$ surface
as the future horizon and the $T=-X$ surface as the past horizon. Also note that
the explicit transformations to $(T,X)$ given in  Table \ref{table:metricprop} corresponds to $l>0$ and 
the right wedge, $\mathcal{R}$. Changing $l $ to $-l$ in these equations with $l<0$ will take 
care of the left wedge, $\mathcal{L}$. The future and past regions will require interchange of
$\sinh$ and $\cosh$ factors. These are direct generalisation of the transformations in 
Eq.~(\ref{expone}) and Eq.~(\ref{exptwo}).

The simplest case corresponds to flat spacetime in which $(T,X)$ are the 
Minkowski coordinates and $(t,l)$ are the Rindler coordinates. 
The range of coordinates extends to $(-\infty, \infty)$. 
The $g_{00}$ does not go to $(-1)$ at spatial infinity 
 in $(t,l)$ coordinates and the horizon is at $l=0$. 

The second case  is that of a Schwarzschild 
black hole. The full manifold is described in the $(T,X)$ coordinates, (called the Kruskal 
coordinates, which are analogous to the inertial  coordinates in flat spacetime) but
the metric is \emph{not} static in terms of the Kruskal time $T$.
The horizon at  $X^2 = T^2$ divides the black hole manifold into the four regions $\mathcal{R,L,F,P}$.
 In terms of the Schwarzschild coordinates,
   the metric is independent of $t$ and the horizon is at $l=2M$ where $M$ is the mass of the black hole.
   The standard Schwarzschild coordinates $(t,l)$ is a 2-to-1 map from the Kruskal coordinates $(T,X)$.
   The region $l>2M$  which describes the exterior of the black hole
    corresponds to $\mathcal{R}$ and $\mathcal{L}$
   and the region $0<l<2M$, that  describes
    the interior of the  black hole,
    corresponds to $\mathcal{F}$ and $\mathcal{P}$. The transverse coordinates are now ($\theta,\phi$) and the surfaces
   $t=$ constant, $l=$ constant are  2-spheres.  
   
  In the case of a black hole formed due to gravitational collapse,  the Schwarzschild solution is applicable to the region outside the collapsing matter, if the collapse is spherically symmetric.  The surface of the collapsing matter will be a timelike curve cutting through $\mathcal{R}$ and $\mathcal{F}$, making 
  the whole of $\mathcal{L}$,$\mathcal{P}$ (and part of $\mathcal{R}$ and $\mathcal{F}$) irrelevant since they will be inside the collapsing matter. In this case, the past horizon does not exist and we are only interested in the future horizon. Similar considerations apply whenever the actual solution corresponds only to part of the full manifold.
   
   There are five crucial differences between the Rindler and Schwarzschild coordinates:
   (i) The Rindler coordinates represents flat spacetime which is a non singular manifold.
   The Schwarzschild coordinates describe a black hole manifold which has a physical singularity at
   $l=0$ corresponding to  $T^2 - X^2=16M^2$.  Thus a world line $X=$ constant, crosses the horizon and hits the singularity in finite
$T$.
The region $T^2-X^2>16M^2$ is treated as physically irrelevant
   in the manifold.
   (ii) In the Rindler metric, $g_{ab}$ does not tend to $\eta_{ab}$ when $|{\bf x}|\to \infty$
     while in the Schwarzschild metric it does.
    (iii) The Rindler metric is independent of the  $t$ coordinate just as the Schwarzschild metric
   is independent of the $t$ coordinate. Of course, the flat spacetime is static in $T$ coordinate as well
   while the black hole spacetime is not static in the Kruskal coordinates.
   (iv) The surfaces with $t=$ constant, $l=$ constant are 2-spheres with finite area in the case of 
   Schwarzschild coordinates; for example, the horizon at $l=2M$ has the 
   area $16\pi M^2$. In contrast, the transverse dimensions are non-compact in the case
   of Rindler coordinates and the horizon at $l=0$ has infinite transverse area. 
   (v) There is a non trivial, time dependent, dynamics in the black hole manifold
   which is not easy to see in the Schwarzschild coordinates but is obvious in the 
   Kruskal coordinates. The geometrical structure of the full manifold contains two asymptotically
   flat regions connected by a worm-hole like structure \cite{mtw}.

   Because of these
   features, the $(t,l)$ Schwarzschild  coordinate system has an intuitive appeal which Kruskal
   coordinate system lacks, in spite of the mathematical fact that Kruskal coordinate system
   is analogous to the inertial coordinate system while the Schwarzschild coordinate system is 
   like the Rindler coordinate system. 
   
   The third  spacetime listed in  Table~\ref{table:metricprop}
    is the De Sitter spacetime which, again,
   admits a Schwarzschild type coordinate system and a Kruskal type coordinate system.
   The horizon is now at $l=H^{-1}$ and the spacetime is \emph{not} asymptotically flat. 
   There is also a reversal of the roles of ``inside'' and ``outside'' of the horizon in the case of De Sitter spacetime.
   If the Schwarzschild coordinates are used on the black hole manifold, an observer at large
   distances ($l \to \infty$)  from the horizon $(l=2M)$ will be stationed at nearly flat spacetime and will
   be confined to $\mathcal{R}$. The corresponding observer in the De Sitter spacetime is at $l=0$ which is 
  again  in $\mathcal{R}$. Thus the nearly inertial observer in the De Sitter manifold is near
    the origin, ``inside'' the horizon, while the nearly inertial observer in the black hole manifold is 
    at a large distance from the horizon and is ``outside'' the horizon; but both are located in the region
$\mathcal{R}$ in figure \ref{hyperfig} making this figure to be of universal applicability to all these three metrics. The transverse dimensions
    are compact  in the case of De Sitter manifold as well. 
    
    The De Sitter manifold, however, has a high degree of symmetry and in particular, homogeneity
    \cite{Kim:2002uz,Hawking:1973el}.
    It is therefore possible to obtain  a metric of the kind given in Table~\ref{table:metricprop} with any point 
    on the manifold as the origin. (This is in contrast with the black hole manifold where
    the origin is fixed by the source singularity and the manifold is not homogeneous.)
    The horizon is different for different observers  thereby introducing an observer 
    dependence into the description. This is not of any deep significance in the approach we have adopted, since we have always 
defined the horizon
with respect a family of observers.

    It is certainly possible to provide \emph{a} purely geometrical definition  of horizon in some spacetimes
    like, for example, the Schwarzschild spacetime. 
    The boundary of the causal past of the future time-like infinity in Schwarzschild spacetime
    will provide an intrinsic definition of horizon. But there exists time-like curves (like
those of  observers who fall into the black holes) for which this horizon
    does not block information. The comments made above should be viewed in the light
    of whether it is physically relevant and \emph{necessary} to define horizons as geometric
    entities rather than whether it is \emph{possible} to do so in certain spacetimes. 
    In fact,  a purely geometric definition of horizon actually hides
    certain physically interesting features.
    It is better to define horizons with respect to a family of observers (congruence of timelike
    curves) as we have done. 

As an aside, it may be noted that our definition of horizon (``causal horizon'')
is  more general than that used in the case of  black hole spacetimes etc. in the following sense:
 (a) these causal horizons are always present in any space-time for suitable choice of observers and (b) there is no notion
of any ``marginally trapped surfaces'' involved in their definition.
 There is also no restriction on the
topology of the two-dimensional surfaces (suitably defined sections of
the boundary of causal past). Essentially, the usual black hole horizons are
causal horizons but not conversely.  For our purpose, the causal horizon defined 
in the manner
described earlier turns out to be most appropriate. This is because it provides a notion
of regions in spacetimes which are not accessible to a \emph{particular class of observers}
and changes with the class of observers  under consideration. While more geometrical
notions of horizons defined without using a class of observers definitely have
their place in the theory, the causal horizon incorporates structures like Rindler horizon
which, as we shall see, prove to be very useful. We stress that, though causal horizons
depend on the family of time like curves which we have chosen 
--- and thus is foliation dependent ---  it is generally covariant.
Ultimately, definitions of horizons are dictated by their utility in discussing the issue we are 
interested in and for our discussion causal horizon serves this purpose best.

 While the three  metrics in Table \ref{table:metricprop} act as prototypes in our discussion,  with sufficient amount of similarities {\it and 
  differences} between them, most of our results are applicable to more general
  situations. The key features which could be extracted from the above examples are the following:
  There is a Killing vector field which is timelike  in part of the manifold with the components
  $\xi^a =(1,0,0,0)$ in the Schwarzschild-type static coordinates.
   The norm of this field $\xi^a\xi_a$ vanishes on the horizon which 
  arises as a bifurcation surface $\mathcal{H}$.
  Hence, the points of $\mathcal{H}$ are fixed points of the killing field.  
   There exists a spacelike hypersurface $\Sigma$
  which includes $\mathcal{H}$ and is divided by $\mathcal{H}$ into two pieces $\Sigma_R$ and 
 $ \Sigma_L$, the intersection of which is in fact $\mathcal{H}$. 
  (In the case of black hole manifold, $\Sigma$ is the $T=0$ surface, $\Sigma_R$ and $\Sigma_L$
  are parts of it in the right and left wedges and $\mathcal{H}$ corresponds to the $l=2M$ surface.)
  The topology of $\Sigma_R$ and $\mathcal{H}$ depends on the details of the spacetime
  but $\mathcal{H}$ is assumed to have a non-zero surface gravity. Given this structure
  it is possible to generalise most of the results we discuss in the coming Sections.

 The analysis in Section \ref{infinitez} shows that it is possible to associate a temperature with
 each of these horizons.
   In the case of a black hole manifold, an observer at $l=R\gg 2M$ will detect radiation at late 
   times $(t\to \infty)$ which originated from near the horizon $l=2M$ at early times. This radiation will
   have a temperature $T = (\kappa/2\pi) = (1/8\pi M)$ \cite{Hawking:1975sw}. In the case of De Sitter spacetime, an observer 
   near  the \emph{origin}
   will detect radiation at late times which originated from near the horizon at $l= H^{-1}$. The temperature
   in this case will be $T=(H/2\pi)$ \cite{Gibbons:1977mu}. 
   In each of the cases, 
   the temperature of this radiation, $T=\kappa/2\pi$, is determined by the surface gravity of the horizon.

  \section{Quantum field theory in singular gauges and thermal ambience}\label{heuristics}

   Horizons  introduce new features in quantum theory as one proceeds from non-relativistic
   quantum mechanics (NRQM) to relativistic quantum theory.  
    NRQM has a notion of  absolute time 
    $t$ (with only $t\to c_1t + c_2, \;c_1>0$ being the allowed  symmetry transformation)
    and exhibits invariance under
the Galilean group.
In the path integral representation of non-relativistic quantum mechanics,
one uses only the causal paths $X^\alpha(t)$ which ``go forward"
in this absolute time coordinate $t$.

This restriction has to be lifted in special relativity and the corresponding
path integrals use paths $X^a(s)=(X^0(s),X^\alpha(s))$, which go forward
in the proper-time $s$ but either forward or backward in coordinate time $X^0$.
In the path integral, this requires summing over paths which could intersect
the $X^0=$ constant plane on several points, going forwards and backwards.
For such a path, the particle could be located at infinitely many points on the $X^0=$ constant hypersurface,
which is equivalent to having a many-particle state at any given time $X^0$.
So if we demand a description in which causality is maintained and information on the 
$X^0=$ constant hypersurface could be used to predict the future, such a description
should be based on  a system  which
is mathematically equivalent to infinite number of non relativistic point particles,
located at different spatial locations, at any given time. Thus
combining  special relativity, quantum mechanics and 
causality {\it requires} the use of such constructs with infinite number of degrees of freedom
and quantum fields are such constructs (see, for example, \cite{Feynman:1988}).
In the case of a free particle, this result is summarised by:
\begin{equation}
 G_F(Y,X)\equiv\int_0^\infty ds e^{-ims}\int\mathcal{D}Z^a e^{iA[Y,s;X,0]}
=\langle 0|T[\phi(Y)\phi(X)]|0\rangle
\label{gf}
\end{equation}
Here $A[Y,s;X,0]$ is the action for the relativistic particle to propagate from
 $X^a$ to $Y^a$ in the proper time
$s$ and the path integral is over all paths $Z^a(\tau)$ with these boundary conditions. The
integral over all values of $s$ (with the phase factor $\exp( -iEs)=\exp(-ims)$ corresponding to the 
energy $E=m$ conjugate to proper time $s$) gives the amplitude for the particle to propagate from $X^a$ to $Y^a$.
There is no notion of a quantum field in at this juncture;
the second equality shows that the same quantity can be expressed in terms of a field.

It should be stressed  that  $G_F(Y,X)\neq 0$ when $X^a$ and $Y^a$ are  separated by a spacelike interval;
   the propagation amplitude for a relativistic particle to cross  a light cone  (or horizon) is  \emph{non zero}
    in quantum
   field theory. Conventionally, this amplitude  is  reinterpreted in terms of particle-anti particle pairs.
There is a well-defined way of ensuring covariance under Lorentz transformations for this interpretation 
and since all inertial observers see the same light cone structure it is possible to construct a 
Lorentz invariant quantum field theory. 

The  description in Eq.~(\ref{gf}) is (too) closely tied to the existence of a global time coordinate $T$
   (and those obtained by a Lorentz transformation from that). 
   One can decompose the field operator $\phi(T,{\bf X})$ into
   positive frequency modes [which vary as $\exp(-i\Omega T)$] and negative frequency modes
   [which vary as $\exp(+i\Omega T)$] in  a Lorentz invariant manner and use
   corresponding creation and annihilation operators to define the vacuum state.
   Two observers related by a Lorentz transformation will assign different (Doppler shifted) frequencies  
   to the same mode but a positive frequency mode will always be seen as a positive frequency mode
   by any other inertial observer. 
    The quantum state $|0\rangle$ in Eq.~(\ref{gf}), interpreted as the vacuum state, is thus Lorentz invariant.
  There is also a  well defined way of implementing covariance under Lorentz transformation in the
   Hilbert space so that the expectation values are invariant. 
 The standard procedure for implementing a classical symmetry in quantum theory
   is to construct a unitary operator $U$ corresponding to the symmetry and change the states
   of the Hilbert space by $|\psi\rangle  \to U|\psi\rangle $ and change the operators by $O \to UOU^{-1}$
   so that the expectation values are unaltered.  This can be done in the case of  Lorentz transformations.

The next logical step will be 
to extend  these ideas to curvilinear coordinates
in flat spacetime (thereby extending the invariance group from Lorentz group to
general coordinate transformation group) and to curved spacetime.
Several difficulties arise when we try to do this.

  (i) If the background metric depends on time in a given coordinate system, then the quantum 
  field theory reduces to that in an external time dependent potential. In general, this will lead to
  production of particles by the time dependent background gravitational field. 
 On many occasions, like in the case of an expanding universe,
  this is considered a ``genuine'' physical effect \cite{Parker:1968mv,Parker:1969au}.
If, on the other hand, the metric is static
  in a given coordinate system, one would have expected that a vacuum state could be 
  well defined and no particle production can take place. This is true as long as the 
  spacetime admits a global timelike Killing vector field throughout the manifold. 
  If this is not the case, and  the Killing vector field is timelike in one region and 
  spacelike in another, then the situation becomes more complex. The usual examples
  are those with horizons where the norm of the Killing vector vanishes on the bifurcation
  surface which, in fact, acts as the horizon. 
  In general, it is possible to provide different  realizations of the
  algebra of commutators of field operators,  each of which
  will lead to a different quantum field theory. These different theories will
  be (in general) unitarily inequivalent and the corresponding quantum states will be elements of 
  different Hilbert spaces.
 If we want to introduce general covariance as a symmetry in quantum theory, we 
  need unitary operators which could act on the states in the Hilbert space.
  {\it This procedure, however,
   is impossible to implement.} Mathematically, the elements of general coordinate transformation
   group (which is an infinite dimensional Lie group) cannot be handled in the same way as the elements
   of Lorentz group (which can be obtained by exponentiating elements close to identity or as the products of such
exponentials). 
   
(ii) The standard QFT requires analytic continuation into complex plane of 
independent variables   for its definition.
 It is  conventional to provide a prescription such that 
  the  propagator $G_F$ propagates positive frequency modes 
of the field  forward in time and the negative
frequency modes backward in time. This can be done either (a) through  an $i\epsilon$ 
prescription or (b) by defining $G_F$ in the Euclidean sector and analytically continuing to Minkowski
spacetime. 
Both these procedures (implicitly) select a global time coordinate [more precisely
an equivalence class of time coordinates related by Lorentz transformations].
{\it This procedure is not generally covariant.} 
The analytic
continuation $t\to it$ and the general coordinate transformation $t\to f(t',{\bf x}')$ do not commute
and one obtains different quantum field theories in different coordinate systems. 

(iii) One can also  define $G_F$ as a solution to a differential 
 equation, but $ \langle \psi|T[\phi(Y)\phi(X)]|\psi\rangle$ for any state $|\psi>$
satisfies the same differential  equation and the hyperbolic nature of this wave equation requires 
additional prescription to choose
the appropriate $G_F$. This can be done by the methods (a) or (b) mentioned  in 
(ii) above, in case of inertial frames in flat spacetime. But in curvilinear coordinate system 
or in curved spacetime,
this wave operator defining $G_F$ can 
be ill-defined at coordinate singularities (like horizons) and one requires  extra prescriptions to handle this.

We shall now study several explicit manifestations of these difficulties, their resolutions and the physical consequences.

\subsection{Singular gauge transformations and horizon}\label{qftstart}

  In many  manifolds with horizon, like those discussed in Section  \ref{stexample},
  one can usually introduce a global coordinate system covering the full manifold
  in which the metric is non singular though (possibly) not static. A clear example is the
  Kruskal coordinate system in the black hole manifold in which the metric depends on the
  Kruskal time coordinate.
  Quantum field theory in such a coordinate system will require working with a time dependent
  Hamiltonian; no natural vacuum state exists on such a global manifold because of this
  time dependence. 
  
  Many of these manifolds also allow transformation to another coordinate system
  (like the Schwarzschild coordinate system) in which the metric is independent of the new 
  time coordinate. There exists a well defined family of observers who will be using this coordinate
system and the question arises as to how they will describe the quantum field theory.
 The metric in the new coordinates is singular on the horizon and we need to ask how that singularity
needs to regularised and interpreted.
 These singular coordinate transformations require careful, special
  handling since they cannot be obtained by ``exponentiating'' infinitesimal, non-singular
  coordinate transformations.

 To see this issue clearly, it is better to use the concept of gauge transformations
rather than coordinate transformations. In the standard language of general relativity, one has a manifold with a metric
and different choices can be made for the coordinate charts on the manifold. When one changes from a coordinate chart $x^i$
to $\bar x^i$, the metric coefficients (and other tensors) change in a specified manner. In the language of particle physics,
 the same effect will be phrased differently. The coordinate chart and a back ground metric can be fixed at some fiducial value at first; the theory is then seen to be invariant under some infinitesimal
  transformations $g_{ij}\to g_{ij}+\delta g_{ij}$ where $\delta g_{ij}$ can be expressed in terms of four gauge functions $\xi^a(x)$ by 
$\delta g_{ij} = - \nabla_i\xi_j - \nabla_j\xi_i$. The translation between the two languages is effected by noticing that
the infinitesimal coordinate transformation $x^i\to x^i+\xi^i(x)$ will lead to the same $\delta g_{ab}$ in the general relativistic language. 

It is now clear that
  there are two separate types of gauge (or coordinate) transformations which we need to consider:
the {\it infinitesimal} ones and the {\it large} ones.
 The {\it infinitesimal} gauge transformations of the theory,
 induced by the four gauge functions $\xi^i$  have the form
$\delta g_{ij} = - \nabla_i\xi_j - \nabla_j\xi_i$.
For example, the transformation induced by $\xi^a_{(R)}=(-\kappa XT,-(1/2)\kappa T^2,0,0)$
changes the flat space-time metric $g_{ab}=(-1,1,1,1)$ to the form
$g_{ab}=(-(1+2\kappa X),1,1,1)$, up to first order in $\xi$. This could be naively thought of as the infinitesimal
version of the transformation to the accelerated frame. (It  is naive because the ``small'' parameters
here are $(\kappa X, \kappa T)$ and  we run into 
trouble at large $(X,T)$.)  Obviously, 
one cannot describe a situation in which $N\to 0$
within the class of infinitesimal transformations.

The classical theory is also invariant under finite transformations,
which  are more ``dangerous".
Of particular importance are the {\it large gauge transformations},
which are capable of changing $N>0$ in  a non-singular coordinate system 
to a nontrivial function $N(x^a)$ that vanishes on a hypersurface.
The transformation
 from the $(T,{\bf X})$ to the
Schwarzschild type coordinates belongs to precisely this class.
In particular, the coordinate transformation which
changes the metric from $g_{ab}=(-1,1,1,1)$ to $g_{ab}=(-(1+\kappa X)^2,1,1,1)$ is the ``large" version of
the infinitesimal version  generated by $\xi^a_{(R)}$.
Given such large gauge transformations,
we can discuss regions arbitrarily close to the $N=0$ surface.

 A new issue, which is conceptually important, comes up while doing quantum field theory in  a spacetime
  with a $N=0$ surface.
  All physically relevant results in the spacetime will depend on the 
  combination $Ndt$ rather than on the coordinate time $dt$. The Euclidean rotation $t\to t e^{i\pi/2}$
  can equivalently be thought of as the rotation $N \to N e^{i\pi/2}$. This procedure 
  becomes ambiguous on the horizon at which $N=0$. But the family of observers with a horizon, {\it will}
indeed be using a comoving co-ordinate system in which $N\to 0$ on the horizon.
 Clearly we need a new physical principle to handle quantum field theory as seen by this family of observers.

  To resolve this ambiguity, it is necessary to work in   complex plane in which the metric 
  singularity can be avoided.
  This, in turn, can be done either by analytically continuing in the time coordinate $t$
  or in the space coordinate $x$. The first procedure of analytically continuing in $t$ is well known
  in quantum field theory but not the second one since one rarely works with space dependent
  Hamiltonian in standard quantum field theory. We shall briefly describe these two procedures
  and use them in the coming Sections.

 Let us consider what happens to the coordinate transformations 
  in Eq.~(\ref{keytransform}) and the metric near the horizon, when the  analytic continuation
 $T\to T_E= T e^{i\pi/2}$ is performed. The 
      hyperbolic trajectory in Eq.~(\ref{trajectory}) for $N=1$ (for which $t$ measures the proper time), is given in parametric form as $\kappa T=\sinh\kappa  t, \kappa  X= \cosh \kappa t$. 
This becomes a 
      circle, $\kappa T_E=\sin \kappa  t_E, \kappa X= \cos \kappa t_E$ with, $-\infty<t_E<+\infty$
      on analytically continuing in
     both $T$ and $t$. 
      The  mapping $\kappa T_E=\sin\kappa  t_E$ is many-to-one
and limits the range of $\kappa T_E$  to $\kappa |T_E|\le1$ for $(-\infty < t_E < \infty)$.

Further, the \emph{complex plane probes the region which is classically inaccessible} to the family of observers on $N=$ constant trajectory.
The transformations in (\ref{trajectory}) with $N>0, -\infty<t<\infty$ cover \emph{only} the right
    hand wedge [$|X|>|T|, X>0$] of the Lorentzian sector; one needs to take $N<0, -\infty<t<\infty$ to cover the
    left hand wedge  [$|X|>|T|, X<0$]. 
     Nevertheless, \emph{both $X>0$ and $X<0$} are covered by different ranges of the ``angular" coordinate $t_E.$ The range $(-\pi/2) < at_E <(\pi/2)$ covers $X>0$ while the range $(\pi/2) < at_E <(3\pi/2)$
    covers $X<0$.  
     The light cones of the inertial frame $X^2=T^2$ are mapped into the origin of the $T_E-X$ plane. The region ``inside" the horizon
    $|T|>|X|$ simply \emph{disappears} in the Euclidean sector.  
   Mathematically, 
    Eq.~(\ref{keytransform}) shows that
     $\kappa t \to\kappa  t-i\pi$ changes $X$ to $-X$, ie., the
       complex plane 
        contains information about the physics beyond the horizons
        through   imaginary values of $t$. 

This fact is used in one way or another in several derivations of the temperature associated with the horizon 
\cite{Damour:76,Hartle:1976tp,Gibbons:1977mu,Gibbons:1977ue,Parikh:1999mf,Shankaranarayanan:2000gb,Shankaranarayanan:2000qv,Schutzhold:2000ju}.
     Performing this operation twice shows that 
     $\kappa t \to \kappa t - 2i\pi$ is an identity transformation implying periodicity in 
     the imaginary time $i\kappa t =\kappa  t_E$. 
    More generally, all the 
 events $\mathcal{P}_n\equiv(t=(2\pi n/\kappa ),{\bf x})$  [where $n = 
    \pm 1,\pm 2, ...$] which
    correspond to {\it different} values of $T$ and $X$  will be mapped to the {\it same} point in the 
    Euclidean space. 
    
    This 
    feature arises naturally
   when we analytically continue in the time coordinate $t$ to the Euclidean sector. If we take $t_E=it$, then
   the metric near the horizon becomes:
   \begin{equation}
   ds^2 \approx N^2 dt_E^2 + (dN/\kappa)^2 +dL^2 
   \label{horizonmetric}
   \end{equation}
   Near the origin of the $t_E-N$ plane, this is the metric on the surface of a cone.
   The conical singularity at the origin can be regularised  by assuming that $t_E$ is an angular
   coordinate with $0<\kappa t_E\leq 2\pi$. 
   When we analytically continue in $t$ and map 
   the $N=0$ surface  to the origin of the Euclidean plane,
  the ambiguity of defining $Ndt$ on the horizon  becomes similar to the ambiguity in defining the $\theta$ direction of the polar coordinates
  at the origin of the plane. This  can be resolved by imposing the periodicity in the angular coordinate
  (which, in the present case, is the imaginary time coordinate). 
  
  This procedure of mapping $N=0$ surface to the origin of Euclidean plane will play an important role in later discussion (see Section \ref{thermoroute}). To see its role in a broader context,
 let us consider a class of observers who have a horizon. 
A natural interpretation of general covariance will require that these observers will be able to formulate quantum field theory entirely in terms of an ``effective''
spacetime manifold  made of  regions which are accessible to them. Further, since the quantum field theory is well defined only in the Euclidean sector [or with an $i \epsilon$ prescription] it is necessary
to construct an effective spacetime manifold \emph{in the Euclidean sector} by removing the part
   of the manifold which is hidden by the horizon. For a wide class of metrics with horizon, the metric
   close to the horizon can be approximated by Eq.~(\ref{horizonmetric}) in which
   (the region inside) the horizon is reduced to a point which we take to be the origin. The region
   close to the origin can be described in Cartesian coordinates (which correspond to
   the freely falling observers)
or in polar coordinates  (which would correspond to observers at rest in a Schwarzschild-type coordinates) in the Euclidean space.
   The effective manifold
   for the observers with horizon can now be thought to be the Euclidean manifold with the
   origin removed. This principle is of very broad validity since it only uses the form of the 
   metric very close to the horizon where it is universal. 
   The structure of the metric far away from the origin can be quite complicated (there could even
   be another horizon elsewhere) but the key topological features are independent of this structure.
  It seems reasonable, therefore, to  postulate that the physics of the horizons need to be tackled by using an effective manifold, the topology of which is non trivial because a point (corresponding to the region blocked by the horizon) is removed. We will pursue this idea further in Section \ref{thermoroute} and show how it leads to a deeper understanding of the link between gravity and thermodynamics.
   
There is a second, equivalent, alternative for defining the theories in singular static manifolds. This is to note that
  the Euclidean rotation is equivalent to the $i\epsilon$ prescription in which one uses the transformation
  $t \to t (1+i\epsilon)$ which, in turn, translates to $N\to N(1+i\epsilon)$. Expanding this out, we get 
  \begin{equation}
  N \to N + i \epsilon \ {\rm sign}(N)
  \end{equation}
  Near the origin, the above transformation
  is equivalent to 
 $l\to l(1+i\epsilon)$ [since $l\propto N^2$]. Hence,
  \begin{equation}
   l \to l + i \epsilon \ {\rm sign}(l)
  \label{prescriptionx}
  \end{equation}
  This procedure involves analytic continuation in the \emph{space} coordinate $N$
  while the first procedure uses analytic continuation in the time coordinate. Both the 
  procedures will lead to identical conclusions but in different manners. We shall now explore
  how this arises.
 
 \subsection{Propagators in singular gauges}\label{propsing}
 
\begin{figure}[htbp] 
\begin{center}
\includegraphics[scale=0.5]{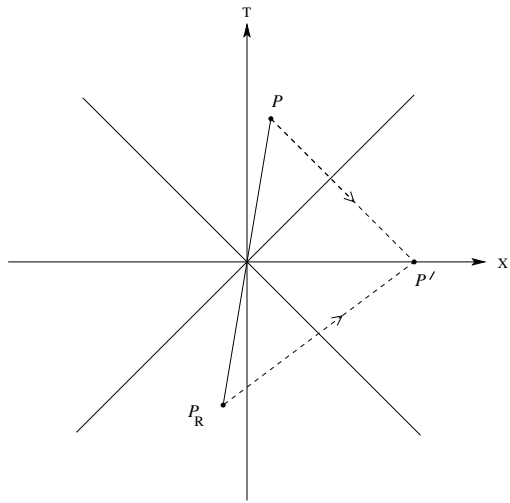}
    \caption{The relation between absorption and emission probabilities across the horizon. See text for details.}
\label{complexhorizon}
\end{center}
\end{figure}

  Let us begin by 
   computing 
     the amplitude for a particle to propagate from an event $ P $
     to another event $ P '$ with an energy $E$ \cite{Hartle:1976tp}. From the general 
     principles of quantum mechanics, this is given by the Fourier transform
     of the Green's function $G_F[ P  \to  P ']$ with respect to 
     the time coordinate. The vital question, of course, is which time coordinate is used as 
     a conjugate variable to energy $E$. Consider, for example, 
     the flat spacetime situation with $ P '  $ being some point on $T=t=0$ axis
     in $\mathcal{R}$ and $ P  $ being some event in $\mathcal{F}$
     with the \emph{Rindler} coordinates ($t, l,0,0$). 
     The amplitude $G_F[ P  \to  P ']$
     will now correspond to a particle propagating from the inside of the horizon
     to the outside. (See Fig. \ref{complexhorizon}; the fact that this amplitude  is non zero in quantum field theory is a
     necessary condition for the rest of the argument.)
     The amplitude for this propagation to take place with the particle
     having an energy $E$ --- when measured with respect to the 
      \emph{Rindler time coordinate} ---  is given by 
     \begin{equation}
     \mathcal{Q}(E; P  \to  P ') = \int_{-\infty}^\infty dt \, e^{-iEt} 
     G_F[ P  (t,{\bf y})\to  P ' (0,{\bf x})]
    \end{equation}
    (The notation in the left hand side should be interpreted as being {\it defined} by the right hand side; obviously, the \emph{events} $P$ and $P'$ can be specified only when the time coordinate is fixed but we are integrating over the time coordinate to obtain the corresponding amplitude in the energy space. The minus sign in $\exp(-iEt)$ is due to the fact that $t$ is the time coordinate of the
    {\it initial} event $ P $.)
    Shifting the integration by $t\to t-i(\pi/\kappa )$ in the integral we will pick up
    a pre-factor $\exp(-\pi E/\kappa )$; further, the event $ P $ will become the event
    $ P _R$ obtained by reflection at the origin of the inertial coordinates
     [see Eqs.~(\ref{expone}),(\ref{exptwo})].
    We thus get 
     \begin{equation}
     \mathcal{Q}(E; P  \to  P ') = e^{-\pi E/\kappa }  \int_{-\infty}^\infty dt \, e^{-iEt} 
     G_F[ P _R\to  P ' ] 
     = e^{-\pi E/\kappa } \mathcal{Q} (E; P _R \to  P ')
    \end{equation}
    The reflected event $ P _R$ is  in the region $ \mathcal{P} $; the amplitude
    $\mathcal{Q} (E ; P _R \to  P ')$ corresponds to the emission of a particle
    by the past horizon (``white hole'' in the case of Schwarzschild spacetime) into the region
    $\mathcal{R}$. By time reversal invariance, the corresponding probability is also 
    the same as the probability $P_{\rm abs}$ for the black hole to absorb  a particle.
    It follows that the probability for emission and absorption of a particle with energy $E$ across
    the horizon are related by 
    \begin{equation}
    P_{\rm em} = P_{\rm abs} \exp \left(-\frac{2\pi E}{\kappa }\right)
    \end{equation}
     This result can be 
    directly generalised to any other horizon since the ingredients which we have used are
    common to all of them. The translation in time coordinates $t\to t-i(\pi/\kappa )$ requires
    analyticity in a strip of width ($\pi/\kappa $) in the complex plane but this can be proved
    in quite general terms.

    The fact that the propagation amplitudes between two events in flat spacetime
    can bear an exponential relationship is quite unusual. The crucial feature  is that
    the relevant amplitude is defined at constant energy $E$, which in turn involves
    Fourier transform of the Green's function with respect to the Rindler time coordinate $t$.
    It is this fact which leads to the Boltzmann factor in virtually every derivation we will
    discuss.

  To see this result more explicitly, 
let us ask how the amplitude in Eq. (\ref{gf}) in flat spacetime will be viewed by observers following the trajectories in Eq.~(\ref{trajectory}) for  $N=1$. For mathematical simplicity, let us consider a massless particle, for which $G_F(Y,X)=- (4\pi^2)^{-1} [s^2(Y,X)- i\epsilon]^{-1}$ where
$s(Y,X)$ is the spacetime  interval between the two events. Consider now $G_F(Y,X)$ between two events along the trajectory in Eq.~(\ref{trajectory}) with $N=1$. Treating $G_F(Y,X)$ as a scalar, we find that
\begin{equation}
G_{\rm F}(Y(t),X(t'))=-\frac{1}{4\pi^2}
 \frac{(\kappa/ 2)^2}{
{\rm sinh}^{2}\left[\kappa (t-t')/2\right]
-i\epsilon}
\label{scalaramp}
\end{equation}
The first striking feature of this amplitude is that it is periodic in the imaginary time under
the change  $it\to it+2\pi/\kappa $ which arises from the fact that Eq.(\ref{trajectory}) has this property. In the limit of $\kappa\to 0$, the $G_F$ is proportional to  $[(t-t')^2 -i\epsilon]^{-1}$
which is the usual result in inertial coordinates. 
Next,
using the series expansion for ${\rm cosech}^2z$, we see that the
 propagator in Eq.~(\ref{scalaramp})  can be expressed as a series:
\begin{equation}
 G_F(\tau)=-
\frac{1}{4\pi^2}
\sum_{n=-\infty}^{n=\infty} 
\left[\left(\tau+2\pi i n\kappa^{-1}\right)^{2}
-i\epsilon\right]^{-1}
\label{trouble}
\end{equation}
where $\tau=(t-t')$. The $n=0$ term corresponds to the inertial propagator (for $\kappa=0$) and the other
terms describe the new effects.
 If we interpret the Fourier transform of $G(t-t')$ 
as the amplitude for propagation in energy space,  Eq.(\ref{trouble}) will give an amplitude
\begin{equation}
\Delta G(|E|)\equiv\int_{-\infty}^{+\infty}d\tau e^{iE\tau }\Delta G(\tau) 
= \frac{1}{2\pi} \frac{|E|}{\exp(\beta |E|) -1}
\label{planckone}
\end{equation}
in which the $\Delta$  indicates that the $n=0$ term has been dropped.
  
  The new feature which has come about is the following: In computing $G_F( P , P ')$ 
  using Eq.~(\ref{gf}) we sum over paths which traverses all over the $X-T$ plane
  even though the two events  are in the right wedge.
  The paths which have contributed in Eq.~(\ref{gf})
  do criss-cross the horizon several times even though the region beyond the horizon
  is inaccessible to the observers following the trajectories in Eq.~(\ref{trajectory}).
  The net effect of paths crossing the horizon leads to the extra term in Eq.~(\ref{planckone}).
  In fact, the $n>0$ terms in Eq.~(\ref{trouble}) contribute for $E<0$, while  $n<0$ terms contribute for
  $E>0$. The result in Eq.~(\ref{planckone}) also shows that 
  \begin{equation}
    \frac{\Delta G(|E|)}{\Delta G(-|E|)}=\exp(-\beta |E|)
    \end{equation}
    which can be interpreted as the probability for a particle to cross the horizon in two different directions.

  These features emerges more dramatically in the Euclidean sector \cite{Christensen:1978tw,Troost:1978yk,Padmanabhan:2003dc}. The Euclidean Green's function is $G_E\propto R^{-2}$
  where $R^2$ is the Euclidean distance between the two points. To express the same
  Euclidean Green's function in terms of $t$ and $t'$, we need to analytically continue
  in $t$ as well by  $t\to t_E= t e^{i\pi/2}$.
 The Green's function now becomes, in terms of $t_E,t_E'$, 
    \begin{equation}
G_{\rm E}(Y_E(t_E),X_E(t_E'))
=\frac{1}{4\pi^2}
 \frac{(\kappa  /2)^2}{\sin^{2}\left[\kappa (t_E-t_E')/2\right]}
\end{equation}
 and can be expressed as a series:   
    \begin{equation}
G_E(t_E-t_E')=
\left(\frac{\kappa^2}{4\pi^2}\right)
\sum_{n=-\infty}^{n=\infty} 
\left[\theta-\theta'+2\pi  n
\right]^{-2}
\label{troublee}
\end{equation}
with $\theta\equiv at_E$. Clearly, each term in the sum can be interpreted as due to a loop which
winds $n$ times around the circle of radius $x=1/\kappa $ in the $\theta$ direction.  
But note that
these winding paths go over the $X<0$ region of Minkowski space.    
 Paths which wind around the origin
   in the 
  Euclidean sector contains information about the region beyond the horizon
    (the left wedge) even though $x>0$. 

As we said before, these results emerge naturally once we realise that the physical theory (in this case the quantum field theory) should be formulated in an effective Euclidean manifold from which the region inaccessible to the chosen family of observers are removed. Here, this family is made of $N=$ constant observers and the inaccessible region corresponds to the origin of the Euclidean plane. The winding numbers for different paths as well as the fact that these paths probe the region beyond the horizon make the quantum field theory nontrivial.

 \subsection{Going around the horizon: complex plane}\label{complexplane}
  
  The above analysis involved analytic continuation in the time coordinate $t$ 
which allowed one to probe the region beyond the horizon, that was classically inaccessible in Re-$t$. As we discussed in Section \ref{qftstart}, the same results must also be obtainable from analytic continuation in $N$ since only the combination $Ndt$ is physically relevant. However, because $N\to 0$ on the horizon, we know
(see Section \ref{ftdim})
that the modes  which vary as $\exp[-i\omega t]$ 
   diverge on  the horizon. The analytic continuation in $N$ should  regularise and  interpret this behaviour
meaningfully.
  In particular, Eq.(\ref{planckone}) suggests that  the probability for a particle 
   with energy $E$ to go from $l=-\delta$ to $l =\delta$ should have an exponential dependence 
   in $\beta E$. It is interesting to see how this result can be interpreted in the ``bad'' coordinates $(t,x)$.
   
 This amplitude, for the outgoing mode $\phi_-$ in Eq.~(\ref{twotwo}), is given by  the ratio 
   $\mathcal{Q}=[\phi_-(\delta)/\phi_-(-\delta)]\approx (-1)^{-i\omega/\kappa}$
   which depends on the nature of the regulator used for defining this quantity.  
   For $l<0$, our prescription in Eq.~(\ref{prescriptionx}) requires us
   to interpret $l$ as having a small, negative imaginary part: $(l-i\epsilon)$.
   (The out-going mode with positive frequency $\phi_-= \exp -i\omega(t-\xi)\propto\exp i\omega\xi$
   is analytic in the upper half of complex-$\xi$ plane and will pick up contributions only from poles in the
   upper half; to obtain nonzero contribution we need to shift the pole
    from $l=0$ to $l=i\epsilon$ which is precisely the interpretation used above). This is same as
   moving along the $l$-axis in the lower half of the complex plane so that 
   $(-1)$ becomes $\exp (- i\pi)$. Then $\mathcal{Q}=\exp(-\omega(\pi/\kappa  ))$ and the probability
   is  $|\mathcal{Q}|^2=\exp(-\omega(2\pi/\kappa  ))=\exp(-\beta\omega)$
   which is the Boltzmann factor that  we would have expected.   
  
    More formally, the above result can be connected up with the concept of anti-particles
    in field theory being particles traveling ``backward in time'' \cite{Damour:76}. If we take $\phi_{-} $ as
    the outgoing particle state with positive frequency, then analytic continuation
    can be used to provide the corresponding anti-particle state.  The standard field theory
    rule is that, if $\phi_-(l)$   describes a particle state 
     $\phi_-(l-i\epsilon)$ will yield an anti particle state. Using the result 
     \begin{equation}
     (l-i\epsilon)^{-i\omega/\kappa }= l^{-i\omega/\kappa }\theta(l)+ |l|^{-i\omega/\kappa }e^{-\pi\omega/\kappa }\theta(-l)
     \label{xeps}
     \end{equation}
     this procedure 
     splits the wave into 
     two components which could be thought of as a particle-anti particle pair.
     The square of the relative weights of the two terms in the above equation, $e^{-2\pi\omega/\kappa }$ gives the Boltzmann factor. 
     In fact, this relation can be used to interpret the amplitude for a particle to go from inside the 
     horizon to outside in terms of a pair of particles being produced just outside the horizon
     with one falling into the horizon and the other escaping to infinity. 
     
     The analyticity arguments used above  contain the gist of thermal behaviour of horizons.  Since the positive frequency mode $\exp(-i\Omega U)$ (with $ \Omega>0$) is analytic in the lower half of complex $U\equiv (T-X)$ plane,
     any arbitrary superposition of  such modes with different (positive) values of $\Omega$ 
     will also be analytic
     in the  lower half of complex $U$ plane. Conversely, if we construct a mode which is 
     analytic in the lower half of complex $U$ plane, it can be expressed as a superposition of purely positive 
frequency modes \cite{Unruh:1976db}.
   From the transformations in Eq.~(\ref{keytransform}), we find that the 
    positive frequency  wave mode near the horizon, $\phi=\exp(-i\omega u)$ can be expressed as
    $\phi\propto U^{i\omega/\kappa }$ for $U<0$. If we interpret this mode as $\phi\propto (U-i\epsilon)^{i\omega/\kappa }$
    then, this mode is analytic throughout the lower half of complex $U$ plane. Using Eq.~(\ref{xeps})
    with $l$ replaced by $U$, we can interpret the mode as
    \begin{equation}
     (U-i\epsilon)^{i\omega/\kappa } =
     \begin{cases}
    e^{[i(\omega/\kappa)\ln U]}&  \text{(for $U>0$)}\\
    e^{\pi\omega/\kappa}e^{[(i\omega/\kappa)\ln |U|) }&   \text{(for $U<0$)}
    \end{cases}
    \end{equation}  
    This interpretation
    of $\ln (-U)$ as $\ln|U| -i\pi=\kappa u-i\pi=\kappa t-\xi-i\pi$ is consistent with 
    the procedure adopted in Section \ref{propsing}, viz., using $\kappa t\to \kappa t-i\pi$ to go from $X>0$
    to $X<0$.

     Similar results arise in a more general context for \emph{any} system described by 
     a wave function $\Psi(t,l; E) = \exp[iA(t,l; E)]$ in the WKB approximation \cite{Padmanabhan:2004tz}.
      The dependence of the quantum mechanical
     probability $P(E) =|\Psi|^2$ on the energy $E$ can be quantified in terms of the derivative
     \begin{equation}
     \frac{\partial \ln P}{\partial E} \approx -\frac{\partial}{\partial E}2(\textrm{Im} A) = 
     -2 \textrm{Im} \left(\frac{\partial A}{\partial E}\right)
     \label{pofe}
     \end{equation}
     in which the dependence on $(t,l)$ is suppressed. 
     Under normal circumstances, action will be real in the leading order approximation
     and the imaginary part will vanish. (One well known example is in the case of tunneling
     in which the action acquires an imaginary part; Eq.~(\ref{pofe}) correctly describes
     the dependence of tunneling probability on the energy.)
     For any Hamiltonian system, the quantity $(\partial A/\partial E)$ can be set to a constant
     $t_0$ thereby determining the trajectory of the system: $(\partial A/\partial E)=-t_0$.
     Once the trajectory is known, this equation determines $t_0$ as a function of 
     $E$ [as well as $(t,l)$]. Hence we can write 
     \begin{equation}
     \frac{\partial \ln P}{\partial E} \approx 2\textrm{Im} \left[ t_0 (E)           \right]
     \label{imto}
     \end{equation}
     From the trajectory Eq.~(\ref{photonpath}) we note that $t_0(E)$
     can pick up an imaginary part if the trajectory of the system crosses the 
     horizon. In fact, since $\kappa t \to\kappa  t-i\pi$ changes $X$ to $-X$ 
     [see Eqs.~(\ref{expone},\ref{exptwo},\ref{keytransform})], the imaginary
     part is given by $(-\pi/\kappa  )$ leading to  $(\partial \ln P/\partial E) = -2\pi/\kappa $.
     Integrating, we find that  the probability for the trajectory of any system to cross the horizon,
     with the energy $E$ will be given by the Boltzmann
     factor 
     \begin{equation}
     P(E) \propto \exp \left[- \frac{2\pi}{\kappa } E\right] = P_0\exp \left[- \beta  E\right] 
     \end{equation}
      with temperature $T=\kappa /2\pi$. (For special cases of this general result see \cite{Keski-Vakkuri:1997xp} and references
      cited therein.)
      
      In obtaining the above result, we have treated $\kappa$ as a constant (which is determined by the background geometry) independent of $E$. A more interesting situation develops if the surface gravity of the horizon changes when some amount of energy crosses it. In that case, we should treat $\kappa=\kappa(E)$ and the above result generalises to
      \begin{equation}
     P(E) \propto \exp -\int  \frac{2\pi dE}{\kappa(E)} \equiv P(E_0)\exp [- (S(E)-S(E_0)] 
     \label{pofefromsofe}
     \end{equation}
 where $dS\equiv (2\pi/\kappa(E))dE=dE/T(E)$ is very suggestive of an entropy function. An explicit example in which this situation arises is in the case of a spherical shell of energy $E$ escaping from a
 a black hole of mass $M$. This changes the mass of the black hole to $(M-E)$ with the corresponding change in the surface gravity. The probability for this emission will be governed by the difference in the entropies
 $S(M)-S(M-E)$. When $E\ll M$ we recover the old result with 
 $S(M)-S(M-E)\approx (\partial S/\partial M)E=\beta E$. 
 (We shall say more about this in Section \ref{entropyhorizon}.)
  
  Finally, it is interesting to examine how these results relate to the more formal approach to quantum field theory.
       The relation between quantum field theories in 
   two sets of coordinates $(t,{\bf x})$ and $(T,{\bf X})$,
   related by  
        Eq.~(\ref{keytransform}),  with the metric being 
     static in the $(t,{\bf x})$ coordinates can be 
     described as follows:  
  Static nature suggests a natural decomposition of wave modes
     as 
     \begin{equation}
     \phi(t,{\bf x}) = \int d\omega  [a_\omega  f_\omega  ({\bf x}) e^{-i\omega t} 
     + a_\omega ^\dagger f_\omega ^* ({\bf x}) e^{i\omega t}]
     \end{equation}
     in $(t,{\bf x})$ coordinates.
     But, as we saw in Section \ref{ftdim},
     these modes are going to behave badly (as $N^{\pm i\omega/\kappa }$)
     near the horizon since the metric is singular near the horizon in these coordinates. 
     We could, however, expand $\phi(t,{\bf x})$ in terms of some other set of 
     modes $F_\nu (t,{\bf x})$ which are well behaved at the horizon. This could, for example, 
     be done by solving the wave equation in $(T,{\bf X})$ coordinates and rewriting the solution
     in terms of $(t,{\bf x})$. This gives an alternative expansion for the field:
     \begin{equation}
     \phi(t,{\bf x}) = \int d\nu [A_\nu F_\nu (t,{\bf x})  + A_\nu^\dagger F_\nu^* (t,{\bf x}) ]
     \end{equation}
     Both these sets of creation and annihilation operators define two 
     different vacuum states $a_\omega|0\rangle _a =0, A_\nu |0\rangle _A=0$.
     The modes $F_\nu(t,{\bf x})$ will contain both positive and negative frequency components
     with respect to $t$
     while the modes $f_\omega ({\bf x}) e^{-i\omega t}$ are pure positive frequency components.
     The positive and negative frequency components of $F_\nu(t,{\bf x})$ 
      can be extracted through the Fourier transforms:
        \begin{equation}
     \alpha_{\omega \nu} = \int_{-\infty}^\infty dt \ e^{i\omega t} F_\nu (t, {\bf x}_f); \quad
      \beta _{\omega \nu} = \int_{-\infty}^\infty dt \ e^{-i\omega t} F_\nu (t, {\bf x}_f)
     \label{bogo}      
     \end{equation}
     where ${\bf x}_f$ is some convenient fiducial location far away from the horizon.
     One can think of $|\alpha_{\omega \nu}|^2$ and $|\beta_{\omega \nu}|^2$ as
     similar to  unnormalised transmission and 
     reflection coefficients.  
    (They are very closely related to the Bogoliubov coefficients
     usually used to relate two sets of creation and annihilation operators.)
     The $a-$particles in the $|0\rangle _A$ state is determined by the quantity
     $|\beta_{\omega \nu}/\alpha_{\omega \nu}|^2$. If the particles are uncorrelated,
     then the normalised flux of out going particles will be 
     \begin{equation} 
     \mathcal{N} = \frac{|\beta_{\omega \nu}/\alpha_{\omega \nu}|^2}{1-|\beta_{\omega \nu}/\alpha_{\omega \nu}|^2}
     \label{bogon}
     \end{equation}
     If the $F$ modes are chosen to be regular near the horizon, varying as
     $\exp(-i \Omega  U)$ etc., then Eq.~(\ref{keytransform}) shows that
     $F_\nu(t,{\bf x}_f) \propto \exp(-i\Omega q e^{-\kappa t})$ etc.  The integrals
     in Eq.~(\ref{bogo}) again reduces to the Fourier transform of an exponentially
     redshifted wave and we get $|\beta_{\omega \nu}/\alpha_{\omega \nu}|^2 = e^{-\beta \omega}$
     and Eq.~(\ref{bogon}) leads to the Planck spectrum. 
     This is the quantum mechanical version of Eq.~(\ref{expodamp}) and Eq.~(\ref{planck}).

   When  we can
     use WKB approximation we can also set  $F_\nu(t,x) = \exp[iA_{\nu} (t,x)]$ in the integrals
     in Eq.~(\ref{bogo}) and use the saddle point approximation. The saddle
     point is to be determined by the condition 
     \begin{equation}
     \pm \omega + \frac{\partial A_{\nu}}{\partial t} =0
     \end{equation}
     where the upper sign is for $\alpha_{\omega\nu}$ and the lower sign is for $\beta_{\omega\nu}$.
     The upper sign corresponds to a saddle point trajectory with 
     energy $E=\omega$ but, for $\beta_{\omega\nu}$ we get the condition $E=-\omega$
     so that the trajectory has negative energy. Writing the  saddle
     point trajectory as  $x_{\pm} (t)$ 
     it is easy to show that 
     \begin{equation}
     |\alpha_{\omega\nu}|^2 = \exp\left[ -\textrm{2Im} \int_{x_+0}^{x_f} p_+(x) dx\right];\quad 
      |\beta_{\omega\nu}|^2 = \exp\left[ -\textrm{2Im} \int_{x_-0}^{x_f} p_-(x) dx\right]
      \label{alphabetakk}
      \end{equation}
      This result contains essentially the same mathematics as Eq.~(\ref{imto}) since
      one can relate the imaginary part of $t_0$ to the imaginary part of $p = (\partial A/\partial x)$
      through the HJ equation. 
     Since positive energies are allowed while negative energies are 
     classically forbidden, this will often lead to $|\alpha|^2 \approx 1$ and 
     $|\beta|^2$ to be an exponentially small number. 
     
     The same result arises when one studies the problem of 
     over-the-barrier reflection in the $(1/x^2)$ potential --- to which
     the field theory near the horizon can be mapped because of scale invariance --- using the method of complex paths [see, e.g., Eq. (A36) of \cite{Srinivasan:1998ty}]. While the literature in this subject often uses the term ``tunneling''
      [see eg., \cite{Parikh:1999mf,Medved:2002zj}]
     to describe the emergence of an imaginary part to $p, A$ etc., in the 
     context of horizons it is more appropriate to think of this process
     as ``over-the-barrier reflection''. 
     Both the processes are governed by an exponential involving an integral
     of $p(x)$ over $dx$.  In tunneling, $p(x)$ becomes imaginary when
     $p^2(x) \propto E-V(x)$ becomes negative. In the over the barrier 
     reflection, $E>V$ and the transmission coefficient remains close to unity
     because the process is classically allowed. The imaginary part, leading to an
     exponentially small reflection coefficient,  arises
     because one needs to analytically continue $x$ into the complex plane
     just as we have done \cite{llthree}. 
     In  Eq.~(\ref{imto})
     as well as in Eq.~(\ref{alphabetakk}) the imaginary part arises because the path
     $x(t)$ needs to be deformed into the complex plane \cite{Srinivasan:1998ty} rather than because the momentum
     $p$ becomes complex.

   \section{ Thermal Density Matrix from tracing over modes hidden by Horizon}\label{thermalden}

In the previous few Sections, we have derived the thermality of horizons from the geometry of the line element in the 
Euclidean spacetime. The key idea has been the elimination of the region inaccessible in Re-$t$
 to a family of  observers (the origin in the
Euclidean plane) and using Im-$t$ to probe these regions. 
If these ideas are consistent, the same effect should arise, when we construct the quantum field theory in the accessible region
(in $N>0$, say) by integrating out the information contained in $N<0$. That is, one family of observers may describe the quantum state in terms of a wave function $\Psi(f_L,f_R)$ which depends on the field modes both on the ``left'' ($N<0$) and
``right'' ($N>0$) sides of the horizon while another family of observers will describe the same system by a density matrix
obtained by integrating out the modes $f_L$ in the inaccessible region. We shall now show that this is indeed the case using
an adaptation of the analysis by 
\cite{lee}(also see, \cite{Unruh:1983ac}).
  
On the $T=t=0$ hypersurface
one can define a vacuum state  $|{\rm vac}\rangle $ of the theory by giving the field
configuration for the whole of $-\infty<X<+\infty$. This field configuration, however, separates
 into two disjoint sectors when one uses
the $(t,N)$ coordinate system.
Concentrating on the $(T,X)$ plane and suppressing $Y,Z$ coordinates in
the notation for simplicity, 
we now need to specify the field configuration
$\phi_R (X)$  for $X>0$ and $\phi_L (X)$ for $X<0$ to match the initial data in
the global coordinates; given this data, the vacuum state is specified by
the  functional $\langle {\rm vac}|\phi_L,\phi_R\rangle $.

\begin{figure}[htbp]
\begin{center}
\includegraphics[scale=0.5]{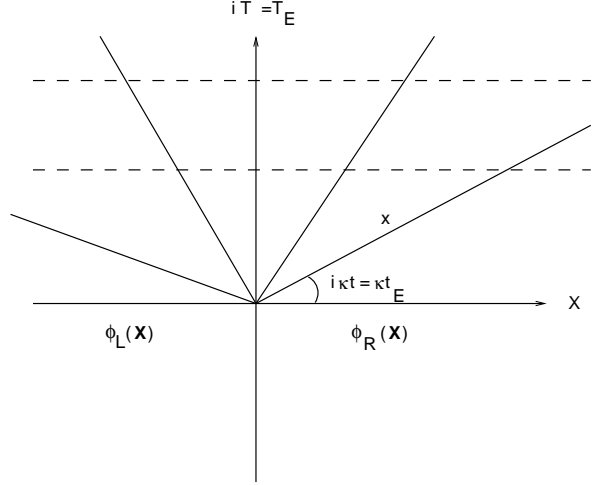}
\caption{Thermal effects due to a horizon; see text for a discussion.}
\label{fig:horizon}
\end{center}
\end{figure}

Let us next consider the {\it Euclidean} sector corresponding to the $(T_E,X)$ plane
where $T_E=iT$. The QFT in this plane can be defined along standard lines. The
analytic continuation in $t$, however, is a different matter; we see from Eq.~(\ref{horizonmetric}) that
 the coordinates
$(\kappa t_E=i\kappa t,x)$ are like polar coordinates in $(T,X)$ plane with $t_E$ having a periodicity of $(2\pi/\kappa )$. 
Figure~\ref{fig:horizon}  now shows that evolution
in $\kappa t_E$ from $0$ to $\pi$ will take the system configuration from $X>0$ to $X<0$. This allows one to prove that $\langle {\rm vac} |\phi_L , \phi_R \rangle \propto \langle \phi_L \vert e^{-\pi H/\kappa } \vert \phi_R \rangle$; normalisation now fixes the proportionality constant, giving 
\begin{equation}
\langle {\rm vac} |\phi_L , \phi_R \rangle = \frac{\langle \phi_L \vert e^{-\pi H/\kappa } \vert \phi_R \rangle}{ \left[ {\rm Tr}(e^{-2 \pi H/\kappa })\right]^{1/2} } \label{eqn:fifteen}
\end{equation}

To provide a simple  proof of this relation, let us consider
the ground state wave functional $\langle {\rm vac} |\phi_L , \phi_R\rangle $ in the extended spacetime
expressed as a path integral. The ground state wave functional can be represented
as a Euclidean path integral of the form
\begin{equation}
\langle {\rm vac} |\phi_L , \phi_R \rangle \propto \int_{T_E=0;\phi=(\phi_L,\phi_R)}^{T_E=\infty;\phi=(0,0)}
\mathcal{D}\phi e^{-A}\label{euclpath}
\end{equation}
  where $T_E=iT $ is the Euclidean time coordinate. From Fig.~\ref{fig:horizon}  it is obvious that
  this path integral could also be evaluated in the polar coordinates by varying the angle
  $\theta =\kappa  t_E$ from 0 to $\pi$. When $\theta =0$ the field configuration corresponds to 
  $\phi = \phi_R$ and when $\theta = \pi$ the field configuration corresponds to $\phi = \phi_L$.
  Therefore
  \begin{equation}
\langle {\rm vac} |\phi_L , \phi_R \rangle \propto \int_{\kappa t_E=0;\phi=\phi_R}^{\kappa t_E=\pi;\phi=\phi_L}
\mathcal{D}\phi e^{-A}
\label{rindact}
\end{equation}
   But in the
    Heisenberg picture, this path integral can be expressed as a matrix element of the Hamiltonian $H_R$
   (in the $(t,N)$ coordinates)  giving us the result:   
    \begin{equation}
\langle {\rm vac} |\phi_L , \phi_R \rangle \propto \int_{\kappa t_E=0;\phi=\phi_R}^{\kappa t_E=\pi;\phi=\phi_L}
\mathcal{D}\phi e^{-A} 
= \langle \phi_L|e^{-(\pi/\kappa )H_R}|\phi_R\rangle
\end{equation}
    Normalising the result properly gives  Eq.~(\ref{eqn:fifteen}).

This result, in turn, implies that for operators $\mathcal{O}$ made out of  variables 
having support on $\mathcal{R}$,   the vacuum expectation values 
$\langle{\rm vac} \vert \, \mathcal{O} (\phi_R)\vert {\rm vac}\rangle$
 become
thermal expectation values. This arises from straightforward algebra of inserting
a complete set of states appropriately:
\begin{eqnarray}
 \langle{\rm vac} \vert \, \mathcal{O} (\phi_R)\vert {\rm vac}\rangle 
 &=& \sum\limits_{\phi_L} \sum\limits_{\phi^1 _R, \phi^2_R} \langle {\rm vac} \vert \phi_L, \phi^1_R \rangle \langle \phi^1_R \vert \mathcal{O} (\phi_R)\vert \phi^2_R \rangle \langle \phi^2_R , \phi_L \vert {\rm vac} \rangle 
 \nonumber \\
& = &\sum\limits_{\phi_L} \sum\limits_{{\phi^1_R}, \phi^2_R} \frac{\langle \phi_L \vert e^{\frac{-\pi H_R}{\kappa }}\vert \phi^{1}_R \rangle \langle \phi^{1}_R \vert \mathcal{O} \vert \phi^2_R \rangle \langle \phi^2_R \vert e^{\frac{-\pi H_R}{\kappa} } \vert \phi_L \rangle}{Tr(e^{-2\pi H_R/\kappa })}\nonumber\\
&=& \frac{Tr (e^{-2\pi H_R/\kappa } \mathcal{O}) }{Tr(e^{-2\pi H_R/\kappa })} \label{eqn:twentytwo}
\end{eqnarray}
Thus, tracing over the field configuration $\phi_L$ behind the horizon leads to a thermal density
matrix $\rho \propto \exp[-(2\pi/\kappa )H]$ for observables in $\mathcal{R}$.

The main ingredients which have gone into this 
   result are the following. (i) The singular behaviour of the $(t,x)$ coordinate system
   near $x=0$ separates out the $T=0$ hypersurface into two separate regions.
   (ii) In terms of \emph{real} $(t,x)$ coordinates, it is not possible to distinguish between the points
   $(T,X)$ and $(-T,-X)$ but the \emph{complex} transformation $t\to t\pm i\pi$ maps the point
   $(T,X)$ to the point $(-T,-X)$. As usual, a rotation in the complex  plane (Re $t$, Im $t$)
   encodes the information contained in the full $T=0$ plane.

  The formalism developed above can be used to express $|{\rm vac}\rangle $ formally in terms of quantum states defined in $\mathcal{R}$ and $\mathcal{L}$.  
  It can be easily shown that
  \begin{equation}
  |{\rm vac}\rangle=\prod_{k_\perp,\omega}\sqrt{1-e^{-2\pi\omega/\kappa }}\sum_{n=0}^\infty|n\rangle_{\mathcal{R}}
|n\rangle_{\mathcal{L}}\;e^{-\pi n\omega/\kappa }
\label{minkvac}
\end{equation}
The result  in Eq.~(\ref{minkvac}) shows that when the vacuum state $|{\rm vac}\rangle $ is ``partitioned" by the horizon at $x=0$,
it can be expressed as a highly correlated combination of states defined in $\mathcal{R}$ and $\mathcal{L}$. While this result is suggestive, it is --- unfortunately --- somewhat formal. One can rigorously prove \cite{Gerlach:1989rz} that the states $|n\rangle $ on
either $\mathcal{R}$  or $\mathcal{L}$ are orthogonal to all the states of the standard Fock space of Minkowski quantum field theory.

   The  results in Eq.~(\ref{eqn:fifteen}) and Eq.~(\ref{eqn:twentytwo}) are completely general
   and we have not assumed any specific Lagrangian for the field.
   For \emph{free} field theories in static spacetimes, it is possible to give a more explicit demonstration
  of the fact that  the vacuum state appears as a thermal density matrix. To do this, we begin by noting
  that in any spacetime, with a metric which is independent of the time coordinate and
  $g_{0\alpha} =0$, the wave equation for a massive  scalar field $(\Box - m^2) \phi =0$
  can be separated in the form $\phi(t,{\bf x}) =  \psi_\omega({\bf x})e^{-i\omega t}$ with the 
  modes $\psi_\omega({\bf x})$  satisfying the equation  
  \begin{equation}
  \frac{|g_{00}|}{ \sqrt{-g}}\partial_\alpha (\sqrt{-g}\, g^{\alpha\beta} \partial_\beta \psi_\omega) =
  - \omega^2 \psi_\omega
  \end{equation}
  The normalisation may be chosen using the conserved scalar product:
  \begin{equation}
  (\psi_\omega, \psi_\nu) \equiv \int d^3x\, \sqrt{-g} |g^{00}| \psi_\omega \psi^*_\nu = \delta_{\omega\nu}
  \end{equation}
  Using this relation in the field equation, it can be easily deduced that 
  \begin{equation}
  \int d^3x\, \sqrt{-g} \partial_\alpha \psi_\omega^* \partial^\alpha \psi_\nu = \omega^2 \delta_{\omega\nu}
  \end{equation}
  Expanding the  field as $\phi(t,{\bf x}) =\sum_\omega q_\omega(t) \psi_\omega({\bf x})$ 
  and substituting into the free
  field  action,
  we find that the action reduces to that of a sum of harmonic oscillators: 
  \begin{equation}
  A =-\frac{1}{2} \int \sqrt{-g} ~dt~ d^3x~( \partial_a \phi \partial^a \phi+ m^2 \phi^2) \
   = \frac{1}{2} \sum_\omega \int dt 
  \left[ |\dot q_\omega|^2 - (\omega^2+m^2)|q_\omega|^2\right]
  \end{equation}
  Let us now apply this result to  the quantum field theory decomposed into oscillators in:
  (i) the   $(T,X)$  space
  as well as in (ii) the $(t,x)$ coordinate system on the right and (iii) the left hand side.
  
   On the $T=0$ surface,
  we expand the field in terms of a set of mode functions $F_\Omega(X,{\bf X}_\perp )$ with
  coefficients $Q_\Omega$; that is, $\phi= \sum_\Omega Q_\Omega F_\Omega(X,{\bf X}_\perp )$.
  Similarly, the field can be expanded in terms of a set of modes in  $\mathcal{R}$ and $\mathcal{L}$:
\begin{equation}
  \phi(X>0, {\bf X}_\perp) = \sum_\omega a_\omega f_\omega(X,{\bf X}_\perp );\quad
  \phi(X<0, {\bf X}_\perp) = \sum_\omega b_\omega g_\omega(X,{\bf X}_\perp ).
\end{equation}
  The functional integral in Eq.~(\ref{euclpath}) now reduces to product over a set of independent harmonic
   oscillators and thus the ground state wave functional can be
   expressed in the form
   \begin{equation}
   \label{gsomega}
  \Psi[Q] = \langle {\rm vac}|\phi({\bf X})\rangle 
  = \prod_\Omega \langle {\rm vac}|Q_\Omega\rangle 
  \propto \exp \left[- \sum_\omega A_E ( T_E = \infty, 0; T_E=0, Q_\Omega)\right] 
   \end{equation}
   where $A_E$ is the Euclidean action with the boundary conditions as indicated. 
   On the other hand, we have shown that this ground state functional is the same as 
   $\langle \phi_R, \kappa t_E=\pi|\phi_L,\kappa t_E=0\rangle $. Hence 
   \begin{equation}
    \label{gsfunc}
  \Psi[a,b]=\langle {\rm vac}|\phi({\bf X})\rangle 
     \propto  \exp\left[- \sum_\Omega A_E ( \kappa t_E=\pi, a_\omega; \kappa t_E=0, b_\omega)\right] 
   \end{equation}
   The Euclidean action for a harmonic oscillator  $q$ with boundary conditions $q=q_1$ at
$t_E=0$ and $q=q_2$ at $t_E=\beta$ is given by 
   \begin{equation}
   A_E(q_1,0; q_2,\beta) =\frac{\omega}{2} \left[ \frac{\cosh \omega\beta}{\sinh \omega\beta} (q_1^2 + q_2^2) - \frac{2q_1q_2}{\sinh \omega \beta}\right]
   \label{eucaction}
   \end{equation}
   Equation  (\ref{gsomega}) corresponds to $\beta =\infty, q_2 =0, q_1=Q_\Omega$ giving
   $A_E ( T_E= \infty, 0; T_E=0, Q_\Omega)] =(\Omega/2) Q_\Omega^2$ leading
   to the standard ground state wave functional.
   The more interesting one is, of course, the one in Eq.~(\ref{gsfunc}) corresponding to
   $\beta =(\pi/\kappa ), q_1=a_\omega, q_2 = b_\omega$.  This gives
   \begin{equation}
    A_E(a_\omega,0; b_\omega,(\pi/\kappa ))
     =\frac{\omega}{2} \left[ \frac{\cosh (\pi\omega/\kappa ) }{ \sinh (\pi\omega/\kappa )}
     (a_\omega^2 + b_\omega^2) - \frac{2a_\omega b_\omega }{ \sinh (\pi \omega/\kappa )}\right]
     \label{eaction}
   \end{equation}
  An observer confined to $\mathcal{R}$ will have observables made out of $a_\omega^,s$. Let
$ \mathcal{O} (a_\omega)$ be any such observable. The expectation value of $ \mathcal{O} $ 
 in the
state $ \Psi$ is given by
\begin{eqnarray}
 \langle \mathcal{O} \rangle  &=&\int \prod_\omega  da_\omega \int \prod\limits_\omega 
db_\omega \Psi^* (a_\omega, b_\omega) \mathcal{O} \Psi (a_\omega, b_\omega) 
\equiv\int \prod\limits_\omega da_\omega \rho (a_\omega, a_\omega) \mathcal{O} (a_\omega) \nonumber\\
&=&\textrm{Tr} (\rho \mathcal{O} ) 
\end{eqnarray}
 where
\begin{eqnarray}
 \rho(a'_\omega, a_\omega) &\equiv& \int \prod_\omega db_\omega \Psi^* (a'_\omega, b_\omega )
\Psi (a_\omega, b_\omega ) \\
& =&C  \exp - \sum_\omega  \left\{
\frac{\omega}{2} \left[ \frac{\cosh (2\pi\omega/\kappa ) }{ \sinh (2\pi\omega/\kappa )}
     (a_\omega^2 + a_\omega'^2) \right.\right.
      - \left.\left.\frac{2a_\omega a_\omega' }{\sinh (2\pi \omega/\kappa )}\right]
     \right\}\nonumber
     \label{rho}
\end{eqnarray}
 is a thermal density matrix corresponding to the temperature $T = (\kappa / 2 \pi )$. 
 
 The fact that the exponential in the density matrix in Eq.~(\ref{rho}) is similar to that in  Eq.~(\ref{eaction}),
 with $\pi$ replaced by $2\pi$,
 is noteworthy and this result can be obtained more directly from an alternative argument. 
  The matrix element of  $\rho $ can be expressed as the integral 
  \begin{equation}
  \langle \phi_R' | \rho |\phi''_R\rangle  = \int \mathcal{D} \phi_L \langle \phi_L\phi_R'|0\rangle  \langle 0|\phi_L\phi_R''\rangle 
  \label{rhonew}
  \end{equation}
   Each of the two terms in the integrand can be expressed in terms of $A_E$ using
   Eq.~(\ref{rindact}). In one of them, we shall take $\kappa t_E = \epsilon$ 
   (with $\epsilon$ being infinitesimal and positive)
   at the lower
   limit of the integral and in the other, we will take $\kappa t_E =-\epsilon$ at the 
   lower limit of the integral. Hence the product which occurs in the integrand of 
   Eq.~(\ref{rhonew}) can be thought of as evolving the field from a configuration $\phi_R''$
   at $\kappa t_E =+\epsilon$ to a configuration $\phi_R'$ at $\kappa t_E = -\epsilon$ 
   rotating in $\kappa t_E$ 
   in the anti clockwise
   direction from $\epsilon$ to $(2\pi -\epsilon)$. In the limit of $\epsilon \to 0$, this is same as 
   evolving the system  by  the angle $\kappa t_E = 2\pi$. 
   So we can set 
    $\beta = (2\pi/\kappa ), q_1 = a_\omega, q_2 = a'_\omega$ in Eq.~(\ref{eucaction}) leading to Eq.~(\ref{rho}).
   In arriving at equation Eq.~(\ref{eaction})
   we have evolved the same system from  $\kappa t_E =0$ to   $\kappa t_E = \pi$ in order to go from $x>0$ to $x<0$.
   This explains the correspondence between Eq.~(\ref{rho}) and  Eq.~(\ref{eaction}).

 To avoid misunderstanding, we stress that the temperature associated to a horizon
   is not directly related to  the question of what a given  non-inertial detector will measure. 
   In the case of 
   a uniformly accelerated detector in flat spacetime, it turns out that the detector results
   will match with the temperature of the horizon \cite{Davies:1975th,Unruh:1976db,Dewitt:1979}. There are, however,
   several other situations in which these two results do not match
\cite{Letaw:1981ik,Letaw:1981yv,Padmanabhan:1982apsci,Sriramkumar:1999nw}. The physics of a non 
   inertial detector is  well understood and there are no unresolved issues
   \cite{Grove:1983rp,Padmanabhan:1985}.

\section{ Asymptotically static horizons and Hawking radiation}\label{hawkrad}   

 The association of a temperature with a horizon, by itself, does not mean that the horizon radiates energy in an irreversible manner or that a black hole ``evaporates". In fact,  the metrics mentioned in 
Section~\ref{stexample} (leading to horizons and temperature) are all 
 trivially invariant under $t\to -t$.  
 The horizons in these  spacetimes exist ``forever''; the most natural vacuum states of the theory share this invariance  and describe a situation in thermal equilibrium. There is no net radiation flowing to regions far away from the horizon.
 
 A completely different class of physical
phenomena arises if the spacetime metric is time dependent, like, for example, in the case of an expanding universe.
 Then the natural choice of mode functions and the corresponding vacuum states at $t\to -\infty$
   and $t\to \infty$, usually called $|{\rm in}\rangle$ and $|{\rm out}\rangle$,
   will be different and the $|{\rm in}\rangle$  vacuum will contain
   ``out-particles''. In general,
the spectrum of particles produced will depend on the detailed nature of the time evolution.
The result will not have the same kind of universality as the results we have discussed so far and each case needs to be 
addressed separately. 

One important exception to this general rule is when the metric (in some coordinate system)
evolves
   from a geometry which has no horizon in the asymptotic past ($t\to -\infty$) to a geometry
   with a horizon in the asymptotic future ($t\to +\infty$). Then the late time behaviour of modes, in a coordinate system
appropriate for the family of observers who has a horizon, is exponentially redshifted and will lead to a thermal spectrum of particles. It must be stressed that we are now dealing with an explicitly time dependent situation, the physics of which is different from the static horizons discussed in the previous Sections.  Time reversal invariance
need not hold and there could be a genuine flow of created particles from one region to another.
  This can arise in different contexts, three of which are of primary interest to us because of their connection with the corresponding static metrics:
   
   (a) One can introduce coordinate systems in flat spacetimes which smoothly interpolates
   between inertial coordinates at $t\to -\infty$ to the Rindler coordinates at $t\to +\infty$.
   Such a coordinate system will appropriately describe a family of observers with time dependent 
   acceleration. The clock time $t$ of this observer with variable acceleration will match with
   inertial time coordinate in the asymptotic past and with the Rindler time coordinate in the
   asymptotic future and the metric will be static in both the limits. It is straightforward to show
   that the vacuum state in the asymptotic past, $|in\rangle$ will contain a thermal distribution
   of out-particles.
   
   (b) A spherically symmetric distribution of matter, collapsing and forming a black hole,
   represents another case in which the horizon develops asymptotically. A family of  observers at  
   constant (large) radii outside will notice a horizon forming as $t\to \infty$. The 
   vacuum state of the asymptotic past will be populated by a thermal distribution of out-particles
   in the future. 
   
   (c) The De Sitter spacetime also allows a time dependent generalisation which is most easily
   obtained by using the cosmological (Friedmann) coordinates to describe the De Sitter metric.
   In these coordinates, the dynamics of the spacetime is described in terms of an expansion
   factor $a(t)$. If $a(t)$ has a power law behaviour at small and moderate $t$ and evolves
   into $a(t) \to \exp(Ht)$ as $t\to \infty$, the geometry will describe a universe which is 
    asymptotically De Sitter. [There is some observational evidence to suggest
   that our universe is indeed evolving in this manner; for a review, see e.g., \cite{Padmanabhan:2002ji}.]
   
   Most of the techniques used in the previous Sections are not applicable when the spacetime is explicitly
   time dependent but the results based on infinite redshift will survive. We have seen
   in Section \ref{infinitez} that a wave mode undergoing exponential redshift
   can lead to a thermal distribution of particles. At late times and far away from the horizon,
   only modes which emanate from near the horizon at early times will contribute significantly.
   These modes would  have undergone exponential redshift in all the three cases described above
and will lead to a thermal spectrum.

   \subsection{Asymptotically Rindler observers in flat spacetime}\label{rindlerflat}
   
  Let us begin with the case of a time dependent Rindler  metric 
   in flat spacetime, which corresponds to 
    an observer who is moving with a
   variable acceleration \cite{Padmanabhan:1982apsci,Padmanabhan:2002ha}.
The transformation from the flat inertial coordinates $(T,{\bf X})$
to the proper coordinates $(t,{\bf x})$ of an observer with variable acceleration 
is effected by $Y=y, Z = z$ and
\begin{equation}
X = \int^{\prime} \sinh \mu (t) dt + x \cosh \mu (t); \quad
T = \int^{\prime} \cosh \mu (t) dt + x \sinh \mu (t)
\label{eqn:fiftyfive}
\end{equation}
where the function $\mu(t)$ is related to the time dependent acceleration
$g(t)$ by
$g(t) = (d \mu/ dt)$.
The form of the metric in the accelerated frame is remarkably simple:
\begin{equation}
ds^2 = - (1+g(t)x)^2 dt^2 + dx^2 + dy^2 + dz^2 
\label{eqn:fiftyseven}
\end{equation}
We will treat $g(t)$ to be an arbitrary function except for the limiting behaviour $g(t)\to 0$ for $t\to -\infty$ and $g(t)\to g_0$=constant for $t\to +\infty$. Hence, at early times, the line element in Eq.~(\ref{eqn:fiftyseven}) represent the standard inertial coordinates and the positive frequency modes $\exp(-i\omega t)$ define the standard Minkowski vacuum, $|in\rangle$. At late times, the metric goes over to the Rindler coordinates  and we are interested in knowing how the initial vacuum state will be interpreted at late times. The wave equation  $(\Box -m^2)\phi =0$ for a massive scalar field
can be separated in the transverse coordinates as
$ \phi(t,x,y,z)=f(t,x)e^{ik_y y} e^{ik_zz} $
where $f$ satisfies the equation
\begin{equation}
- \frac{1}{(1 +g(t)x)} \frac{\partial}{ \partial t} \left( \frac{1 }{ (1 + g(t)x)} \frac{\partial f }{ \partial x} \right) = \chi^2f \label{eqn:sixty}
\end{equation}
with $\chi^2\equiv m^2+k_y^2+k_z^2$. It is  possible to solve this partial differential equation 
with the ansatz
\begin{equation}
f(x, t) = \exp i \left( \int \alpha(t) dt + \beta (t) x \right)\label{eqn:sixtyone}
\end{equation}
where $\alpha$ and $\beta$ satisfy the equations
$\alpha^2(t) - \beta^2(t) = \chi^2 ; 
\dot \beta  = g(t) \alpha; \dot \alpha  = g(t) \beta; $
these are solved uniquely in terms of $\mu(t)$ to give
$ \alpha (t) = \chi \cosh [\mu (t) - \eta]; 
\beta (t) = \chi \sinh [\mu (t) - \eta ] $
where $\eta$ is another constant. The final solution  for the mode labelled by $({\bf k}_\perp,\eta)$ 
is now given by 
\begin{equation}
f_{k_yk_z \eta}(x, t)
 = \exp -i \chi \left[ \int \cosh (\mu - \eta) dt + x \sinh (\mu - \eta) \right] \label{eqn:sixtysix}
\end{equation}
For the limiting behaviour we have assumed for $g(t)$, we see that $\mu(t)$ vanishes at early times and varies as $\mu(t)\approx (g_0t+$constant) at late times. Correspondingly, the mode $f$ will behave as
\begin{equation}
f(x, t)
\rightarrow \exp -i \chi \left[ t \cosh \eta  - x \sinh \eta \right]  \label{eqn:sixtyseven}
\end{equation}
at early times $(t\to-\infty)$ which is just the standard Minkowski positive frequency mode with
$\omega=\chi \cosh\eta, k_x=\chi\sinh\eta$.  At late times the mode evolves to
\begin{equation}
f(x, t)\rightarrow \exp -i\left[ (\chi/2g_0)  (1+g_0x)e^{g_0t}\right] \label{eqn:sixtyeight}
\end{equation}
We are once again led to a wave mode with exponential blueshift at any given $x$. The metric
is static in $t$ at late times and the out-vacuum will be defined in terms of
modes which are positive frequency with respect to $t$. The Bogoliubov transformations between the mode in Eq.~(\ref{eqn:sixtyeight}) and modes which vary as $\exp(-i\nu t)$
will involve exactly the same mathematics as in   Eq.~(\ref{planck}). We will get a 
thermal spectrum at late times.

\subsection{Hawking radiation from black holes}\label{hawkradbh}

  The simplest model for the formation of the black hole is based on a  spherical distribution  of mass
  $M$ which collapses under its own weight to form a black hole. 
  Since only the exponential redshift of the modes at late times is relevant as far as the thermal spectrum
  is concerned, the result should be independent of the detailed nature of the collapsing matter
  \cite{Hawking:1975sw,Gerlachbh,Unruh:1976db}.
  Further, the angular coordinates do not play a significant role
  in this analysis, allowing us to work in the two dimensional $(t,r)$ subspace. The line element exterior 
  to the spherically symmetric distribution of matter
  can be taken to be $ds^2 = -C(r) du dv$ where
  \begin{equation}
  \xi  = \int dr\, C^{-1}; \quad u=t-\xi +R^*_0;\quad v=t+\xi -R^*_0
  \end{equation} 
   and $R_0^* $ is a constant. In the interior, the line element is taken to be $ds^2 = -B(U,V) dUdV$
   with 
   $U=\tau- r+R_0,\ V=\tau+ r-R_0$ and $R_0$ and $R_0^*$ are related in the same manner as
   $r$ and $\xi $.  Let us assume that, for $\tau<0$, matter was at rest with its surface at 
   $r=R_0$ and for $\tau>0$, it collapses inward along the trajectory $r=R(\tau)$. The coordinates
   have been chosen so that at the onset of collapse ($\tau=t=0$) we have $u=U=v=V=0$
   at the surface. Let the  coordinate transformations between the interior and exterior be given by
   the functional forms $U=f(u)$ and $v=h(V)$. Matching the geometry along the trajectory
   $r=R(\tau)$, it is easy to show that 
   \begin{equation}
   \frac{dU}{du} = (1-\dot R)C \left( \left[ BC ( 1-\dot R^2) + \dot R^2\right]^{1/2} - \dot R\right)^{-1}
   \label{fofu}
   \end{equation}
   \begin{equation}
  \frac{dv}{dV} = \frac{1}{C(1+\dot R)}\left( \left[ BC ( 1-\dot R^2) 
  + \dot R^2\right]^{1/2} + \dot R\right)
  \label{hofv}
   \end{equation}
  As the modes propagate inwards they will reach $r=0$ and re-emerge as out-going modes.
  In the $(t,r)$ plane, this requires reflection of the modes on the $r=0$ line, which 
  corresponds to $V=U-2R_0$. 
  The solutions to the two dimensional wave equations $\Box \phi =0$
  which (i) vanish on the line $V=U-2R_0$ and (ii) reduce to standard exponential form in the 
  remote past, can be determined by noting that, along $r=0$ we have
  \begin{equation}
  v=h(V)=h[U-2R_0] = h[f(u)-2R_0]
  \end{equation}
  Hence the solution is 
  \begin{equation}
  \Phi = \frac{i}{\sqrt{4\pi \omega}} \left( e^{-i\omega v} - e^{-i\omega h[f(u)-2R_0]}\right)
  \label{modes}
  \end{equation}
  (The second term, which is the ``reflected wave'' at $r=0$ can, in fact, be entirely interpreted
  in terms of Doppler shift arising from a fictitious moving surface having the trajectory $r=0$.)
  Given the trajectory $R(\tau)$, one can integrate  Eq.~(\ref{fofu}) to obtain $f(u)$ and use
  Eq.~(\ref{modes}) to completely solve the problem. This will describe time-dependent particle production
  from some collapsing matter distribution and --- in general --- the results will depend on the details of the collapse.
  
  The analysis, however, simplifies considerably and a universal
  character emerges if the collapse proceeds to form a horizon on which $C\to 0$. Near $C=0$, equations
  (\ref{fofu}) and (\ref{hofv}) simplifies to 
  \begin{equation}
  \frac{dU}{du} \approx\frac{\dot R -1}{2\dot R} C(R); \quad \frac{dv}{dV} \approx\frac{B(1-\dot R )}{2\dot R} 
  \label{nearhori}
  \end{equation}
  where we have used the fact that $(\dot R^2)^{1/2}=-\dot R$ for the collapsing
  solution. Further, near $C=0$, we can expand $R(\tau)$ as 
  $R(\tau) = R_h +\nu(\tau_h - \tau) + \mathcal{O} [(\tau_h - \tau)^2]$
  where $R=R_h$ at the horizon and $\nu=-\dot R(\tau_h)$. Integrating Eq.~(\ref{nearhori}) treating $B$
  approximately constant, we get
  \begin{equation}
   a  u \approx - \ln |U+R_h - R_0 -\tau_h| + \textrm{const}
  \label{uofU}
  \end{equation}
  where $\kappa =(1/2)(\partial C/\partial r)_{R_h}$ is the surface gravity.
  and
  \begin{equation}
  v \approx \textrm{constant} - B V(1+ \nu) / 2\nu 
  \label{vofV}
  \end{equation}
  It is clear that: (i) The relation between $v$ and $V$ is linear and hence holds no surprises; it also depends on $B$. (ii) The relation between $U$ and $u$, which can be written as $U\propto \exp(- \kappa   u)$ is universal (independent of $B$) and signifies the exponential  redshift we have alluded to several times. The late time
  behaviour of out-going modes can now be determined using Eq.~(\ref{uofU}) and Eq.~(\ref{vofV}) in Eq.~(\ref{modes}).
  We get:
  \begin{equation}
  \Phi \cong  \frac{i}{\sqrt{4\pi \omega}} \left( e^{-i\omega v} 
  - \exp\left(i\omega\left[ ce^{- \kappa  u} + d\right]\right)\right)
  \label{latephi}
  \end{equation}
  where $c,d$ are constants. This mode with exponential redshift, when expressed in terms of $\exp(-i\nu u)$
  will lead to a thermal distribution of particles with temperature $T=\kappa  /2\pi$. For the case of a black hole,
  if we take $\kappa   =1/4M$, then the Bogoliubov coefficients are given by
  \begin{equation}
  \alpha_{\omega\nu} = \frac{-2 i M \nu e^{-i\omega d } e^{i\nu t_0}}{2\pi \sqrt{\nu \omega}} 
  \left( \frac{-e^{-(t_0 + d)/4M}}{\omega c} \right)^{-4 iM \nu} 
   \  e^{2M\pi \nu} \Gamma (-4 iM \nu)
  \end{equation}
  and 
  $\beta _{\omega\nu} = e^{-4 M \pi \nu } \alpha^*_{\omega \nu}$.
  Note that these quantities {\it do}  depend on $c,d,t_0$ etc; but the modulus 
  \begin{equation}
  |\beta_{\omega\nu}|^2=\frac{1}{2} \frac{4M}{\left[\exp(8\pi M\nu) -1)\right]}
  \end{equation} 
  is independent of these factors. [The mathematics is essentially the same as in Eqs.(\ref{powernu}),(\ref{planck}).]
 This shows that the vacuum state at early times will be interpreted as containing 
  a thermal spectrum of particles at late times.

  \subsection{Asymptotically De Sitter spacetimes}
  
  The De Sitter universe is a solution to Einstein's equations $G^a_b = 8\pi T^a_b$
  with a source given by $T^a_b= \Lambda \delta^a_b$. The spacetime metric given in 
  Table \ref{table:metricprop} is given in terms of the parameter $H^2 =\Lambda/3$ and is useful for
  providing easy comparison with Schwarzschild and Rindler metrics. But this coordinate system 
  hides the symmetries of the De Sitter manifold. Since the source is homogeneous, isotropic
  and constant in space and time, the metric can be cast as a section of a maximally symmetric
  manifold. Using the Friedmann-Robertson-Walker coordinates, which is appropriate for describing
  maximally symmetric 3-space, one can express the De Sitter spacetime in the form
  \begin{equation}
  ds^2 = -dt^2 + a^2(t) \left[ \frac{dr^2}{1- kr^2} + r^2 (d\theta^2 + sin^2 \theta d\phi^2 )\right]
  \end{equation}
  with  $k=0, a(t) = \exp(Ht)$ or with $k=1, a(t) =H^{-1}\cosh (Ht)$. (There is also a solution with $k=-1$
which we do not need).
  
To proceed from such an ``eternal'' De Sitter universe, 
  to an asymptotically De Sitter universe, we only have to add normal matter or radiation
  to the source of the Einstein's equations. At sufficiently late times the energy densities of matter
  or radiation will be diluted exponentially leading to the De Sitter solution at late times.
  (This occurs in a wide class of dark energy models \cite{Padmanabhan:2002trctp,Choudhury:2003tj}.)
  Mathematically, this will correspond to $a(t)$ which is a power law at small $t$ tending to
  $a(t) \propto \exp(Ht)$ for $Ht \gtrsim 1$.
  
 In the asymptotic future, one can introduce the static Schwarzschild coordinates in the manifold and define
  a vacuum state. However, it is not possible to assign a natural (or unique) vacuum state in the asymptotic past 
  if $a(t)$ is time dependent and one needs to invoke some extra prescription to define a vacuum
  state.   This issue has been extensively discussed in the literature and several possible prescriptions 
  based on different criteria have been explored (see e.g., \cite{Birrel:bkqft}). 
  One of the simplest choices will be to choose  modes which vary as $\exp(-i\omega t)$ near
  $t\approx t_0$ will define an instantaneous vacuum state around $t\cong t_0$. 
  At late times, the frequency of the wave mode will vary as $\omega(t) \propto a(t)^{-1} \propto 
  e^{-Ht}$ in the WKB approximation. Fourier transforming these modes with respect to another
  instantaneous vacuum state defined through the modes which vary as $\exp(-i\omega t)$
  near $t \to + \infty$, one can recover a thermal spectrum of particles at late times in the 
  initial vacuum state \cite{Gibbons:1977mu}.
  It is clear from this discussion that the asymptotically De Sitter spacetime requires a somewhat
  different approach compared to the other two cases because of explicit time dependence.
   
\section {Expectation values of Energy-Momentum tensor}\label{exptab}

  The flow of radiation at late times, away from the horizon, is the new feature which arises
  when horizon forms in the asymptotic future. 
A  formal way of describing this result is to use the expectation value 
$\langle \psi|T_{ab}|\psi\rangle $ of the energy momentum tensor  of the matter field $T_{ab}$. If the quantum state is time reversal invariant, then expectation values of flux, $\langle T^\alpha_0\rangle $, will vanish, though the expectation value of 
energy density, $\langle T^0_0\rangle $, can be nonzero and correspond to thermal radiation at some equilibrium temperature, related to the surface gravity of horizon. 

It is clear that a new element, the quantum state $|\psi\rangle $,
 has entered the discussion. In a given spacetime with a horizon, one can, of course, make different choices for this state, even if we nominally decide it should be a ``vacuum state". 
  The expectation value of various operators, including $T_{ab}$ will be quite different 
  in each of these states and there is no assurance that they will even be finite near the horizon (or 
  at infinity) in an arbitrary state. Similarly, if the mode functions  are not  invariant under time reversal,
  then the expectation value of energy-momentum tensor in the corresponding vacuum
  state may show a flux of radiation. 
  
  This new feature allows us to mimic the effects of formation of asymptotic horizons by choosing
  a quantum state which is not time reversal invariant. That is, we can identify quantum states
  which will contain flux of radiation emanating from the horizon at late times
  even though we are working in a static spacetime with a metric which is invariant under
  time reversal. This is possible only because the late time behaviour in the case of spacetimes with
  asymptotic horizons (discussed in the previous Section)  is independent of the details of the 
  metric during the transient phase. We shall now see how such quantum states and the 
  expectation values of $T_{ab}$ in those states can be constructed.
   
   \subsection{The $\langle T_{ab}\rangle $ in two-dimensional field theory}

  A purely technical difficulty in such an approach arises from the fact that the mode functions
  in four dimensional spacetime are fairly complicated in form and the expectation value 
  $\langle T_{ab}\rangle $ is usually not tractable analytically.  However, the situation simplifies enormously
  in two dimensions and since the results in two dimension capture the essence of physics,
  we shall use this approach to explain the choice of vacuum states and the corresponding 
  results.

    In the (1+1) dimension, the metric has three independent 
   components while the freedom
   of two coordinate transformations allows us to impose two conditions on them. 
Hence we can reduce any two dimensional
   metric to a conformally flat form locally.
   Consider such a  spacetime with signature $(-,+)$ and line element expressed  as 
   \begin{equation}
   ds^2 = -C (x^+,x^-) dx^+dx^-; \qquad x^{\pm} = t\pm x
   \label{one}
   \end{equation}
   A massless  scalar field in this spacetime is described by the action
   \begin{equation}
   A = -\frac{1}{ 2} \int d^2 x \sqrt{-g}g^{ab} \partial_a \phi \ \partial_b \phi = \frac{1}{ 2} \int dt dx [ \dot \phi^2 - \phi^{'2}]
   \end{equation}
   since $\sqrt{-g}g^{ab} = \eta^{ab}$ for the metric in Eq.~(\ref{one}).
   The field  equation 
  $ (\partial^2\phi/ \partial x^+\partial x^-) =0$
   has the general solution:
   \begin{equation}
   \phi(x^+,x^-)= \phi_1(x^+)+\phi_2(x^-)
   \label{newnine}
   \end{equation}
   with $\phi_1(x^+) = \phi_1(t+x)$ being the   `in-going' (or `left moving') mode and
   $\phi_2(x^-) = \phi(t-x)$ being the `outgoing' (or `right moving') mode.
   The  expansion of the scalar field, in terms of the normalised plane wave mode functions,
   is given by
   \begin{equation}
   \phi = \int_{-\infty}^\infty \frac{dk}{ \sqrt{2\pi |k|}} \left[a(k) e^{i(kx - |k|t)} + h.c.\right]
   \label{three}
   \end{equation}
   It is more convenient to rewrite this expansion in terms of  the  in-going and outgoing modes
    (as in Eq.~(\ref{newnine}))  and label
   them by the frequency $\omega = |k|$. This is easily done by separating the integration
   range in Eq.~(\ref{three}) into $(-\infty, 0)$ and $(0,\infty)$ and changing the variable of 
   integration from $k$ to $-k$ in the first range.  This gives 
   \begin{equation}
    \phi  =\int_{0}^\infty \frac{d\omega}{ \sqrt{4\pi \omega}} \left[a(\omega) e^{-i\omega x^-} + b(\omega) e^{-i\omega x^+} +h.c.\right]
    \label{four}
   \end{equation}
   which is of the form in Eq.~(\ref{newnine}).
    There is a direct correspondence between the set of  modes [$\exp(-i\omega x^\pm)$]
   and the vacuum state $|x^\pm\rangle $ annihilated by the operators $a(\omega), b(\omega)$.

 The stress-tensor for the scalar field is given by
  $ T_{ab}=\partial_a \phi\, \partial_b\phi - (1/2) g_{ab} ( \partial^c \phi \, \partial_c \phi)$. To evaluate its expectation
value, it is convenient to relate it to the Feynman Green function in this vacuum state 
 $G_F(x^\pm,y^\pm)=\langle x^\pm|T(\phi(x^\pm)\phi(y^\pm))|x^\pm\rangle $, by:
   \begin{equation}
   \langle T_{ab}(x)\rangle  =\lim_{x\to x'} \left[ \partial_a\partial_b' - \frac{1}{ 2}g_{ab} \partial^c\partial_c'\right] G_F(x,x')
   \end{equation}
   Using this procedure, it can be shown that the (regularised)  expectation values are given by (see e.g.,\cite{Birrel:bkqft})
   \begin{equation}
   \langle x^\pm|T_{\pm\pm}|x^\pm\rangle  = -\frac{1}{12\pi} C^{1/2}\partial^{2}_{\pm}C^{-1/2}
   \label{inertmet}
   \end{equation}
\begin{equation}
\langle T_{+-}\rangle  = \frac{C}{ 96\pi} R; \quad R=4C^{-1}\partial_+\partial_- \ln C
\label{finaltab}
\end{equation}   
We shall now use the results in  Eq.~(\ref{inertmet}), Eq.~(\ref{finaltab}) to evaluate $\langle T_{ab}\rangle $
in spacetimes with horizons.

  \subsection{Vacuum states and $\langle T_{ab}\rangle $ in the presence of Horizons} 
   
  Since the mode functions are plane waves in 
conformally flat (1+1) spacetime, we can immediately identify two 
natural sets of modes and corresponding vacuum states.  The out-going and in-going modes of the 
form given by 
$(4\pi \omega)^{-1/2} [\exp(-i\omega u), \exp(-i\omega v)]$
define a static vacuum state (called Boulware vacuum in the 
case of Schwarzschild black hole \cite{Boulware:1975dm} but can be defined in any other spacetime) natural to the $(t,x)$ or $(t,l)$ coordinates. 
The modes of the kind 
$(4\pi \omega)^{-1/2} [\exp(-i\omega U), \exp(-i\omega V)]$
define another  vacuum state [called Hartle-Hawking vacuum in the 
case of Schwarzschild black hole \cite{Hartle:1976tp}] natural to the $(T,X)$ coordinates. 
(Note that these two coordinate frames $(T,X)$ and $(t,x)$ are related by Eq.~(\ref{keytransform}).)
Finally, the modes of the kind
$(4\pi \omega)^{-1/2} [\exp(-i\omega U), \exp(-i\omega v)]$
 define a third vacuum state [called  the  Unruh vacuum \cite{Unruh:1976db}] which
is natural to the situation in which a horizon forms asymptotically, as in the case of gravitational collapse.
This is obvious from the discussion in Section \ref{hawkradbh} [see 
 Eq.~(\ref{latephi})] which shows how these modes originate in the collapse scenario.
 
 Using the result that, in any conformally flat coordinate 
 system of the form $ds^2= - C(x^+,x^-)dx^+dx^-$, the expectation values
 of the stress-tensor component are given by Eq.~(\ref{inertmet}), Eq.~(\ref{finaltab}), 
 we can explicitly evaluate the various expectation values. In the cases of interest to us the conformal factor only depends on the tortoise coordinate $\xi=(1/2)(x^+ - x^-)$.
 For example, in the Boulware vacuum we get 
 \begin{equation}
 \langle B|T_{--}|B\rangle =\langle B|T_{++}|B\rangle  = \frac{1}{ 96\pi}\left[C C''-\frac{1}{ 2}(C')^2\right]
 \end{equation}
(where the prime denotes derivative with respect $\xi$)
  while in the Hartle-Hawking vacuum we get 
  \begin{equation}
  \langle HH|T_{--}|HH\rangle =\langle HH|T_{++}|HH\rangle 
  =\langle B|T_{--}|B\rangle  +  \frac{ \kappa ^2}{48\pi}.
  \end{equation}
   In both these cases, there is  no flux since $\langle T_{xt}\rangle =0$.
  Near the horizon, we have 
  \begin{equation}
  \langle B|T_{\pm \pm}|B\rangle \approx - \frac{\kappa  ^2}{48\pi};\quad \langle HH|T_{\pm \pm}|HH\rangle \approx 0. 
  \end{equation}
  The coordinate system used by an inertial observer near the horizon
will have $U$ instead of $u$ and hence the actual values measured by an inertial observer near the horizon will vary as $  
 \langle B|T_{uu}|B\rangle (du/dU)^2$ and will \emph{diverge} on the horizon if we choose the vacuum state $|B\rangle $. 
  
  A more interesting situation arises in the case of Unruh vacuum which differs from the
  Boulware vacuum only in the outgoing modes. If the coordinate $x^-$ is replaced by
  $X^-\equiv F(x^-)$, the conformally flat nature of the line element is maintained and 
  the only stress tensor component which changes is $\langle T_{--}\rangle $. Using this fact,  we find that
 \begin{equation}
  \langle U|T_{--}|U\rangle =\langle HH|T_{--}|HH\rangle ; \quad
  \langle U|T_{++}|U\rangle =\langle B|T_{++}|B\rangle  
  \end{equation}
   thereby making
  $\langle U|T_{--}|U\rangle \ne \langle U|T_{++}|U\rangle  $. This leads to a flux of radiation
  with
  \begin{equation}
  \langle U|T_{\xi t}|U\rangle =-(\kappa ^2/48\pi)
  \end{equation}
   It is also clear that the energy density, as measured by inertial observers, is finite near the future horizon
   in $|U\rangle $.

In the case of eternal black hole (or eternal De Sitter), there are {\it two} horizons in the full manifold corresponding to $T=\pm X$. So far we have discussed the behaviour near the future horizon, $T=X$ (in global coordinates). One can perform a similar analysis at the past horizon $T=-X$ for each of these quantum states. The stress-tensor expectation value in $|HH\rangle$ is finite at both  horizons. In contrast, the
expectation value in $|B\rangle$ diverges at both  horizons while the expectation value in $|U\rangle$ [which is finite at the future horizon ($T=X$)]  diverges in the past horizon ($T=-X$).  Since we require the expectation values to be finite at both horizons, $|HH\rangle$ is a suitable choice in the case of eternal black hole etc. However, when a black hole forms due to gravitational collapse, the past horizon does not exist since it is covered by the internal metric of the collapsing matter. Therefore, both $|HH\rangle$ and $|U\rangle$ are acceptable choices for a black hole formed due to gravitational collapse.  The (time symmetric) Hartle-Hawking state describes thermal equilibrium and zero flux and  the (time-asymmetric) Unruh vacuum describes a state with a flux of radiation.

  In the case of  a Schwarzschild black hole, the explicit formulas for the stress-tensor
  expectation value are given by
  \begin{equation}
  \langle T_{++}\rangle_U = \langle T_{++}\rangle_B=\langle T_{--}\rangle_B = \frac{\pi}{12} T_H^2 
  \left[ \frac{48M^4}{r^4} - \frac{32 M^3}{r^3}\right]
  \end{equation} 
   \begin{equation}
  \langle T_{--}\rangle_U  = \frac{\pi}{12} T_H^2 
  \left[ 1-\frac{2M}{r} \right]^2 \left[ 1+ \frac{4 M}{r} + \frac{12 M^2}{r^2} \right]
  \end{equation}
  where $T_H = (\kappa /2\pi) = (1/8\pi M)$. At $r\to \infty$, there is a constant flux of magnitude
  $(\pi/12) T_H^2$ which is the  flux at the temperature $T_H$. 
  
  Though these results are valid only in $(1+1)$ spacetime, the results for
  the four dimensional spacetime in the $r-t$ sector can be  approximated by
  $\langle T_{ab}^{4D} \rangle \approx(1/4\pi r^2) \langle T_{ab}^{2D}\rangle$.
  Since the net flux across a spherical surface of constant $r$ in $4D$ is given by
  $4\pi r^2 \langle T_{ab}^{4D} \rangle$, we can directly interpret $\langle T_{ab}^{2D} \rangle$
  as the net flux in the $4D$ case. Our results then imply that the energy flowing to infinity
  per second is given by $(\pi/12) T_H^2$.

While the above results are generally accepted and is taken to imply the radiation of energy from a collapsing black hole
to infinity at late times, 
 there are some serious unresolved issues related to situations with asymptotic horizons. These issues are particularly
important for the general case rather than for black hole since in the latter the asymptotic flatness of the spacetime
helps to alleviate the problems somewhat. We shall now briefly discuss these issues.

   We saw in Section \ref{hawkrad} that one can construct a coordinate system even in flat spacetime
   such that certain quantum states exhibit a flux of radiation away from the horizon. 
But in De Sitter or Rindler spacetimes there is {\it no}
natural notion of  ``energy source", analogous to the mass of the black-hole,  which  could  decrease as the radiation flows away from the horizon.
The conventional view is to assume that:  (1) In the case of black-holes, one considers the collapse scenario as ``physical" and the natural quantum state is the Unruh vacuum. The notion of evaporation,  etc. then follow in a concrete manner. The eternal black-hole (and the Hartle-Hawking vacuum state) is taken to be just a mathematical construct not realized in nature.  (2) In the case of Rindler, one may like to think of a time-symmetric vacuum state as natural and treat the situation as one of thermal equilibrium. This forbids using quantum states with outgoing radiation in the  Minkowski spacetime.  

The real trouble  arises for spacetimes which are asymptotically De Sitter.  Does it ``evaporate" ? 
The analysis in the earlier Sections show that it is imperative to associate a temperature with the
 De Sitter horizon but the idea of the cosmological constant changing due to evaporation of the De Sitter spacetime seems too radical.
  Unfortunately, there is no clear mathematical reason for  a dichotomous approach as regards a collapsing black-hole and an asymptotically De Sitter spacetime, since
 the mathematics is  identical. 
  Just as collapsing black hole leads to an asymptotic event horizon, a universe
  which is dominated by cosmological constant at late times will also lead to a horizon.
  Just as we can mimic the time dependent effects in a collapsing black hole by
  a time asymmetric quantum state (say, Unruh vacuum), we can mimic the late time behaviour of 
  an asymptotically De Sitter universe by a corresponding time asymmetric quantum state.
  Both these states will lead to stress tensor expectation values in which there will be a flux
  of radiation. 
 The energy source for expansion at early times (say, matter or radiation) is irrelevant just as the collapse details are irrelevant in the case of a black-hole. 

If one treats the De Sitter horizon as a  `photosphere' with temperature $T=(H/2\pi)$ and area $\mathcal{A}_H=4\pi H^{-2}$,
then the radiative luminosity will be $(dE/dt)\propto T^4\mathcal{A}_H\propto H^2$. If we take $E=(1/2)H^{-1}$ (which will br justified in Section \ref{thermoid}), 
this will
lead to a decay law \cite{Padmanabhan:2002sh} for the cosmological constant of the form:
\begin{equation}
\Lambda(t)=\Lambda_i\left[1+k (L_P^2\Lambda_i)(\sqrt{\Lambda_i}(t-t_i))  \right]^{-2/3}\propto (L_P^2t)^{-2/3}
\end{equation}
where $k$ is a numerical constant and the second proportionality is for $t\to \infty$. It is interesting that  this naive model leads to a late time cosmological constant which is independent of the initial value ($\Lambda_i$). Unfortunately, its value is still far too large.
These issues are not analysed in adequate detail in the literature and might have important implications for the cosmological constant problem. (For some recent work and references to earlier literature,
see  \cite{Davis:2003ye,Davies:2003me}; for an interesting connection between thermality in
Rindler  and DeSitter spacetime, see \cite{Deser:1997ri,Deser:1998xb}.)

\subsection{Spacetimes with multiple horizons}\label{multihorizon}

A new class of mathematical and conceptual 
difficulties emerge when the spacetime has more than one horizon. For example,
metrics in the form in 
Eq.~(\ref{standardhorizon})
 with $f(r)$ having  simple zeros at 
 $r=r_i , i=1,2,3, ...$, exhibit coordinate singularities at $r=r_i$. The coordinate $t$ alternates
 between being timelike and spacelike when each of these horizons are crossed.
 Since all curvature invariants are well  behaved at the horizons, it will be possible to introduce
 coordinate patches such that the metric is also well behaved at the horizon. 
 This is done exactly as in Eq.~(\ref{keytransform}) near each horizon $r=r_i$ with $\kappa$ replaced by 
 $a_i =N'(r_i) = f'(r_i)/2$.
  
 When there is more than one
 horizon,  we need to introduce one Kruskal like coordinate patch
 for \emph{each} of the horizons; the $(u,v)$ coordinate system is unique in the manifold but
 the $(U_i,V_i)$ coordinate systems are different for each of the horizons since the 
 transformation in Eq.~(\ref{keytransform}) depends explicitly on $a_i$'s which are (in general)
 different for each of the horizons. In such a case,
 there will be regions of the manifold in which more than one Kruskal like patch
 can be introduced. The compatibility between these coordinates  leads to new constraints.

    Consider, for example, the region between two consecutive horizons $r_n <r<r_{n+1}$ in which $t$ is timelike. The coordinates
    $(U_i, V_i)$ with $i=n, n+1$ overlaps in this region.
    Euclideanisation of the metric can be easily effected in the region
    $r_n<r<r_{n+1}$ by taking $\tau =it$. This will lead to the transformations 
    \begin{eqnarray}
    U_{n+1} &=& -U_n \exp [(a_{n+1} + a_n)(-i\tau -\xi)];\nonumber \\
     V_{n+1} &=& -V_n \exp[ -(a_{n+1} + a_n)(-i\tau+\xi)]
    \end{eqnarray}
    Obviously, single valuedness can be maintained only if the period of $\tau$ is an integer multiple
    of $2\pi / (a_{n+1}+a_n)$. 
    More importantly, we get    from Eq.~(\ref{keytransform}) the relation 
    \begin{equation}
    U_i +  V_i  = \frac{2}{ a_i} \exp \left(a_i \xi\right) \sinh \left(-ia_i \tau\right)
    \end{equation}
    which shows that $(U_i, V_i)$ can be used to define values of $\tau$ only up to integer multiples of
    $2\pi / a_i $  in each patch. But since $(U_n, V_n)$ and $(U_{n+1}, V_{n+1})$ are to be 
    well defined coordinates in the overlap, the periodicity $\tau \to \tau +\beta$ which 
    leaves both the sets $(U_n, V_n)$ and $(U_{n+1}, V_{n+1})$
     invariant must be such that $\beta $ is an integer
    multiple of both $2\pi / a_n $ and $2\pi / a_{n+1}$. 
    This will require $\beta = 2\pi n_i/a_i$ for all $i$ with $n_i$ being a set of integers. This,
    in turn, implies that $a_i/a_j =n_i/n_j$ making the ratio between any two surface 
    gravities a rational number, which is  the condition 
    for a non singular Euclidean extension to exist.
  
  These issues also crop up when one attempts to develop a quantum field theory
  based on different mode functions and vacuum states (see, for example, \cite{Choudhury:2004ph}).
    It is easy to develop the quantum field theory in the $t-r$ plane if we treat it 
    as a $(1+1)$ dimensional spacetime. In a region between two consecutive horizons $r_n<r<r_{n+1}$,
    we can use (at least) three sets of coordinates: $(u,v), (U_n,V_n), (U_{n+1},V_{n+1})$ all of 
    which maintain  the conformally flat
    nature of the $(1+1)$ dimensional metric, allowing us to
      define suitable mode functions and 
    vacuum state in a straightforward manner.    
    The outgoing and in-going modes of the kind 
$(4\pi \omega)^{-1/2} $ $[\exp(-i\omega u), \exp(-i\omega v)]$
define a static (global) Boulware vacuum state. 
The modes of the kind 
$(4\pi \omega)^{-1/2} [\exp(-i\omega U_i), $ $\exp(-i\omega V_i)]$ with $i=(n, n+1)$ define \emph{two 
different}  Hartle-Hawking vacua.
As regards the Unruh type vacua, we now have three different choices. The mode functions $\mathcal{U}_n
= (4\pi \omega)^{-1/2}  [\exp(-i\omega U_n), $ $\exp(-i\omega v)]$
 define the analogue of Unruh vacuum for the horizon at $r=r_n$. Similarly, 
    $\mathcal{U}_{n+1}= (4\pi \omega)^{-1/2}  [\exp(-i\omega u), $ $\exp(-i\omega V_{n+1})]$
    define another vacuum state corresponding to the horizon at $r=r_{n+1}$. What is more, 
    we can now also define another set of modes and a vacuum state based on 
    $\mathcal{U}_{n,n+1}= (4\pi \omega)^{-1/2}  [\exp(-i\omega U_n), $ $\exp(-i\omega V_{n+1})]$.
    The physical meaning of these three vacua can be understood from the radiative 
    flux $\langle\psi|T_{\xi t}|\psi\rangle$
    in each of these states. We find that  $\langle \mathcal{U}_n|T_{\xi t}|\mathcal{U}_n\rangle =-(a_n^2/48\pi)$;
     $\langle \mathcal{U}_{n+1}|T_{\xi t}|\mathcal{U}_{n+1}\rangle =(a_{n+1}^2/48\pi)$ and
     \begin{equation}
     \langle \mathcal{U}_{n,n+1}|T_{\xi t}|\mathcal{U}_{n,n+1}\rangle =\frac{a_{n+1}^2 - a_n^2}{48\pi}
     \end{equation}
     It is clear that the quantum state $|\mathcal{U}_{n,n+1}\rangle$ corresponds to one with
     radiative flux at two different temperatures arising from the two different horizons;
     in the case of Schwarzschild-De Sitter spacetime, one flux  will correspond
     to radiation flowing outward from the black hole horizon and the other to  radiation
     flowing inward from the De Sitter horizon. A detector kept between the horizons will
     respond as though it is immersed in a radiation bath containing two distinct Planck
     distributions with different temperatures \cite{Markovic:1991ua}.
     
     In addition to the coordinate systems we have defined, it is also possible to
     introduce a global non singular coordinate system for the SdS metric.  (The method
     works  for several
     other metrics with similar structure, but we shall concentrate on SdS for definiteness.)
     Let the horizons be at $r_1$ and $r_2$ which are the roots of $(1-2M/r -H^2r^2) =0$
    with surface gravities $\kappa_1, \kappa_2$. We introduce  the two sets of Kruskal-like coordinates
     $(U_1,V_1), (U_2,V_2)$ by the usual procedure. The global coordinate system in which the metric
     is well behaved at both the horizons is given by
     \begin{equation}
     \bar{U} = \frac{1}{\kappa_1} \tanh \kappa_1 U_1 + \frac{1}{\kappa_2} \tanh \kappa_2 U_2 ;\quad
     \bar{V} = \frac{1}{\kappa_1} \tanh \kappa_1 V_1 + \frac{1}{\kappa_2} \tanh \kappa_2 V_2
     \end{equation}
     in the region I ( $U_1 < 0, V_1 > 0, U_2 > 0, V_2 < 0$). Similar definitions can be introduced
     in all other regions of the manifold \cite{Choudhury:2004ph,Tadaki:1990cg,Tadaki:1990aa} maintaining 
     continuity and smoothness of the metric. The resulting metric in the $\bar{U}, \bar{V}$ coordinates
     has a fairly complicated form  and depends explicitly on the time coordinate
     $\bar{T} = (1/2)( \bar{U}+\bar{V})$. In general, the metric coefficients are \emph{not} periodic
     in the imaginary time; however, if the ratio of the surface gravities is rational
     with $\kappa_2/\kappa_1 =n_2/n_1$, then the metric is periodic in the imaginary time
     with the period $\beta = 2 \pi n_2/\kappa_2 = 2 \pi n_1/\kappa_1$.
     Since the  physical basis for such a condition  is unclear, it is difficult to attribute a
    single temperature to spacetimes with multiple horizons.
    This demand of $\kappa_2/\kappa_1 =n_2/n_1$ is related to
    an expectation of thermal equilibrium which is violated in spacetimes with 
    multiple horizons having different temperatures. Hence, such spacetimes
    will not --- in general --- have a global notion of temperature.

\section{Entropy of Horizons}\label{entropyhorizon}

  The analytic properties of spacetime manifold in the complex plane directly
  lead to the association of a temperature with a generic class of horizons. In Section \ref{hawkrad} 
  we also saw that there exist quantum states  in which a flux of thermal radiation will flow away from the 
  horizon if the horizon forms asymptotically. Given these results, it is natural to enquire whether 
  one can attribute other thermodynamic variables, in particular entropy, to the horizons.
  We shall now discuss several aspects of this important --- and not yet completely
  resolved --- issue. 
  
  The simplest and  best understood situation arises in the case of a Schwarzschild
  black hole formed due to gravitational collapse of matter. In this case, one can rigorously demonstrate
  the flow of thermal flux of radiation to asymptotic infinity at late times, which  can be collected
  by observers located in (near) flat spacetimes at $r\to\infty$. 
  Given a temperature and a change in energy, one can invoke classical thermodynamics
  to define 
 the {\it change in} the entropy  via $dS=dE/T(E)$. Integrating this equation will
lead to the function $S(E)$ except for an additive constant which needs to be
determined from additional considerations. 
  In the Schwarzschild spacetime, which is asymptotically flat, it is  possible to 
  associate an energy $E=M$ with the black-hole.
 Though the  calculation  was done in a metric with a fixed value of energy $E=M$,
  it seems reasonable to assume that ---
as the energy flows to infinity at late times ---  the mass of the black hole
will decrease.  {\it If} we make this assumption,  {\it then} one can integrate the equation $dS=dM/T(M)$ to obtain the
entropy of the black-hole to be 
\begin{equation}
S=4\pi M^2=\frac{1}{4}\left(\frac{\mathcal{A}_H}{L_P^2}\right)
\label{entropyarea}
\end{equation} 
where $\mathcal{A}_H=4\pi (2M)^2$
is the area of the event horizon and $L_P=(G\hbar/c^3)^{1/2}$ is the Planck length. 
This integration  constant is   fixed by  the additional assumption
 that $S$ should vanish when $M=0$.\footnote{One may think that this assumption is eminently 
 reasonable since the Schwarzschild metric reduces to the Lorentzian
metric when $M\to 0$.  But note that in the same limit of $M\to 0$, 
the temperature of the black-hole diverges. Treated as a limit of Schwarzschild spacetime, 
normal flat spacetime has infinite --- rather than zero --- temperature.}

   The fact that entropy of the Schwarzschild black hole is \emph{proportional} to the
   horizon area was conjectured \cite{Bekenstein:1972tm,Bekenstein:1973ur,Bekenstein:1974ax}
   even before it was known that black holes have a temperature. The above analysis 
   fixes the proportionality constant between area and entropy to be (1/4) in Planck units.
   It is  also obvious that the entropy is purely a quantum mechanical effect and diverges
   in the limit of $\hbar \to 0$. Nevertheless, even in the {\it classical} processes involving black holes, the horizon area does
   act in a manner similar to entropy. For example, when two black holes coalesce and settles down to a final
   steady state (if they do),  the sum of the areas of horizons does not decrease. Similarly, in some
   simple processes
   in which energy is dumped into the black hole, one can prove an analogue for first law of thermodynamics
   involving the combination $TdS$. While both $T$ and $S$ depend on $\hbar$ the combination $TdS$
   is independent of $\hbar$ and can be described in terms of classical physics.
   
   The next natural question is whether the entropy defined by Eq.~(\ref{entropyarea}) is the same as ``usual
   entropy". If so, one should be able to show that for any processes involving matter and black holes,
   we must have $d(S_{BH}+S_{matter})/dt\geq 0$ which goes under the name generalised 
   second law (GSL).  One simple example in which the area (and thus the entropy) of the black hole 
   \emph{decreases} is the Hawking evaporation; but the GSL holds since the thermal radiation produced in the process has entropy.
   It is generally believed that GSL always holds though a completely general proof
   is difficult to obtain. Several thought experiments, when analysed properly, uphold this law 
   \cite{Unruh:1982ic} 
   and a proof is possible under certain restricted assumptions regarding the initial state 
   \cite{Frolov:1993fy}.  
   All these suggest that the area of the black hole 
   corresponds to an entropy which is  same  as the ``usual entropy''. 
   
   In the case of normal matter, entropy can be provided
   a statistical interpretation  as the logarithm of the number of 
   available microstates that are consistent with the macroscopic parameters which are
   held fixed. That is,  $S(E)$ is related to the 
degrees of freedom (or phase volume) $g(E)$ by $S(E)=\ln g(E)$. Maximisation of the phase 
volume for systems which can exchange energy
will then lead to equality of the quantity $T(E)\equiv (\partial S/\partial E)^{-1}$ for the systems. 
It is usual to identify this variable as the thermodynamic temperature. 
(This definition works even for self-gravitating systems in microcanonical ensemble; 
see eg., \cite{Padmanabhan:1990gm}.)

Assuming that the entropy of the black hole should have a similar interpretation, one is 
led to the conclusion that the density of states for a black hole of energy $E$ should vary as
   \begin{equation}
   g(E) \propto \exp \left(\frac{1}{4}\frac{\mathcal{A}_H}{L_P^2}\right) = \exp\left[ 4\pi \left(\frac{E}{E_P}\right)^2\right]
   \label{gbhofe}
   \end{equation}
  Such a growth implies, among other things, that the Laplace transform of $g(E)$ 
  does not exist so that no canonical partition function can be defined
  (without some regularization).

 This brings us to the next question: What are the microscopic states by counting which
 one can obtain the result in Eq.~(\ref{gbhofe}) ? That is, what are the degrees of freedom
 (or the missing information content)  which lead to this entropy ? 
 
 There are two features that need to be stressed regarding  these questions. First, classically,
 the black hole is determined by its charge, mass and angular momentum and hence has ``no hair"
(for a review, see e.g., \cite{Bekenstein:1998aw}). 
 Therefore,  the degrees of freedom which could presumably account for all the information contained in the initial (pre-collapse)
 configuration cannot be classical.  Second, the question is intimately related to what happens to the matter that collapses to form the black hole. If the matter is ``disappearing" in a singularity then the information content of the matter can also ``disappear". But since singularities are unacceptable in physically correct theories, we expect the classical singularity to be replaced by some more sophisticated description in the correct theory. Until we know what this description is, it is impossible to answer in a convincing manner what happens to the information and entropy which is thrown into the black hole or  was contained in the initial pre collapse state. 
 
 In spite of this fact, several attempts have been made in the literature to understand features related to entropy of black holes. 
 A statistical mechanics derivation of entropy was originally attempted in \cite{Gerlach:1976hbh};
 the entropy  has been interpreted as the logarithm of: (a)  the number of 
      ways in which black hole might have been formed \cite{Bekenstein:1973ur,Hawking:1976de};
     (b) the number of internal black hole states consistent with a 
      single black hole  exterior  \cite{Bekenstein:1973ur,Bekenstein:1975tw,Hawking:1976de} and
     (c)  the number of horizon quantum states \cite{Wheelerbk:1990,'tHooft:1990fr,Susskind:1993if}.
      There are also  other approaches which are more mathematical ---  like
      the ones based on Noether charge \cite{Wald:1993nt,Jacobson:1994vj,Visser:1993nu},
      deficit angle related to conical singularity \cite{Banados:1994qp,Susskind:1993ws}, 
      entanglement entropy  \cite{Dowker:1994fi,Israel:1976ur,Callan:1994py}
      and thermo field theory and related approaches \cite{Frolov:1998vs,Frolov:1997up,Frolov:1997aj}.
      Analog models for black holes which might have some relevance to this question 
      are discussed in \cite{Unruh:1995je,Jacobson:1993hn,Brout:1995wp}.
       There are also attempts to compute the entropy using the Euclidean gravitational  action
    and canonical partition function \cite{Gibbons:1977ue}. However, since we know that canonical partition function does not
    exist for this system these calculations require a non trivial procedure for their interpretation.
    In fact, once the answer is known, it seems fairly easy to come up with very imaginative derivations
    of the result. 
We shall comment on a few of them.

 To begin with, the thermal radiation surrounding the black hole has an entropy 
   which one can attempt to compute. It is fairly easy to see that this entropy will 
   proportional to the horizon area but will
   diverge quadratically.
   We saw in Section \ref{ftdim} 
 that, near the horizon,  the field becomes free 
and solutions are simple plane waves. It is the  existence of such a continuum
of wave modes which leads to infinite phase volume for the system.
More formally, the number of modes  $n(E)$ for a scalar field $\phi$ with vanishing
boundary conditions at two radii $r=R$ and $r=L$ is given by
\begin{equation}  
n(E) 
 = \frac{2}{ 3\pi} \int^L_R \frac{r^2 dr}{ \left( 1 - 2M/r\right)^2}\left[ E^2 - 
 \left( 1 -\frac{2M}{ r}\right)  m^2\right]^{3/2} \label{eqn:qnofe}
\end{equation}
in the WKB limit. [This result is essentially the same as the one contained in Eq.(\ref{hjmodes}); see
\cite{'tHooft:1985re,Padmanabhan:1986rs}.] This expression diverges as $R\to 2M$ showing that
a scalar field propagating  in a black hole spacetime has infinite phase volume. The corresponding entropy
computed using the standard relations:
\begin{equation} 
S = \beta \left[ \frac{\partial}{ \partial \beta} - 1 \right] F; \qquad
F = - \int_0^\infty dE \frac{n(E)}{e^{\beta E} - 1 },
\end{equation}
is quadratically divergent:  $S = (\mathcal{A}_H/l^2)$ with $l \to 0$. 
The divergences described above occur around any infinite
redshift surface and is a geometric (covariant) phenomenon. 
  
   The same result can also be obtained from what is known as
   ``entanglement entropy'' arising from the quantum correlations 
    which exist across the horizon. 
    We saw in Section \ref{thermalden} that if
     the field configuration inside the horizon
    is  traced over in the vacuum functional of the theory, then one obtains a density matrix $\rho$
    for the field configuration outside [and vice versa]. The entropy $ S=-Tr(\rho \ln \rho)$
    is usually called the entanglement entropy. This is essentially the same as the previous 
    calculation and, of course, $S$ diverges quadratically on the horizon
    \cite{Frolov:1993ym,Zurek:1985gd,Dowker:1994fi,Israel:1976ur,Callan:1994py}. Much of this can be done 
    without actually bringing in gravity anywhere; all that is required is a spherical
    region inside which the field configurations are traced out \cite{Bombelli:1986rw,Srednicki:1993im}.  
Physically, however,  it does not seem reasonable to integrate over all modes without any cut off in these calculations.
By cutting off the mode at 
   $l \approx L_P$ one can obtain the ``correct'' result but in the absence of a 
   more fundamental argument for regularising the modes, this result is not of 
   much significance.  The cut off can be introduced in a more sophisticated manner by changing the dispersion relation near Planck energy scales but again there are different prescriptions that are available 
\cite{Padmanabhan:1998jp,Padmanabhan:1998vr,Unruh:1994zw,Corley:1996ar} and none of them are really convincing.

   The entropy computed using any non gravitational degrees of 
    freedom will scale in proportion with the number, $g_s$, of the species of fields
    which exist in nature. 
   This  does not cause a (separate) problem since one can
    re-absorb it in the renormalisation of gravitational constant $G$. In any calculation of effective action for a quantum field  in curved spacetime,
     one will obtain a term proportional to $R$ with a quadratically divergent coefficient. This 
     coefficient is absorbed by renormalising the gravitational constant and this procedure will also take care of $g_s$.

    In conventional description, entropy is also associated with the amount of missing information
    and one is tempted to claim that information is missing inside the horizon of black hole thereby
   leading to the existence of non zero entropy.  It is important to distinguish carefully the separate roles played by the horizon and singularity in this case; let us, for a moment, ignore the black hole singularity inside the horizon. Then
   the fact that a horizon hides information is no different from the fact
   that the information contained in a room is missing to those who refuse to enter the room.
   The observers at $(r, \theta, \phi) =$ constant in the Schwarzschild metric do not venture into
   the horizon and hence cannot access the information at $r<2M$. Observers who are comoving with
   the collapsing matter, or even those who plunge into the horizon later on, can access 
   (at least part of) the  information
   which is not available to the standard Schwarzschild observers at $r>2M$. In this respect, there is no difference
   between a Rindler observer in flat spacetime and a $(r, \theta, \phi) =$ constant observer
   in the Schwarzschild spacetime (see figure \ref{hyperfig}) and it is irrelevant what happens to the information content of matter which has collapsed inside the event horizon. 
   The information missing due to a horizon is observer dependent 
   since --- as we have stressed before --- the horizon is defined with respect to
   a congruence of timelike curves (``family of observers''). If one links the black hole entropy
   with the missing information then the entropy  too will become observer dependent.

In the examples which we have discussed in the previous Sections, the  thermal density matrix and temperature of the 
   horizon indeed arose from the integration of modes which are hidden by the horizon.
In the case of a black hole formed by collapse, there is a well defined, non singular, description of physics 
in the asymptotic past. As the system evolves, the asymptotic future is made of two parts. One part is outside the horizon and the other part (classically) hits a singularity  inside the horizon. The initial quantum state has now evolved to a correlated state with one component inside the horizon and one outside. If we trace over the states inside the horizon, the outside will be described by a density matrix. None of this is more mystifying than the usual phenomenon in quantum theory of starting with a correlated quantum state of a system with two parts (say, two electrons each having two spin states), spatially
separating the two components and tracing over one of them in describing the (spatially) localised measurements made on the other.
 There is no real information loss paradox in such systems.  
  
 In the case of the black hole there is an additional complication that the matter
   collapses to a singularity classically taking the information along with it. In this description, some of the information will be missing even to those observers who dare to plunge inside the event horizon. But, as we said before, this issue cannot be addressed until the problem of final singularity is solved.
   We have no idea what happens to the matter (or the wave modes of the 
   quantum field) near the singularity and as such it is not possible to do a book keeping on the 
   entropy content of matter inside the black hole. 
   As the black hole evaporates, its mass will decrease but such a semi-classical calculation
   cannot be trusted at late stages. 
   
   There is considerable discussion in the literature on the 
   ``information loss problem'' related to this issue. Broadly speaking, this problem arises
   because the evolution seems to take a pure quantum state to a state with significant amount
   of thermal radiation. It is, however, difficult even to attempt to tackle this problem properly
   since physics loses its predictive power at a singularity. One cannot meaningfully ask
   what happens to the information encoded in the matter variables which collapses to a singularity.
   So to tackle this question, one needs to know the correct theory which replaces the singularity.
   If for example, a Planck size remnant is formed inside the event horizon then one needs to ask
   whether a freely falling observer can retrieve most of the information at late stages
   from this remnant.
   [Some of the discussions in the literature also mixes up results obtained in different 
   domains with qualitative arguments for the concurrent validity. For example, one key assumption
   in the information loss paradox is that the initial state is pure. It is far from obvious that in a fully quantum
   gravitational context a pure state will collapse to form a black hole \cite{Myers:1997qi}].

 One immediate consequence, of linking entropy of horizons to the information hidden by them, is that all horizons must be attributed an 
   entropy proportional to its area, with respect to the observers who perceive this horizon. More 
   precisely,  given  a congruence of timelike curves in a spacetime we define the horizon to be
   the boundary of the union of the causal  pasts of the congruence. Assuming this is non trivial
   surface, observers moving on this congruence will attribute a constant entropy per unit area
   $(1/4L_P^2)$ to this horizon. (We shall  say more about this in Section \ref{thermoroute}.)
 The analysis given in Section \ref{complexplane} [see Eq.~(\ref{pofefromsofe})] shows that
   whenever a system crosses the horizon with energy $E$, the probability picks
   up a Boltzmann factor related to the entropy. In the case of a spherically symmetric
   horizon, one can imagine thin shells of matter carrying some amount of energy being 
   emitted by the horizon. This will lead to the correct identification of entropy for the 
   horizon. It is conceivable that similar effect occurs whenever a packet of energy crosses the 
   horizon even though it will be difficult to estimate its effect on the surface gravity of the horizon.
   Naive attempts to  compute the corresponding results for other geometries will not work and a
      careful formalism using the entropy density of horizons --- which is currently 
   not available --- will be required.

\subsection{Black hole entropy in quantum gravity models}\label{qgbh}

The above discussion highlights the fact that any model for quantum gravity, which has something to say about the black hole singularity, will also make definite predictions about the entropy of the black hole. There has been considerable amount of work
in this direction based on different candidate models for quantum gravity. We will summarise some aspects of this briefly.
[More extensive discussions as well as references to original literature can be found in the reviews \cite{Rovelli:1998yv,Das:2000su}].
The central idea in any of these approaches is to introduce microscopic degrees of freedom so that one can 
 attribute  large number of microscopic states to solutions that could be taken to represent a classical black hole configuration. By counting
these microscopic states, if one can show that $g(E)\propto\exp(\alpha E^2)$, then it is usually accepted as an explanation
 of black hole entropy.

In standard string theory this is done as follows: There are certain special states in string theory, called BPS states
\cite{Peet:2000hn}, that contain electric and magnetic charges which are equal to their mass. Classical supergravity has these states as classical solutions, among which are the extremal black holes with electric charge equal to the mass (in geometric units). These solutions can be expressed as a Reissner-Nordstrom metric with both the roots of $g_{00}=0$ coinciding: obviously, the surface gravity at the horizon, proportional to $g_{00}'(r_H)$ vanishes though the horizon has finite area. Thus these black holes,
classically, have zero temperature but finite entropy. Now, for certain compactification schemes in string theory (with
$d=3,4,5$ flat directions), in the limit of
$G\to 0$, there exist BPS states which have the same mass, charge and angular momentum of an extremal black hole in $d$
dimensions. One can explicitly count the number of such states in the appropriate limit and one finds that the result gives the exponential of black hole entropy with correct numerical factors \cite{Strominger:1996sh,Das:2000su,Breckenridge:1996sn}.
 This is done in the weak coupling limit
and a duality between strong coupling and weak coupling limits \cite{Sen:1994yi,Vafa:1994tf,Sen:1999mg}
is used to argue that the same result will arise in the 
strong gravity regime. Further, if one perturbs the state slightly away from the BPS limit, to get a {\it near} extremal black hole
and construct a thermal ensemble, one obtains the standard Hawking radiation from the corresponding near extremal black hole
 \cite{Breckenridge:1996sn}.

While these results are intriguing, there are several issues which are still open: First, the extremality or near extremality was used crucially in obtaining these results. We do not know how to address the entropy of a normal Schwarzschild black hole which is far away from the extremality condition. Second, in spite of significant effort, we do not still have a clear idea of
 how to handle the classical singularity or issues related to the information loss paradox. This is disappointing since one might have hoped that these problems are closely related. Finally, the result is very specific to black holes. One does not get any insight into the structure of other horizons, especially De Sitter horizon, which does not fit the string theory structure in a natural manner.

The second approach  in which some success related to black hole entropy is claimed, is in the  loop
quantum gravity (LQG). While string theory tries to incorporate all interactions in a unified manner, loop quantum gravity 
\cite{Rovelli:1998gg,Thiemann:2001yy}
 has the limited goal of providing a canonically quantised version of Einstein gravity. One key result which emerges from
this programme is a quantisation law for the areas. The variables used in this approach are like
a gauge field $A^i_a$ and the Wilson lines associated with them. The open Wilson lines carry a quantum number $J_i$ with them
and the area quantisation law can be expressed in the form: $\mathcal{A}_H=8\pi G\gamma\sum \sqrt{J_i(J_i+1)}$ where $J_i$ are spins defined on the links $i$ of a spin network and $\gamma$ is free parameter called Barbero-Immirizi parameter.
The $J_i$ take half-integral values if the gauge group used in the theory is SU(2) and take integral values if the gauge group is SO(3).
These quantum numbers, $J_i$, which live on the links that intersect a given area, become undetermined if the area refers to a horizon. Using this, one can count the number of microscopic configurations contributing to a given horizon area and estimate the entropy.
One gets the correct numerical factor (only) if $\gamma=\ln m/2\pi \sqrt{2}$ where $m=2$ or $m=3$ depending on whether the gauge
group SU(2) or SO(3) is  used in the theory \cite{Krasnov:1998wc,Rovelli:1996dv,Ashtekar:2000eq,Ashtekar:1998yu}.

Again there are several unresolved issues. To begin with, it is not clear how exactly the black hole solution arises in this approach since it has been never easy to arrive at the low energy limit of gravity in LQG. Second, the answer depends on the Immirizi parameter $\gamma$ which needs to be adjusted to get the correct answer, if we know the correct answer from elsewhere.
Even then, there is an ambiguity as to whether one should have SU(2) with  $\gamma=\ln 2/2\pi \sqrt{2}$ or SO(3) with
 $\gamma=\ln 3/2\pi \sqrt{2}$. The SU(2) was the preferred choice for a long time, based on its close association with fermions
which one would like to incorporate in the theory. However, recently there has been some rethinking on this issue due to the following consideration: For a classical black hole, one can define a class of solutions to wave equations called quasi normal modes [see e.g.,\cite{Corichi:2002ty,Kokkotas:1999bd,Berti:2003jh,Cardoso:2003vt}]. 
These modes have discrete frequencies which are complex, given by
\begin{equation}
\omega_n=i\frac{n+(1/2)}{4M}+\frac{\ln(3)}{8\pi M}+\mathcal{O}(n^{-1/2})
\end{equation}
The $\ln(3)$  in the above equation is not negotiable
\cite{Padmanabhan:2003fx,Choudhury:2003wd,Motl:2003cd,Motl:2002hd}. If one chooses SO(3) as the gauge group, then one can connect up the frequency 
of quanta emitted by a black hole when the area changes by one quantum in LQG with the quasi normal mode frequency
\cite{Dreyer:2002vy,Hod:1998vk}. It is not clear whether this is a coincidence or of some significance. Third, most of the details of the LQG are probably not relevant to the computation of the entropy. Suppose we have {\it any} formalism of quantum gravity in
which there is a minimum quantum for length or area, of the order of $L_P^2$. Then, the horizon area $\mathcal{A}_H$ 
can be divided into $n=(\mathcal{A}_H/c_1L_P^2)$ patches where $c_1$ is a numerical factor. If each patch has $k$ degrees of freedom
(due to the existence of a surface field), then the total number of microscopic states are $k^n$ and the resulting entropy is
$S=n\ln k=(4\ln k/c_1)(\mathcal{A}_H/4L_P^2)$ which will give the standard result if we choose $(4\ln k/c_1)=1$. The essential ingredients
are only discreteness of the area and existence of certain degrees of freedom in each one of the patches. 

Another key issue in counting the degrees of freedom is related to the effective dimensionality. If we repeat the above argument with the volume $V\propto M^3$ of the black hole then one will get an entropy proportional to the volume rather than area. It is clear that, near a horizon, only a region of length $L_P$ across the horizon contributes the microstates so that in 
the expression $(V/L_P^3)$, the relevant $V$ is $M^2 L_P$ rather than $M^3$. It is possible to interpret this as due to the entanglements of modes across the horizon over a length scale of $L_P$, which --- in turn --- induces a nonlocal coupling between the modes on the surface of the horizon. Such a field will have one particle excitations, which have the same density of states as black hole \cite{Padmanabhan:1998jp,Padmanabhan:1998vr}. While this is suggestive of why we get the area scaling rather than volume scaling, a complete understanding is lacking. 
The area scaling of entropy has also led to different proposals of holographic bounds [see, e.g. \cite{Bousso:2002ju}]
 which is beyond the scope of this review.

\section{The thermodynamic route to gravity}\label{thermoroute}

Given the fact that entropy of a system
     is closely related to accessibility of information, it is inevitable that there will be
    some connection between gravity and thermodynamics. To bring this out, 
   it is useful to recollect the way Einstein handled the principle of 
    equivalence and  apply it in the present context. Einstein did not attempt to ``derive" principle of equivalence in the conventional 
    sense of the word. Rather, he accepted it as a key feature which must find expression in the 
    way gravity is described --- thereby obtaining  a geometrical description of gravity. 
    Once the geometrical interpretation of gravity is accepted, it follows that there {\it will} arise
    surfaces which act as one-way-membranes for information and  will thus lead to some connection
    with thermodynamics. It is, therefore, more in tune with the spirit of Einstein's analysis
    to {\it accept} an inevitable connection between gravity and thermodynamics and ask 
    what such a connection would imply. We shall now describe this procedure in detail.

 The existence of a class of observers with limited access to spacetime regions, because of the existence 
of horizons, is a generic feature. This, a priori, has nothing to do with the dynamics of general relativity or gravity; such examples exist even in flat spacetime. But when the spacetime is flat, one can introduce an additional ``rule'' that only the inertial coordinates  must be used to describe physics. While this appears to be artificial and 
ad hoc, it is logically tenable. It is the existence of gravitational interaction, which makes spacetime curved, that removes this option and forces us to consider different curvilinear coordinate systems. Further, gravity  makes these  phenomena
related to horizons appear more natural in certain contexts, as in the case of black holes. A
 region of spacetime, described in some coordinate system with a non-trivial metric tensor $g_{ab}(x^k)$,
can then have a light cone structure such that information about one sub-region
is not accessible to observers in another region.

Such a limitation  is \emph{always}  dependent on the family of observers with respect to which the horizon is defined. 
To appreciate this fact, let us note that the  freedom of choice
of the coordinates allows 4 out of 10 components of the metric tensor to be
pre-specified, which we shall take to be $g_{00}=-N^2,~g_{0\alpha}=N_\alpha$.
These four variables allow us to characterise the observer-dependent information.
For example,  with the choice $N=1,N_\alpha=0,g_{\alpha\beta}=\delta_{\alpha\beta}$, 
the ${\bf x}=$ constant trajectories
correspond to a class of inertial observers in flat spacetime  while with $N=(ax)^2,N_\alpha=0,g_{\alpha\beta}=\delta_{\alpha\beta}$ the  ${\bf x}=$ constant trajectories represent a class of accelerated observers with a horizon at $x=0$. We  only need to change the form 
 of $N$ to make this transition in which a class of time-like trajectories, ${\bf x}=$ constant, acquire a horizon.  Similarly observers plunging into a black
hole  will find it natural to describe  the Schwarzschild metric 
in the synchronous gauge
with  $N=1,N_\alpha=0$ (see e.g.,  \cite{lltwo}) in which they  can indeed access 
the information contained inside the horizon.
The less masochistic observers will use a more standard foliation which has $N^2=(1-2M/r)$ and the surface $N=0$ will act as the horizon which restricts the flow of information from $r<2M$ to the observers at $r>2M$.

This aspect, viz. that different observers (defined as different families of timelike curves)
may have access to different
regions of space-time and hence differing amount of information, introduces
a very \emph{new} feature into physics.
It is now necessary to ensure  that
\emph{physical theories in a given coordinate system are formulated
entirely in terms of the variables that an observer using that coordinate
system can access}  \cite{Padmanabhan:2003ub}.
This ``principle of effective theory'' is analogous to the renormalisation group arguments
used   in high energy physics
which
``protects" the low energy theories from the unknown complications of the
high energy sector. For example, one can use QED to predict results
at, say, $10$ GeV without worrying about the structure of the theory at
$10^{19}$ GeV, as long as one uses coupling constants and variables
defined around $10$ GeV and determined observationally.
In this case, one invokes the effective field theory approach in the
momentum space.
We can introduce the same reasoning in coordinate space 
and demand---for example---that the observed physics outside a black hole
horizon must not depend on the unobservable processes beyond the horizon.

In fact, this  is a natural extension of a more
conventional procedure used in flat spacetime physics.
Let us recall that, in standard description of 
flat spacetime physics, one often
divides the spacetime by a space-like surface $t=t_1$=constant. Given the necessary information
on this surface, one can predict the evolution for $t>t_1$ {\it without knowing the details at} $t<t_1$.
In the case of curved spacetime with horizon, similar considerations apply. For example, if the
 spacetime contains a Schwarzschild black hole, say, then the light cone structure guarantees
that the processes inside the black hole horizon cannot affect the outside events {\it classically}.
What makes our demand non trivial is the fact that the 
situation in {\it quantum theory} is quite different.  Quantum fluctuations of fields 
will  have
nontrivial correlations across the horizon 
which is indicated by the fact that the propagators do not vanish for spacelike separations.
(Alternatively, QFT in the Euclidean sector probes the region beyond the horizon.)
 Our principle of effective theory states that it must be possible to
``protect" the physical processes outside the horizon from such effects influencing it across the horizon.

   For a wide class of horizons which we have 
   discussed, the region inside the horizon (essentially the $\mathcal{F}$ and $\mathcal{P}$
   of the maximally extended Kruskal-type coordinate systems) disappears ``into'' the origin
   of the Euclidean coordinate system. The principle of effective theory requires that one should
   deal with the corresponding effective manifold in which the region that is inaccessible to a family
   of observers is removed. In the examples studied in the earlier Sections, this required removing 
   a point (say,  the origin) from the $X-T$
  plane  in the Euclidean manifold. 
The standard results of quantum field theory in coordinate systems with static horizons can be obtained from this approach. We shall now proceed to study \emph{gravity} from this approach.

In the case of gravity, the information regarding the region inside the horizon will now manifest in two different forms.
  First, as a periodicity in the imaginary time coordinate and non trivial winding number 
  for paths which circle the   point which is removed.
  Second, as a boundary term in the Euclidean action for gravity, since the 
   Euclidean action needs to be defined carefully taking into 
  account any contribution which arises from an infinitesimal region around the point which is 
  removed. 
  
  The origin in the Euclidean spacetime translates to the horizon surface in 
  the Lorentzian spacetime. If we choose to work entirely in the Lorentzian spacetime,
  we need to take care of the above two effects by: 
  (i) restricting the time integration to a suitable (finite) range in defining the action and
  (ii) having a suitable surface term to the action describing gravitational dynamics which
  will get a contribution from the horizon. 
Since the horizon surface is the only common element to the inside and outside regions, the effect of 
the quantum  entanglements across a horizon can only  appear as a surface term in the action. So
  it is an inevitable consequence of principle of equivalence that 
   the action functional describing gravity \emph{must} contain certain
boundary terms which are capable of encoding the information equivalent
to that present beyond the horizon.  
  We shall now see that this surface term can be determined from  general principles and, 
  in fact, one can deduce the form of the full action for gravity using this approach \cite{Padmanabhan:2003grg}.
  
  Before we begin the detailed discussion,  we mention related approaches exploring the connection between
  thermodynamics and gravity at different levels. Many people have attempted to relate the 
  thermodynamics of gravity and matter systems to the Euclidean action
 \cite{Gibbons:1977ue,York:1986it,Braden:1990hw,Brown:1991fk,Martinez:1989hn,Hayward:1990zm,Hayward:1991}. 
  Some of these papers also discuss the derivation of laws of thermodynamics as applicable to
  matter coupled to gravity. An attempt to derive Einstein's equations from thermodynamics,
  which is closer in spirit to the discussion presented here, was made by \cite{Jacobson:1995ab}
  but this work did not unravel the structure of gravitational action functional. Several
  intriguing connections between not only gravitational systems but even other field 
  theoretic phenomena and condensed matter systems have been brought out by
  \cite{Volovik:2000ua,Volovik:2003fe}.
  
Let us now proceed with our programme.
In order to provide a local, Lagrangian, description of gravitational physics,
this boundary term must be expressible as an integral of a four-divergence, allowing us to write the action functional for gravity
formally as
\begin{equation}
A_{\rm grav} = \int d^4 x \sqrt{-g}\, L_{\rm grav} = \int d^4 x \sqrt{-g} \, \left( L_{\rm bulk}+ \nabla_i U^i \right)
 = A_{\rm bulk} + A_{\rm sur}
\label{firsteqn}
\end{equation}
where $L_{\rm bulk}$ is quadratic in the first derivatives of the metric and we are using the
convenient  notation $\nabla_i U^i\equiv (-g)^{-1/2}\partial_i[(-g)^{1/2}U^i]$ irrespective of whether $U^i$ is a genuine four
vector or not.  Since different families of  observers will have different levels of accessibility
 to information, we do expect
$A_{\rm sur}$ to depend on the foliation of spacetime. On the other hand,
since the overall dynamics should be the same for all observers, $A_{\rm grav}$ should be a scalar. 
It follows that neither $A_{\rm bulk}$ nor $A_{\rm sur}$ are covariant  but
their sum should be a covariant scalar.

Let us first determine the form of $A_{\rm sur}$.
The horizon for a class of observers arises in  a specific gauge and resultant $A_{\rm sur}$ 
will in general depend on the gauge variables $N,~N_\alpha$. 
Of the gauge variables $N,N_\alpha$, the lapse function $N$
plays a more important role in our discussion than $N_\alpha$,
and we can set $N_\alpha=0$ without loss of generality. 
The residual gauge (co-ordinate) transformation that keeps $N_\alpha=0$ but changes the other components of the metric is given by the \emph{infinitesimal} space-time transformation
$x^i \to x^i + \xi^i(x^j)$, with the condition
$g_{\alpha\beta}\dot\xi^\beta= N^2 (\partial \xi^0/\partial x^\alpha)$,
which is equivalent to
\begin{equation}
\xi^\alpha = \int dt~N^2 g^{\alpha\beta}
              \frac{\partial\xi^0}{\partial x^\beta} + f^\alpha(x^\beta) ~.
\label{largegauge}
\end{equation}
Such transformations keep $N_\alpha=0$, but change $N$ and $g_{\alpha\beta}$
according to $\delta g_{ij} = - \nabla_i\xi_j - \nabla_j\xi_i$ (see e.g.,  \cite{lltwo}, \S97).

We next introduce a
$(1+3)$ foliation with the standard notation for the metric components
($g_{00}=- N^2,~g_{0\alpha}=N_\alpha$).
Let $u^i=(N^{-1},0,0,0)$ be the four-velocity of observers
corresponding to this foliation, i.e. the normal to the foliation;
 $a^i=u^j\nabla_ju^i$ be the related acceleration; and
 $K_{ab}=-\nabla_a u_b - u_aa_b$ be the extrinsic curvature
of the foliation, with $K\equiv K^i_i = -\nabla_i u^i$.
(With this standard definition, $K_{ab}$ is purely spatial,
$K_{ab}u^a=K_{ab}u^b=0$; so one can work with the spatial components
$K_{\alpha\beta}$ whenever convenient.)

Given this structure,
we can  list all possible vector fields $U^i$ which can be  used in
Eq.~(\ref{firsteqn}). This vector has to be built out of $u^i, g_{ab}$ and
the covariant derivative operator $\nabla_j$ acting only once.
The last restriction arises because the equations of motion
should be of no order higher than two. Given these conditions,
(i) there is only one vector field --- viz., the $u^i$ itself --- which has no derivatives  and (ii) only 
 three vectors
$(u^j\nabla_j u^i, u^j\nabla^i u_j, u^i\nabla^j u_j)$ which are linear in covariant derivative operator. The first one is
the acceleration $a^i=u^j\nabla_j u^i$; the second identically vanishes since $u^j$ has unit norm; the third can be written as
$-u^i K$. Thus 
$U^i$ in the surface term  must be  a linear combination of $u^i, u^i K$ and $a^i$ at the lowest order. The corresponding
term in the action \emph{must} have the form 
\begin{equation}
\label{threedaction}
A_{\rm sur}=\int d^4 x \, \sqrt{-g} \nabla_i U^i 
=
\int d^4 x \, \sqrt{-g} \nabla_i \left[ \lambda_0 u^i+\lambda_1 K u^i + \lambda_2 a^i\right]
\end{equation}
where $\lambda$'s are numerical constants to be determined.

Let the region of integration be a  four volume
$\mathcal{V}$ bounded by two space-like surfaces $\Sigma_1$ and $\Sigma_2$
and two time-like surfaces $\mathcal{S}$ and $\mathcal{S}_1$.
The space-like surfaces are constant time slices with normals $u^i$,
and the time-like surfaces have normals $n^i$ and we shall choose $n_iu^i=0$.
The induced metric on the space-like surface $\Sigma$ is
$h_{ab} = g_{ab} +u_au_b$, while the induced metric on the time-like
surface $\mathcal{S}$ is $\gamma_{ab} = g_{ab} -n_an_b$.
These two surfaces intersect on a two-dimensional surface $\mathcal{Q}$,
with the induced metric
$\sigma_{ab} = h_{ab} - n_an_b = g_{ab} +u_au_b -n_an_b$. 
In this foliation, the first two terms of Eq.~(\ref{threedaction}) contribute only on the $t=$ constant hypersurfaces
($\Sigma_1$ and $\Sigma_2$)
 while the third term 
 contributes on $\mathcal{S}$ and hence on a horizon
  (which we shall treat as the null limit of a time-like surface $\mathcal{S}$, like the limit
$r\to 2M+$ in the black hole spacetime). Hence we get, on the horizon, 
\begin{equation}
A_{\rm sur} = \lambda_2 \int d^4x\sqrt{-g}~\nabla_ia^i
= \lambda_2 \int_\mathcal{S} dt d^2x~N \sqrt{|\sigma|}
(n_\alpha a^\alpha) 
\label{boundary}
\end{equation} 
Further, in any static spacetime with a horizon: (i) The integration over $t$ becomes multiplication by $\beta\equiv 2\pi/\kappa $ where $\kappa$ is the surface gravity of the horizon, 
since there is a natural periodicity in the Euclidean sector.  (ii) As the surface
  $\mathcal{S}$ approaches the horizon, the quantity $N (a_in^i)$ tends to $-\kappa$ 
  which is constant over the horizon. (see e.g., \cite{Brown:1995su} as well as the discussion at the 
  end of Section \ref{stexample}).
\footnote{The minus sign in $(-\kappa)$ depends on the convention 
        adopted for $n_\alpha$. It arises naturally
        under two circumstances. First is  when the region 
        outside the horizon is treated as bounded on one side by the horizon and 
        $n_\alpha$ is the outward normal as perceived from the outside observers. 
        Second, when we take the normal to the horizon to be pointing to the outside 
        (like in the direction of unit vector $\hat{\bf r}$ in Schwarzschild geometry) but we take 
        the contribution to the surface integral from two surfaces (at $r\to\infty$ and $r\to 2M$ 
        in the Schwarzschild spacetime) and subtract one from the another. 
        The horizon contributes at the lower limit of the integration and picks up a minus sign.}
 Using $\beta \kappa=2\pi$, the surface term gives, on the horizon, the 
  contribution
  \begin{equation}
A_{\rm sur} =- \lambda_2 \kappa \int_0^\beta dt\int d^2x\, \sqrt{\sigma} = -2\pi \lambda_2 \mathcal{A}_H
  \label{horizonarea}
  \end{equation}
  where $\mathcal{A}_H$ is the area of the horizon. 

It is interesting to ask  how the above result arises if we choose to work entirely in Euclidean spacetime. Such an exercise is important for two reasons. First, the range of integration for time coordinate has a natural limit only in Euclidean
sector and while obtaining Eq.(\ref{horizonarea}) we have  ``borrowed'' it and used it in the Lorentzian sector; it will be nice to see it in the proper context. Second, in the Euclidean sector, there is no light cone and horizon gets mapped to the origin of the $t_E-x$ plane. In the effective manifold, we would have removed this point and the surface term has to arise from a limiting procedure. It is important to see that it works correctly. We shall now briefly discuss the steps involved in this analysis. 

Consider a simply connected, compact region of the Euclidean manifold $\mathcal{M}$ with two bounding surfaces $\mathcal{S}_0$ and $\mathcal{S}_\infty$, where $\mathcal{S}_0$ encloses a small region around the origin (which corresponds to the horizon in our coordinate system) and $\mathcal{S}_\infty$ is an outer boundary at large distance which 
we really do not care about. We assume that the region $\mathcal{M}$ is foliated by such surfaces and the normal to the surface defines a vector field $u^i$. The earlier arguments now show that the only non-trivial terms we can use in the Lagrangian are again of the form in Eq.(\ref{threedaction}) but the nature of boundary surfaces have now changed.
 We are interested in the  contribution from the inner boundary near the origin, where we can take the metric to be approximately Rindler:
\begin{equation}
ds^2_E\approx (\kappa x)^2 dt_E^2+dx^2+dL_\perp^2
\end{equation}
and the inner surface to be $S^1\times R^2$ where $S^1$ is small circle around the origin in the $t_E-x$ plane and $R^2$ is the transverse plane. While evaluating Eq.~(\ref{threedaction}), the 
  integral of
$\nabla_ia^i$ will now give $a_iu^i=0$ on the boundary while the integral of
$\nabla_iu^i$ will now give $u_iu^i=1$, leading to the area of the boundary. In the limit of the radius of $S^1$ going to zero, this contribution from $\nabla_iu^i$  vanishes. The interesting contribution comes from
the integral of
$\nabla_i(Ku^i)$ term, which will  give the integral of  $K=-\nabla_iu^i$ on the boundary.  Taking $u^i=\delta^i_x$ we get the contribution
\begin{equation}
-\lambda_2 \int d^2x_\perp\int_0^{2\pi/\kappa}dt_E\partial_x(\kappa x)=-2\pi \lambda_2 \mathcal{A}_H
\label{euarea}
\end{equation}
exactly as in Eq.(\ref{horizonarea}). This analysis, once again, demonstrates the consistency of working in an effective manifold with the origin removed.

Treating the action as analogous to entropy, we see that 
 the information blocked by
a horizon, and encoded in the surface term,  {\it must be} proportional to the area of the horizon. Taking into consideration the non compact horizons, like
the Rindler horizon, we may state that the entropy (or the information content) per unit area of the horizon is a constant related to $\lambda_2$. Writing $\lambda_2\equiv -(1/8\pi \mathcal{A}_P)$, where $\mathcal{A}_P$ is a fundamental constant
with the dimensions of area, the entropy associated with the horizon will be $S_H=(1/4)(\mathcal{A}_H/\mathcal{A}_P)$. The numerical factor in $\lambda_2$ is chosen for later convenience; the sign is chosen so that $S\ge 0$.

Having determined the form of $A_{\rm sur}$ we now turn to the nature of $A_{\rm grav}$ and 
$A_{\rm bulk}$. We need to express the Lagrangian $\nabla_i U^i$ as a difference between
two Lagrangians $L_{grav}$ and $L_{bulk}$ such that: (a) $L_{grav}$ is a generally covariant scalar.
(b) $L_{bulk}$ is utmost quadratic in the time derivatives of the metric tensor. 
(c) Neither $L_{\rm grav}$ nor $L_{\rm bulk}$ should  contain  four divergences since such terms are
already taken into account in $L_{\rm sur}$. 
This is 
just an exercise in differential geometry and leads to Einstein-Hilbert action. Thus it is possible to
obtain the full dynamics of gravity purely from thermodynamic considerations
\cite{Padmanabhan:2003grg}. We shall, however, obtain this result in a slightly different manner which throws light on certain peculiar features of  Einstein-Hilbert action, as well
as the role played by local Lorentz invariance.

\subsection{Einstein-Hilbert action from spacetime thermodynamics}\label{grfresults}

 Since
the field equations of gravity are generally covariant
and of second order in the metric tensor,
one would naively expect these equations to be derived from an action
principle involving $g_{ab}$ and its first derivatives $\partial_k g_{ab}$,
analogous to the situation for many other field theories of physics.
The arguments given in the last Section show that the existence of horizons (and the principle of effective
theory) suggest that the gravitational Lagrangian will have a term $\nabla_i U^i$ [see Eq.~(\ref{threedaction})] which contains {\it second} derivative of $g_{ab}$.

While any such Lagrangian can describe the classical physics correctly, there are some  restrictions which quantum theory imposes on Lagrangians with second derivatives. Classically, one can postulate that
the equations of motion are obtained by varying an action with some arbitrary function $f(q,\dot q)$ of $q$ and $\dot q$  held fixed at the end points. Quantum mechanically, however, it is natural to demand that either $q$ or $p\equiv (\partial L/\partial \dot q)$ is held fixed rather than a mixture of the two.  This criterion finds a natural description in the path integral approach to quantum theory. 
If one uses the
coordinate representation  in non-relativistic quantum mechanics, the probability amplitude for the dynamical
variables to change from $q_1 $ (at $t_1$) to $q_2$ (at $t_2$) is given by
\begin{equation}
\psi(q_2,t_2) = \int dq_1 K \left( q_2,t_2;q_1,t_1 \right) \psi(q_1,t_1) ~,
\end{equation}
\begin{equation}
K \left( q_2,t_2;q_1,t_1 \right) = \sum\limits_{\rm paths}
\exp \left[ \frac{i}{\hbar} \int dt~L_q(q,\dot q) \right] ~,
\label{qsopa}
\end{equation}
where the sum is over all paths connecting $(q_1,t_1)$ and $(q_2,t_2)$,
and the Lagrangian $L_q(q,\dot q)$ depends on $(q,\dot q)$. 
It is, however, quite possible to study the same system in momentum space,
and enquire about the amplitude for the system
to have a momentum $p_1$ at $t_1$ and $p_2$ at $t_2$. 
From the standard rules of quantum theory, the amplitude for the particle
to go from $(p_1,t_1)$ to $(p_2,t_2)$ is given by the Fourier transform
\begin{equation}
G \left( p_2,t_2;p_1,t_1 \right) \equiv \int dq_2 dq_1
~K \left( q_2,t_2;q_1,t_1 \right) 
\ \exp \left[ -\frac{i}{\hbar} \left( p_2 q_2 - p_1 q_1 \right) \right]
\label{qftofq} 
\end{equation}
Using Eq.~(\ref{qsopa}) in Eq.~(\ref{qftofq}), we get
\begin{eqnarray}
G \left( p_2,t_2;p_1,t_1 \right)
& = &\sum\limits_{\rm paths}
\int dq_1 dq_2 \exp \left[ \frac{i}{\hbar}
\left\{ \int dt~L_q - \left( p_2 q_2 - p_1 q_1 \right) \right\} \right]
\nonumber \\
&=&\sum\limits_{\rm paths} \int dq_1 dq_2 \exp \left[ \frac{i}{\hbar}
\int dt \left\{ L_q - \frac{d }{ dt} \left( pq \right) \right \} \right]
\nonumber \\
& \equiv& \sum_{\rm paths} {} \exp \left[ \frac{i}{\hbar}
\int L_p(q, \dot q, \ddot q)~dt \right] ~.
\label{lp}
\end{eqnarray}
where
\begin{equation}
L_p \equiv L_q - \frac{d}{ dt} \left( q\frac{\partial L_q }{ \partial\dot q} \right) ~.
\label{lbtp}
\end{equation}
In arriving at the last line of Eq.~(\ref{lp}), we have
(i) redefined the sum over paths to include integration over $q_1$ and $q_2$;
and (ii) upgraded the status of $p$ from the role of a parameter in the
Fourier transform to the physical momentum $p(t)=\partial L/\partial \dot q$.
This result shows that, given any Lagrangian $L_q(q,\partial q)$
involving only up to the first derivatives of the dynamical variables,
it is \emph{always} possible to construct another Lagrangian
$L_p(q,\partial q,\partial^2q)$ involving up to second derivatives,
such that it describes the same dynamics but with different boundary
conditions \cite{Padmanabhan:2002xm,Padmanabhan:2002ma}.
The prescription is given by Eq.~(\ref{lbtp}).
While using $L_p$, one keeps the \emph{momenta}  fixed
at the endpoints rather than the \emph{coordinates}.
This boundary condition is specified by the subscripts on the Lagrangians.
The result generalises directly to multi-component fields and provides a natural interpretation of Lagrangians with second derivatives.

  Thus, in the case of gravity,  the {\it same}  equations
    of motion can be obtained from $A_{\rm bulk}$ or from another (as yet unknown) action:
    \begin{eqnarray}
    \label{aeh}
   A' &=& \int d^4x \sqrt{-g} L_{\rm bulk} - \int d^4x \partial_c \left[ g_{ab}
    \frac{\partial \sqrt{-g} L_{\rm bulk} }{ \partial(\partial_c g_{ab})}
    \right]   \nonumber\\
   &\equiv&  A_{\rm bulk} - \int d^4 x \partial_c (\sqrt{-g}V^c) 
    \end{eqnarray}   
    where $V^c $ is made of $g_{ab} $ and $\Gamma^i_{jk}$. Further, $V^c$ must be linear
    in the $\Gamma$'s since the original Lagrangian $L_{\rm bulk}$ was quadratic in the first derivatives
    of the metric. (This argument assumes that we have fixed the relevant dynamical variables $q$ of the system; in the case of gravity, we take these to be $g_{ab}$).
    Since $\Gamma$s vanish in the local inertial frame and the metric reduces to
    the Lorentzian form, the action $A_{\rm bulk}$ cannot be generally covariant.
    However, the action $A'$ involves the second derivatives  of the metric and 
    we shall see later that that the action $A'$ is indeed generally covariant.

    To obtain a quantity $V^c$, which is linear in $\Gamma$s
     and having  a single index $c$, from $g_{ab} $ and $\Gamma^i_{jk}$,
     we must contract on two of the indices on $\Gamma$
    using the metric tensor. 
    (Note that we require $A_{\rm bulk}$, $A'$ etc. to be Lorentz scalars and $P^c, V^c$ etc.
    to be vectors under Lorentz transformation.)
    Hence the most general choice for $V^c$ is the linear combination
      \begin{equation}
    V^c =  \left(a_1 g^{ck} \Gamma^m_{km} +a_2 g^{ik} \Gamma^c_{ik}\right) 
    \label{defvc}
     \end{equation}
     where $a_1(g)$ and $a_2(g)$ are unknown functions of the determinant $g$ of the metric
  which is the only (pseudo) scalar entity which can be constructed from $g_{ab}$s and $\Gamma^i_{jk}$s.
      Using the identities $\Gamma^m_{km} =\partial_k
     (\ln \sqrt{-g})$, \ $\sqrt{-g}g^{ik}\Gamma^c_{ik} = -\partial_b(\sqrt{-g}g^{bc})$,
     we can rewrite the expression for $P^c \equiv \sqrt{-g}V^c$ as 
     \begin{equation}
    P^c =\sqrt{-g}V^c=
    c_1(g)g^{cb} \partial_b \sqrt{-g} +c_2(g) \sqrt{-g} \partial_b g^{bc}
    \label{defpc}
    \end{equation}
    where $c_1\equiv a_1 - a_2,\  c_2\equiv -a_2$ are two other unknown functions of the determinant $g$. 
    If we can fix these coefficients by using a physically well motivated prescription, then
    we can determine the surface term and --- by integrating --- the Lagrangian $L_{\rm bulk}$.
   
 To do this,
    let us consider a static spacetime in which all $g_{ab}$s are
    independent of $x^0$ and $g_{0\alpha} =0$. Around any given event $\mathcal{P}$
    one can construct a local Rindler frame with an acceleration 
    of the observers with ${\bf x} $ = constant, given by $a^i = (0,{\bf a})$ and
    ${\bf a}= \nabla (\ln \sqrt{g_{00}})$.
    This Rindler frame will have a horizon which is a plane surface normal to the 
    direction of acceleration and a temperature $T=|{\bf a}|/2\pi$ associated with this horizon.
 The result obtained in Eq.(\ref{horizonarea}) shows that
  the entropy $S$ associated with
    this horizon is proportional to its area or, more precisely, 
    \begin{equation}
     \frac{dA_{\rm sur}}{d\mathcal{A}_\perp}= \frac{1}{ 4 \mathcal{A}_P} 
    \label{postulate}
    \end{equation}
    where $\mathcal{A}_P$ is a fundamental constant with the dimensions of area. 
    In particular, this result must hold in flat spacetime in Rindler coordinates.
     In the static Rindler frame, the surface term is
     \begin{equation}
     A_{\rm sur}  =- \int d^4 x \partial_c P^c = -\int_0^\beta dt \int_\mathcal{V} d^3 x
     \nabla \cdot {\bf P} 
     =\beta \int_{\partial\mathcal{V}} d^2 x_\perp \hat{\bf n}\cdot {\bf P}
     \label{keyeqn1}
     \end{equation}
The overall sign in the last equation depends on the choice of direction for $\hat{\bf n}$; we have chosen it
to be consistent with the convention employed earlier in Eq.~(\ref{horizonarea}).
     We  have restricted the time integration to an interval $(0,\beta)$
     where $\beta = (2\pi /|{\bf a}|) $ is the inverse temperature in the Rindler frame,
      since the Euclidean action will be periodic in the imaginary
     time with the period $\beta$. 
    We shall choose the  Rindler frame such that the acceleration is along
    the $x^1=x$ axis.
      The most general form of the metric representing the Rindler frame can be
  expressed in the form 
  \begin{eqnarray}
    \label{rindmetric}
  ds^2 &=& -(1+2al) dt^2 + \frac{dl^2}{(1+2al)} + (dy^2+dz^2) \\
  &=& -\left[ 1+2al(x) \right]dt^2 + \frac{l^{'2}dx^2 }{[1+2al(x)]} + (dy^2+dz^2)\nonumber
  \end{eqnarray}
  where $l(x)$ is an arbitrary function and $l' \equiv (dl/dx)$. Since the acceleration is 
  along the x-axis,  the metric in the transverse direction is unaffected. The first form of the metric is the standard
Rindler frame in the $(t,l,y,z)$ coordinates. We can, however, make any coordinate transformation
from $l$ to some other variable $x$ without affecting the planar symmetry or the static nature of the metric. This leads to the general form of the metric given in the second line, in terms of the $(t,x,y,z)$ coordinates.
  Evaluating the surface term $P^c$ in (\ref{defpc})
  for this metric, we get
  the only non zero component to be
  \begin{equation}
  P^x = 2ac_2(g) + [1+2al(x)] \frac{l''}{l^{'2}} [c_1(g)- 2 c_2(g)]
  \end{equation}
       so that the action in Eq.~(\ref{keyeqn1})
    becomes 
    \begin{equation}
    A_{\rm sur}=\beta P^x \int d^2x_\perp =\beta P^x \mathcal{A}_\perp
    \label{detcone}
    \end{equation}   
    where $\mathcal{A}_\perp$ is the transverse area of the $(y-z)$ plane.
     From  Eq.~(\ref{postulate}) it follows that
     \begin{equation}
     2 a \beta c_2 (g) + \beta [c_1 - 2 c_2] (1+2al) \frac{l''}{l^{'2}} = \frac{1}{4\mathcal{A}_P}
     \label{xxx}
     \end{equation}
    For the expression in the left hand side to be  a constant independent of $x$ for any choice of $l(x)$, 
  the second term must vanish
  requiring $c_1(g) = 2 c_2(g)$. 
  An explicit way of obtaining this result is to consider a class of functions $l(x)$ which satisfy the relation
  $l'=(1+2al)^n$ with $0\leq n\leq 1$. Then
  \begin{equation}
  \beta [c_1(l') - 2 c_2(l')] (1+2al) \frac{l''}{l^{'2}}=2a\beta[ c_1(l') - 2 c_2(l')]n
  \end{equation}
  which can be independent of $n$ and $x$ only if $c_1(g) = 2 c_2(g)$. 
  Further, using $a\beta = 2\pi$,
  we find that  $c_2(g)=(16\pi \mathcal{A}_P)^{-1}$
  which is a constant independent of $g$. 
    Hence $P^c$ has the form 
    \begin{eqnarray}
        \label{pcfix}
    P^c &=&   \frac{1}{16\pi \mathcal{A}_P} \left( 2g^{cb} \partial_b \sqrt{-g} + \sqrt{-g} \partial_b g^{bc}\right) 
    =\frac{\sqrt{-g}}{16\pi \mathcal{A}_P}  \left( g^{ck} \Gamma^m_{km} - g^{ik} \Gamma^c_{ik}\right) \nonumber\\
    &=&- \frac{1}{16\pi \mathcal{A}_P}\frac{1}{\sqrt{-g}} \partial_b(gg^{bc})
    \end{eqnarray}
    The second equality is obtained by using the standard identities mentioned after 
     Eq.~(\ref{defvc}) while the third equality follows directly by combining the 
    two terms in the first expression.

    The general form of $P^c$ which we obtained in Eq.~(\ref{defpc}) is not of any use unless
   we can fix  $(c_1,c_2)$. In general, this will not have 
   any simple form and will involve an undetermined
   range of integration over time coordinate.
   But in the case of gravity, two natural features conspire together to give an elegant form to this
   surface term. First is the fact that Rindler frame has a periodicity in Euclidean time and the range
   of integration over the time coordinate is naturally restricted to the interval $(0,\beta) = (0,2\pi/a)$.
   The second is the fact that the 
   surviving term in the 
   integrand $P^c$ is linear in the acceleration $a$ thereby neatly
   canceling with the $(1/a)$ factor arising from time integration. 
  
    Given the form of $P^c$ we need to solve the 
   equation   
    \begin{equation}
 \left(\frac{\partial \sqrt{-g}L_{\rm bulk}}{\partial g_{ab,c}}g_{ab}\right)=
 P^c= - \frac{1}{ 16\pi \mathcal{A}_P}\frac{1}{\sqrt{-g}} \partial_b(gg^{bc})
 \label{dseq}
\end{equation}
to obtain 
the first order Lagrangian density.
    It is straightforward to show that this equation is satisfied by the Lagrangian
\begin{equation}
\sqrt{-g}L_{\rm bulk}  = 
 \frac{1}{ 16\pi \mathcal{A}_P} \left(\sqrt{-g} \, g^{ik} \left(\Gamma^m_{i\ell}\Gamma^\ell_{km} -
\Gamma^\ell_{ik} \Gamma^m_{\ell m}\right)\right).
\label{ds}
\end{equation}  
    This is the second surprise. The Lagrangian which we have obtained is precisely
    the first order Dirac-Schrodinger Lagrangian for gravity (usually called the $\Gamma^2$
    Lagrangian).
    Note that we have obtained it without introducing the curvature tensor anywhere in the picture.
  
    Given the two pieces, the final second order Lagrangian follows from our Eq.~(\ref{aeh})
    and is, of course, the standard Einstein-Hilbert Lagrangian:    
   \begin{equation}
  \sqrt{-g} L_{grav}=\sqrt{-g}L_{\rm bulk} - \frac{\partial P^c}{\partial x^c} =  \left(\frac{1}{ 16\pi \mathcal{A}_P}\right)R\sqrt{-g}.
      \label{lgrav}
       \end{equation}
    Thus our full second order Lagrangian  {\it turns out} to be the standard 
Einstein-Hilbert Lagrangian.
This result has been obtained,
by relating the surface term in the action  to the entropy per unit area.
This relation uniquely determines the gravitational action principle and gives rise to a generally covariant
action; i.e., the surface terms dictate the form of the Einstein Lagrangian in the bulk. 
The idea that surface areas  encode  bits of information per quantum of area  allows one to determine the nature of gravitational interaction on the bulk, which is an interesting realization of the holographic principle.

The solution to Eq.~(\ref{dseq}) obtained in Eq.~(\ref{ds}) is not unique. However, self consistency requires that the final equations of motion for gravity must admit the line element in  Eq.~(\ref{rindmetric}) as a solution.  It can be shown, by fairly detailed algebra, that this condition makes the Lagrangian in Eq.~(\ref{ds}) to be the only solution.

We stress the fact that there is a very peculiar identity connecting the $\Gamma^2$ Lagrangian $L_{\rm bulk}$ and the
Einstein-Hilbert Lagrangian $L_{grav}$, encoded in  Eq.~(\ref{lgrav}). This relation, which is purely a differential geometric identity, can be stated through the equations:
 \begin{equation}
   L_{grav} =L_{\rm bulk}-\nabla_c\left[ g_{ab}
    \frac{\partial  L_{\rm bulk} }{\partial(\partial_c g_{ab})}
    \right];\quad
      L_{\rm bulk}=L_{grav}-\nabla_c\left[ \Gamma^j_{ab}
    \frac{\partial  L_{grav} }{ \partial(\partial_c \Gamma^j_{ab})}
    \right]
    \label{lageh}
   \end{equation}   
  This  relationship  defies any simple explanation in conventional
   approaches to gravity but arises very naturally in the approach presented here. The first   line in Eq.~(\ref{lageh})
also shows that the really important degrees of freedom in gravity are indeed the surface degrees of freedom. To see
this we merely have to note that at any given event, one can choose the local inertial frame in which $L_{\rm bulk}\sim
\Gamma^2$ vanishes; but the left hand side of the first line in  Eq.~(\ref{lageh}) cannot vanish, being proportional
to $R$. That is, in the local inertial frame all the geometrical information is preserved by the surface term in the
right hand side, which  cannot be made to vanish since it depends on the second derivatives of the metric tensor.
In this sense, gravity is intrinsically holographic.

The approach also throws light on another key feature of the surface term in the Einstein-Hilbert action. To see this, consider the expansion of the action in terms of a graviton field by $g_{ab}=\eta_{ab}+\lambda h_{ab}$
where $\lambda= \sqrt{16\pi G}$ has the dimension of length and $h_{ab}$ has the correct dimension of
(length)$^{-1}$ in natural units with $\hbar=c=1$.  Since the scalar curvature has the structure $R\simeq (\partial g)^2+\partial^2g$, substitution of $g_{ab}=\eta_{ab}+\lambda h_{ab}$ gives to the lowest order:
\begin{equation}
L_{EH}\propto \frac{1}{\lambda^2}R\simeq (\partial h)^2+\frac{1}{\lambda}\partial^2h
\end{equation}
Thus the full Einstein-Hilbert lagrangian is non-analytic in $\lambda$ because the surface term is 
non-analytic in $\lambda$! It is sometimes claimed in literature that one can
obtain a correct theory for gravity by starting with a massless spin-2 field
 $h_{ab}$ coupled to the energy momentum tensor $T_{ab}$ of other matter sources to the lowest 
 order,  introducing self-coupling of $h_{ab}$ to its own energy momentum tensor at the
 next order 
 and iterating the process.  
{\it It will be quite surprising if, starting from $(\partial h)^2$ and doing a honest iteration on $\lambda$, one can obtain a piece which is non-analytic in $\lambda$.} At best, one can hope to get the quadratic part of $L_{EH}$ which gives rise to the $\Gamma^2$ action but not the four-divergence term involving $\partial^2g$. The non-analytic nature of the surface term is vital for it to give a finite contribution on the horizon and the horizon entropy cannot be interpreted in terms of gravitons propagating around Minkowski spacetime. Clearly, there is lot more to gravity than gravitons (for a detailed discussion, see \cite{Padmanabhan:2004xk}).

The  analysis leading to Eq.~(\ref{lgrav}) can also be carried out in the Euclidean sector, starting from Eq,(\ref{euarea}). It is shown in Appendix 
\ref{appn:eh} that the integral of $\partial_c P^c$ with $P^c$ given by Eq.(\ref{pcfix}), can be alternatively thought of as the integral of $K$ over the boundaries [see 
Eq (\ref{pkrel})].
The rest of the analysis is straight forward so we will not discuss it.

In the above discussion we split the Einstein-Hilbert action as a quadratic part and a surface term. There is a  different way of expressing the Einstein-Hilbert action which will turn out to be useful for our later purposes. This is done by introducing the
 $(1+3)$ foliation and writing the  the bulk Lagrangian   as (see Appendix \ref{appn:eh}):
\begin{equation}
R \equiv L_{\rm EH}
= L_{\rm ADM} - 2\nabla_i(Ku^i + a^i) \equiv L_{\rm ADM}+L_{\rm div}
\label{ehandadm}
\end{equation}
where 
\begin{equation}
L_{\rm ADM} = {}^{(3)}\mathcal{R} + (K_{ab}K^{ab} - K^2)
\label{defadml}
\end{equation}
is the ADM Lagrangian \cite{Arnowitt1962} quadratic in $\dot g_{\alpha\beta}$,
and $L_{\rm div} = -2\nabla_i(Ku^i + a^i)$ is a total divergence.
Neither $L_{\rm ADM}$ nor $L_{\rm div}$ is generally covariant.
For example, $u^i$ explicitly depends on $N$,
which changes when one makes a coordinate transformation
from the synchronous frame to a frame with $N\neq 1$.

There is a conceptual difference between the $\nabla_i(Ku^i)$ term
and the $\nabla_i a^i$ term that occur in $L_{\rm div}$ in Eq.(\ref{ehandadm}).
This is obvious in the standard foliation, where $Ku^i$ contributes
on the constant time hypersurfaces, while $a^i$ contributes on the
time-like or null surface which separates the space into two regions
(as in the case of a horizon).
To take care of the $Ku^i$ term more formally, we recall that
the form of the Lagrangian used in functional integrals depends
on the nature of the transition amplitude one is interested in computing,
and one is free to choose a different representation. We shall now switch to the momentum representation of the action functional, as described earlier in the discussion leading to Eq.~(\ref{lbtp}).

Since $L_{ADM}$ is quadratic in $\dot g_{\alpha\beta}$, we can treat
$g_{\alpha\beta}$ as the coordinates and obtain another Lagrangian $L_\pi$
in the momentum representation along the lines of Eq.(\ref{lbtp}).
The canonical momentum corresponding to $q_A=g_{\alpha\beta}$ is
\begin{equation}
p^A = \pi^{\alpha\beta}
= \frac{\partial (\sqrt{-g}~L_{ADM})}{\partial \dot g_{\alpha\beta}}
=- \frac{ \sqrt{-g}}{ N} (K^{\alpha\beta}- g^{\alpha\beta}K) ~,
\end{equation}
so that the term $d(q_Ap^A)/dt$ is just the time derivative of
\begin{equation}
g_{\alpha\beta}\pi^{\alpha\beta}
= - \frac{\sqrt{-g} }{ N}(K-3K) = \sqrt{-g} (2Ku^0) ~.
\end{equation}
Since
\begin{equation}
\frac{\partial}{ \partial t} (\sqrt{-g}~Ku^0)
= \partial_i (\sqrt{-g}~Ku^i) = \sqrt{-g}~\nabla_i (Ku^i) ~,
\end{equation}
the combination
$\sqrt{-g}~L_\pi \equiv \sqrt{-g} [L_{\rm ADM} - 2 \nabla_i(Ku^i)]$
describes the same system in the momentum representation with
$\pi^{\alpha\beta}$ held fixed at the end points. (This result is known in literature \cite{York1988}
and can be derived from the action principle, as done in Appendix \ref{appn:eh}. The procedure adopted here,
which is based on Eq.~(\ref{lageh}) relating the bulk and surface terms, provides a clearer
interpretation.)
Switching over to this momentum representation, the relation between the
action functionals corresponding to Eq.~(\ref{ehandadm}) can now be expressed as 
\begin{equation}
A_{\rm EH} = A_\pi + A_{\rm boun} ~,
\label{ehandpi}
\end{equation}
\begin{equation}
A_\pi \equiv A_{\rm ADM} - \frac{1 }{ 8\pi}\int\sqrt{-g}~d^4x~\nabla_i(Ku^i) ~.
\label{defmomspace}
\end{equation}
Here $A_\pi$ describes the ADM action in the momentum representation, and 
\begin{equation}
A_{\rm boun} =- \frac{1 }{ 8\pi} \int d^4x~\sqrt{-g}~\nabla_ia^i
= -\frac{1 }{ 8\pi} \int dt \int_\mathcal{S} d^2x~N \sqrt{\sigma}
(n_\alpha a^\alpha) 
\label{boundaryone}
\end{equation} 
is the boundary term arising from the integral over the surface.
In the last equality,
$\sigma_{\alpha\beta} = g_{\alpha\beta} - n_\alpha n_\beta$
is the induced metric on the boundary 2-surface with outward normal
$n_\alpha$, and the gauge $N_\alpha=0$ has been chosen.

\subsection{Einstein's equations as a thermodynamic identity}\label{thermoid}

The fact that the information content, entangled across a horizon, is proportional
to the area of the horizon arises very naturally in the above derivation. This, in turn,
shows that the fundamental constant characterising gravity is the 
quantum of area $4\mathcal{A}_P$ which can hold approximately one bit of information.
The conventional gravitational constant, given by
$G = \mathcal{A}_P c^3/\hbar$  will, in fact, {\it diverge} if we take the limit $\hbar \to 0$ with $\mathcal{A}_p =$ constant.
This is  reminiscent of the structure of bulk matter made of atoms.
Though one can describe bulk matter using various elastic constants
etc., such a description cannot be  considered as the  strict $\hbar \to 0$ limit of quantum
mechanics --- since no atomic system can exist in this limit. Similarly, 
spacetime and gravity are inherently quantum mechanical just as 
bulk solids are \cite{Sakharov:1968pk,Padmanabhan:2004kf}.

This suggests that spacetime dynamics is like the thermodynamic limit in solid state physics.
In fact, this paradigm arises very naturally for
any {\it static} spacetime with a horizon \cite{Padmanabhan:2003pk}. Such a spacetime has a metric in Eq.~(\ref{startmetric})
with the horizon occurring at the surface $N=0$ and  its temperature $\beta^{-1}$ determined by the surface gravity on the horizon. Consider a four-dimensional region of spacetime defined as follows:
  3-dimensional spatial region  is taken
to be some compact volume
$\mathcal{V}$ with boundary $\partial\mathcal{V}$. The time integration is restricted to the range $[0,\beta]$ since there is a periodicity in Euclidean time.
We now define the entropy associated with the same spacetime region by:
\begin{equation}
S=\frac{1}{8\pi G}\int\sqrt{-g}d^4x\nabla_i a^i=\frac{\beta}{8\pi G}\int_{\partial\mathcal{V}}\sqrt{\sigma}d^2x(Nn_\mu a^\mu)
\label{defs}
\end{equation}
The second equality is obtained because,
for static spacetimes: (i) time integration reduces to multiplication by $\beta$ and (ii) since only the spatial components of
$a^i$ are non zero, the divergence  becomes a three dimensional one over $\mathcal{V}$
which is converted to an integration over its boundary ${\partial\mathcal {V}}$. 
 If the boundary ${\partial\mathcal {V}}$  is a horizon, 
$(Nn_\mu a^\mu)$ will tend to a constant surface gravity $\kappa$ and the using $\beta\kappa=2\pi$ we get $S=\mathcal{A}/4G$
where $\mathcal{A}$ is the area of the horizon. 
(For convenience, we have chosen the sign of $n^\alpha$ such that $N a_\mu n^\mu \to \kappa$, rather than $-\kappa$.)
Thus, in the familiar cases, this does reduce to the standard expression
for entropy. Similar considerations apply to each piece of {\it any} area element when it acts as a horizon for some Rindler observer. Results obtained earlier
 show that the bulk action for gravity can be obtained from a surface term in the action, if we take the entropy
of any horizon to be proportional to its area with an elemental area $\sqrt{\sigma}d^2x$ contributing an entropy
$dS=(Nn_\mu a^\mu)\sqrt{\sigma}d^2x$. The definition given above  in Eq.~(\ref{defs}) is the integral expression of the same.

 The  total energy $E$ in this region, acting as a source 
 for gravitational acceleration, is given by the Tolman energy  \cite{tolman}  defined by
\begin{equation}
\label{defe}
E=2\int_\mathcal{V}d^3x\sqrt{\gamma}N (T_{ab}-\frac{1}{2}Tg_{ab})u^au^b
\end{equation}
The covariant combination  $2(T_{ab}-(1/2)Tg_{ab})u^au^b$ [which reduces to $(\rho+3p)$ for an ideal fluid]
 {\it is} the correct source for gravitational
{\it acceleration}. For example, this will make geodesics accelerate away from each other in a universe dominated by cosmological
constant, since $(\rho+3p)<0$.
The factor $N$ correctly accounts for the relative redshift of energy in curved spacetime. It is now possible to obtain some interesting relations between these quantities.

In any  space time, there is differential geometric identity (see Eq.~(\ref{rabcd}))
\begin{equation}
R_{bd} u^b u^d 
= \nabla_i(Ku^i +a^i) - K_{ab}K^{ab} + K_a^a K^b_b
\label{idone}
\end{equation}
where $K_{ab}$ is the extrinsic curvature of spatial hypersurfaces and $K$ is its trace. This
reduces to
$\nabla_i a^i=R_{ab}u^au^b$ in static spacetimes with $K_{ab}=0$. Combined with Einstein's equations, this gives
\begin{equation}
\frac{1}{8\pi G}\nabla_i a^i= (T_{ab}-\frac{1}{2}Tg_{ab})u^au^b
\label{poisson}
\end{equation}
This equation deals directly with $a^i$ which occur as the  components of the metric tensor in
Eq.~(\ref{iso}).
We  now integrate this relation  with the measure $\sqrt{-g}d^4x$ over a four dimensional region chosen
as before.
 Using Eq.(\ref{defe}),(\ref{defs}), the integrated form of Eq.(\ref{poisson}) will read quite simply as
\begin{equation}
S=(1/2)\beta E,
\label{sofe}
\end{equation} 

Note that both $S$ and $E$ depend  on the congruence of timelike curves chosen to 
define them through $u^a$.
If these ideas are consistent, then the free energy of the spacetime must have direct geometrical
meaning {\it independent of the congruence of observers used to define the entropy $S$ and $E$}.
It should be stressed that the energy $E$ which appears in Eq.(\ref{defe}) is {\it not} the integral 
\begin{equation}
U\equiv\int_\mathcal{V}d^3x\sqrt{\gamma}N(T_{ab}u^au^b)
\label{defofu}
\end{equation}
based on $\rho=T_{ab}u^au^b$ but the integral of $(\rho+3p)$, since the latter is the source of gravitational acceleration in a region.
 The free
energy, of course, needs to be defined as $F\equiv U-TS$, since pressure --- which is an independent thermodynamic variable ---
 should not appear in the free energy. 
This gives:
\begin{equation}
\beta F\equiv \beta U -S =-S +\beta\int_\mathcal{V}d^3x\sqrt{\gamma}N(T_{ab}u^au^b)
\end{equation}
and using Eqs.(\ref{defs}),(\ref{poisson}) and $R=-8\pi G T$, we find that
\begin{equation}
\beta F=\frac{1}{16\pi G}\int d^4x\sqrt{-g}R
\label{ehfree}
\end{equation}
which is just the Einstein-Hilbert action. The equations of motion obtained by minimising the action can be equivalently thought of as minimising the macroscopic free energy. For this purpose, it {\it is} important that $F$ is generally covariant and is independent of the $u^i$ used in defining other quantities.

The sign of $E$ in Eq.~(\ref{defe}) can be negative if matter with $\rho+3p<0$ dominates in the region $\mathcal{V}$. The sign of $S$
in Eq.~(\ref{defs}) depends on the convention chosen for the direction of the normal to $\partial\mathcal{V}$
but it is preferable to choose this such that $S>0$. Then the sign of $\beta$ will arrange itself so that Eq.~(\ref{sofe}) holds.
(Of course, the temperature is $T=|\beta|^{-1}>0$). 
As an illustration, consider the Schwarzschild spacetime and the De Sitter universe. For spherically symmetric metrics with a horizon, having  $g_{00}=-g^{11},g_{00}(r_H)=0$, we can write
$g_{00}\approx g'_{00}(r_H)(r-r_H)$ near the horizon and $\beta=-4\pi/g'_{00}(r_H)$ in our signature convention. Hence
$\beta=8\pi M>0$ for Schwarzschild while $\beta=-2\pi/H<0$ for de Sitter.
In the first case,
$\beta=8\pi M$ and we can take $E=M$ for any compact two surface $\partial\mathcal{V}$
that encloses the horizon. Since $Na=(M/r^2)$, Eq.(\ref{defs}) gives $S=4\pi(M^2/G)$ for any $\partial\mathcal{V}$.
This result agrees
with Eq.(\ref{sofe}). The de Sitter case is more interesting since it is nonempty. In the static coordinates with $-g_{00}=g^{rr}=(1-H^2r^2)$, let us choose a spherical surface of radius $L<H$. We then have $E=-H^2L^3$ and $S=\pi H L^3$ from (\ref{defe}) and (\ref{defs}).
Once again, equation (\ref{sofe}) holds since $\beta=-2\pi/H$.

 We should, therefore, be able to 
  rewrite Einstein's equations in a form analogous to the 
     $TdS -dU= PdV$ equation \cite{Padmanabhan:2002sh,Padmanabhan:2002jr}. It is fairly straight forward to 
     achieve this in the case of spacetimes of the form:
     \begin{equation}
     ds^2=-f(r)dt^2+f(r)^{-1}dr^2 + r^2(d\theta^2+\sin^2\theta
     d\phi^2)
     \label{basemetric}
     \end{equation}
      with  $f(r)=1-2m(r)/r$. This metric solves the Einstein's equations if the energy density $\rho(r)/8\pi$ and the transverse pressure $\mu(r)/8\pi$ are arranged to give $\rho(r)=(m'/2r^2); \mu(r)=\rho+(1/2)r\rho'(r)$ and the radial pressure is set equal to the energy density. If there is a horizon at $r=a$, with 
$f(a)=0,f'(a)\equiv B$, then the temperature $T$ is determined by 
$T^{-1}=\beta=4\pi/B$. Further,  we  find that, for a spherical region of radius $r=a$,
\begin{equation}
S=\pi a^2; \quad E=\frac{1}{2} a^2 B, \quad |U| = \frac{a}{2}
\end{equation}
These relations hold on the horizon for a class of solutions parametrised by the function $m(r)$ with $a$
determined as the root of the equation $2m(a)=a$.
What is more,  these  relations,  along with the fact that radial pressure is equal
to the energy density, allow us to  write Einstein's' equations as
\begin{equation}
 dU=TdS-PdV
 \end{equation}
where the differentials are interpreted as $dU=(dU/da)da$ etc. 
In these spacetimes, $S\propto U^2$ giving the density of states $g(U)=\exp(c U^2)$ where $c$ is a constant. 

   The above results are of particular importance to a horizon which is {\it not} associated with a black hole,
viz. De Sitter horizon.  In this case, $f(r) = (1-H^2r^2)$, $a=H^{-1}, B=-2H<0$
    so that the temperature --- which should be positive --- is $T=|f'(a)| / (4\pi) =(-B)/4\pi$.
   For horizons with $B=f'(a)<0$ (like the De Sitter horizon) we have
   $f(a)=0, f'(a)<0$, and it follows that $f>0$ for
    $r<a$ and $f<0$ for $r>a$; that is, the ``normal region'' in which $t$ is timelike is inside
    the horizon as in the case of, for example, the De Sitter metric.  
    The Einstein's equations for the metric in Eq.(\ref{basemetric}) evaluated at the horizon $r=a$ reads as:
    \begin{equation}
    \frac{-B}{ 4\pi} d\left( \frac{1}{ 4} 4\pi a^2 \right) + \frac{1}{ 2} da=- T^r_r(a) d \left( \frac{4\pi }{ 3}  a^3 \right)
    =P(-dV)
    \label{tdseqn}
    \end{equation}
    The first term on the left hand side is again of the form $TdS$ (with positive temperature and entropy).
    The term on the right hand side has the correct sign since the
     inaccessible region (where $f<0$) is now outside the horizon and the volume of this region 
    changes by $(-dV)$.
   Once again, we can use Eq.~(\ref{tdseqn}) to identify the entropy and the energy: 
 \begin{equation}
 S=\frac{1}{  4} (4\pi a^2) = \frac{1}{ 4} \mathcal{A}_{\rm horizon}; \quad U=-\frac{1}{ 2}H^{-1}
 \end{equation}
As a byproduct, our approach provides an interpretation of energy for the De Sitter spacetime 
and a consistent thermodynamic interpretation of 
De Sitter horizon. 

Our identification, $U=-(1/2) H^{-1}$ is also supported by the following argument:
If we use the ``reasonable" assumptions $S=(1/4)(4\pi H^{-2}), V\propto H^{-3}$ and $U=-PV$ in the equation
$TdS -PdV =dU$ and treat $U$ as an unknown function of $H$, we get the equation
\begin{equation}
 H^2\frac{dU}{ dH}=-(3UH+1)
 \label{arguetwo}
 \end{equation}
which integrates to give precisely $U=-(1/2)H^{-1}. $ 
Note that we only needed the proportionality, $V \propto H^{-3}$ in this argument since $PdV \propto 
(dV/V)$. The ambiguity between the coordinate and proper volume is  irrelevant.

These results can be stated more formally as follows: In standard thermodynamics, we can consider two equilibrium states of a system differing infinitesimally in  the extensive variables volume, energy and entropy  by $dV,dU$ and $dS$ while having {\it same} values for the intensive variables temperature ($T$) and pressure ($P$). Then, the first law of thermodynamics asserts that $TdS=PdV + dU$ for these states. In a similar vein, we can consider two spherically symmetric solutions to Einstein's equations with the radius of the horizon differing by $da$ while having the same source $T_{ik}$ and  the same value for $B$.   Then the entropy and energy will be infinitesimally different for these two spacetimes; but the fact that both spacetimes satisfy Einstein's equations shows that $TdS$ and $dU$ will be related to the external source $T_{ik}$ and $da$ by Einstein's equations. Just as in standard thermodynamics, this relation could be interpreted as connecting a sequence of quasi-static equilibrium states.

The analysis is  classical except for the crucial periodicity argument which 
   is used to identify the temperature uniquely.  This is again done locally by approximating
the metric by a Rindler metric close to the horizon and identifying the Rindler temperature. This
idea bypasses the difficulties in defining  and normalising Killing vectors in spacetimes
which are not asymptotically flat.

Finally we mention that this   framework also imposes a  strong constraints on the form of action
functional $A_{grav}$ in {\it semi-classical} gravity. It can be shown that,  the area of the horizon,
as measured by \emph{any} observer blocked by that horizon, will be quantised
\cite{Padmanabhan:2003ub}.
In normal units, 
$\mathcal{A}_{\rm horizon}=8\pi m(G\hbar/c^3)=8\pi mL_{\rm Planck}^2$
where $m$ is an integer.
 (Incidentally, this will match with the result from loop quantum gravity, for the high-$j$ modes, if the Immirizi parameter is unity.)
In particular, any flat spatial surface can be made
a horizon for a suitable Rindler observer, and hence all area elements
(in even flat space-time) must be intrinsically quantised.
In the quantum theory, the area operator for one observer
need not commute with the area operator of another observer,
and there is no inconsistency in all observers measuring quantised areas.
The changes in area, as measured by any observer, are also quantised,
and the minimum detectable change is of the order of $L_{\rm Planck}^2$.
It can be shown, from very general considerations, that there is an
operational limitation in measuring areas smaller than $L_{\rm Planck}^2$,
when the principles of quantum theory and gravity are combined \cite{Padmanabhan:1987au};
our result is consistent with this general analysis. (The Planck length plays a significant
role in different approaches which combine the principles of quantum theory and gravity;
see, for example, \cite{Amelino-Camelia:1994vs,Amelino-Camelia:2003uc}.)
While there is considerable amount of literature  suggesting that the area of a \emph{black hole}
  horizon is quantised 
  [for a small sample of references, 
  see \cite{Bekenstein:1974jk,Bekenstein:1973ur,Hod:1998vk,Gour:1999yu,Louko:1996md,Mukhanov:1986me,Kogan:1986yd,Mazur:1987jf,Lousto:1995jd,Peleg:1995gg,Das:2002xb,Bekenstein:1998aw,Danielsson:1993um,Maggiore:1994ww,Bekenstein:1995ju,Kastrup:1996pu} as well as papers cited in Section \ref{qgbh}]
   the result mentioned  above is more general
   and is applicable to any static horizon.

\section{Conclusions and Outlook}

We shall now take stock  of the results discussed in this review from a broader perspective and will
attempt to provide an overall picture.

Combining the principles of  quantum theory with special relativity (and Lorentz invariance) 
 required a fairly drastic
change in the description of physical systems.  Similarly, it is natural for new issues to arise
when we take the next step of combining quantum theory with the concept of general covariance or when we attempt to do quantum field theory in a curved background spacetime. However, one would have 
naively expected these issues 
to be \emph{kinematical} in the sense that they are independent of the field equations or the action for gravity.
Our discussion shows that there is a strong link between the kinematical aspects and the dynamics of gravity
because of the 
  structure of  classical general relativity. 
While it may be convenient to distinguish between the kinematical
  aspects (discussed in Sections \ref{horizonsr} to \ref{exptab}) and the dynamical aspects (discussed in Sections \ref{entropyhorizon} -- \ref{thermoroute}), each by itself can only give a partial picture. The overall structure of the theory should allow a seamless transition
  across these two aspects.

  In this review this was attempted by  (i) noting that one needs to use
the Euclidean sector  to incorporate the new ingredients which arise
  when special relativity is combined with quantum mechanics and (ii) using the 
  fact that when quantum theory is formulated in the Euclidean sector, a  unique structure 
  emerges in the presence of horizons.  Using a congruence of  timelike curves  to define a horizon, one finds that it is possible to incorporate the kinematical effects of 
  (at least static) horizons in a general manner and associate the notion of temperature with the 
  horizons. This is achieved by
  using a coordinate system in which the spacetime region hidden by a horizon 
  is mapped to a single point in the Euclidean sector and constructing an effective manifold  for
  a family of   observers by removing this point.  The resulting non trivial topology leads to the standard
  results of quantum field theory in curved spacetimes with horizons.

  The importance of the above point of view lies in its ability to provide a deeper relationship
  between gravity and thermodynamics, as shown in Section \ref{thermoroute}.  If one accepts the idea
  --- that the 
  physical theory for a class of observers should be formulated in an effective manifold in which
  the region inaccessible to those observers is removed --- then one is led to enquire what it implies
  for the dynamics of gravity. Using the fact that the horizon is the common element between the 
  inaccessible and accessible regions, it is possible to argue that  the action functional for gravity
  must contain (i) a well defined surface term and (ii) a bulk term which is related to the surface term
  in a specific manner. Hence,  this point of view allows one to determine the action functional 
  for  gravity from thermodynamic considerations. What is more, it links the kinematical and 
  dynamical  aspects   of the theory in an interesting manner.
  
 This approach is very similar in spirit to that of renormalisation group theory (RGT) in particle physics.  When  an experimenter does not have information about the model at scales $k>\Lambda$, say, in \emph{momentum space}, the RGT allows one to use an
  effective low energy theory with the coupling constants readjusted to incorporate the missing information. This, in turn, puts restrictions on the nature of the theory as well as the ``running" of the coupling constants.  
  Similarly,  when a given family of observers has limited information because they
  are blocked by a horizon (in \emph{real} space rather than momentum space) it is necessary to add certain boundary terms in the action functional in order to provide a consistent description.   Just as the RGT contains nontrivial information about the low energy sector of the theory, our approach allows us to determine the form of the action in the long wavelength limit of gravity. As far as the loss of information due to a horizon is concerned, there is no need to distinguish between the uniformly accelerated observers in flat spacetime and, say,  the observers located permanently at $r>2M$
  in the Schwarzschild spacetime.

  There are some new insights that arise in this approach which are worth exploring further.
  \begin{itemize}
  \item Einstein's  equations for  gravity can be obtained from a  variety of action functionals,
  any two of which differ by a surface term. In the case of Einstein-Hilbert action, the surface term
  is related in a very specific manner to the bulk term. 
   (See e.g., Eq.~(\ref{lageh}); it is rather intriguing that this relation has  not been
 explored 
in the literature before.)  \emph{This relation is so striking that 
  it demands an explanation} which is indeed provided by the  thermodynamic paradigm
  described in Section \ref{thermoroute}. 
  
  \item The approach makes gravity ``holographic'' in a specific sense of the word. The Einstein-Hilbert
  Lagrangian has the structure $L_{EH} = L_1 + L_2 $ where $L_1 \sim (\partial g)^2, L_2  \sim \partial^2 g$.
  Along any world line, one can choose a coordinate system such that  $(\partial g)^2 \to 0$
  suggesting that the  dynamics of the theory is actually contained in the $L_2 \sim   \partial^2 g$
   term which leads to the surface term in the action. We saw in Section \ref{grfresults} that one could determine
   the bulk term from the surface term under certain assumptions. This fact, that the structure of the
   surface term in an action determines the theory, provides a possible interpretation of holographic
   principle (which is somewhat different from the conventional interpretation of the term).

  \item The approach supports the paradigm that the spacetime is similar to the continuum limit
  of a solid that is obtained when one averages over the underlying   microscopic degrees of freedom \cite{Sakharov:1968pk}. 
   As described in Section \ref{thermoid}, this strongly indicates the possibility
   that gravity is intrinsically quantum mechanical at all scales just as solids cannot exist in the 
   strict $\hbar \to 0 $ limit. Just as  the bulk properties of solids can be described without
   reference to the underlying atomic structure, much of classical and semi classical gravity 
    (including 
   the entropy of black holes) will be independent of the underlying description of the microscopic
   degrees of freedom. Clear signs of this independence emerges from the study of Einstein-Hilbert
   action which contains sufficient structure to lead to many of the results involving the horizon
   thermodynamics. Hence any microscopic description of gravity which leads to Einstein-Hilbert
   action as the long wavelength limit will also incorporate much of horizon physics.
   
   \end{itemize}

  \bigskip

\section*{Acknowledgements}

I thank Apoorva Patel, Ashoke Sen, G. Date, N.K. Dadhich, S. Mathur, S. Nemani, R. Nityananda,   
K. Subramanian, Tulsi Dass
and G.Volovik  for comments on the earlier drafts of the review.

\appendix

  \section{Gravitational Action Functional}\label{appn:eh}

This appendix summarises several aspects of action functionals used in gravity and derives some of the
results not readily available in the literature.

 The conventional action principle for general relativity is  the Einstein-Hilbert action given 
  by 
   \begin{equation}
   A_{\rm EH} \equiv \frac{1}{16\pi} \int R\sqrt{-g} d^4x 
\label{aehbegin}
   \end{equation}  
  Straightforward algebra shows that the scalar curvature can be expressed in the form
  \begin{equation}
  R\sqrt{-g} = \frac{1}{4} \sqrt{-g} M^{abcijk} g_{ab,c} g_{ij,k} - \partial_j P^j 
  \equiv \sqrt{-g} L_{\rm quad}  - \partial_j P^j
  \label{eqnone}
  \end{equation}
  where 
  \begin{equation}
  M^{abcijk} 
  =g^{ck}\left[g^{ab} g^{ij}- g^{ai} g^{bj} \right] + 2 g^{cj} \left[  g^{ai} g^{bk}- g^{ki} g^{ba}\right]
  \end{equation}
  and 
  \begin{equation}
  P^j = \sqrt{-g} g_{ac,i} ( g^{ac} g^{ji}- g^{ia} g^{cj}) \equiv \sqrt{-g} V^j
  \end{equation}
  This result  is  equivalent to a more conventional expression for the gravitational
  action written in terms of Christoffel symbols with: 
    \begin{equation}
    L_{\rm quad} = g^{ab} \left(\Gamma^i_{ja} \Gamma^j_{ib} -\Gamma^i_{ab} \Gamma^j_{ij}\right)
    \label{lquad}
    \end{equation}
      and 
    \begin{equation}
    P^c = \sqrt{-g} \left(g^{ck} \Gamma^m_{km}- g^{ik} \Gamma^c_{ik}\right) =-
     \frac{1}{\sqrt{-g}} \partial_b (g g^{bc})
    \label{defpcone}
     \end{equation}
     
 The manner in which $P^c$ is expressed hides its geometrical interpretation. To bring this out, note that the integral of $\partial_cP^c$ can be evaluated in a given coordinate system, most simply by:
 \begin{eqnarray}
 \int d^4x   \partial_cP^c&=&\int dx^0  dx^1 dx^2 dx^3  ( \partial_0P^0+ \partial_1P^1+\cdots)\nonumber\\
 &=& \int_{x^0}  dx^1 dx^2 dx^3 P^0 +\int _{x^1}  dx^0 dx^2 dx^3  P^1+\cdots
 \end{eqnarray}
where the subscript on the integral indicates the coordinate that is held constant.
 To study the integral of $P^n$ on the $x^n=$ constant surface, let us choose a coordinate system in which the metric has the form
 \begin{equation}
 ds^2=g_{nn}(dx^n)^2+g^{\perp}_{ab}dx^adx^b
 \end{equation}
where $n=0,1,2,3$ and for each choice of $n$ the $a,b$ run over the other three coordinates. (We have assumed that the cross terms vanish  to simplify the computation.) The $P^c$ in this coordinate system can be computed using  the last expression in Eq.~(\ref{defpcone}). We get:
 \begin{equation}
 P^n=-\frac{1}{\sqrt{g_{nn}}}\frac{1}{\sqrt{g^{\perp}}}
 \partial_n\left(g_{nn}g^{\perp}\frac{1}{g_{nn}}\right)=-\frac{2}{\sqrt{g_{nn}}}\partial_n\sqrt{g^{\perp}}
 \end{equation}
 The normal to the surface $x^n=$ constant is given by $n^a=g_{nn}^{-1/2}\delta^a_n$ and the trace of the extrinsic curvature of the $x^n=$ constant is 
 \begin{equation}
 K=-\nabla_a n^a=-\frac{1}{\sqrt{g_{nn}}}\frac{1}{\sqrt{g^{\perp}}}
 \partial_n\left(\sqrt{g_{nn}}\sqrt{g^{\perp}}\frac{1}{\sqrt{g_{nn}}}\right)
=-
 \frac{1}{\sqrt{g^{\perp}}}\frac{1}{\sqrt{g_{nn}}}\partial_n\sqrt{g^{\perp}}
 \end{equation}
 Hence we get the result
 \begin{equation}
 \int_{\mathcal{V}} d^4x   \partial_cP^c=\sum_{\partial\mathcal{V}}2\int K\sqrt{g^\perp} d^3x
 \label{pkrel}
 \end{equation}
 where the sum is over all the bounding surfaces. Thus the total divergence term can be expressed as the sum over the
integrals of the extrinsic curvatures on each boundary and the Einstein-Hilbert action in Eq.(\ref{aehbegin}) becomes
\begin{equation}
A_{\rm EH}=\int_\mathcal{V}\frac{d^4x}{16\pi} \sqrt{-g}L_{\rm quad} -\int_{\partial\mathcal{V}}\frac{d^3x}{8\pi} \sqrt{g^\perp} K
\label{gibbhawk}
\end{equation}

This result can be obtained in a more geometrical fashion, which is instructive.
We foliate the space-time by a series of space-like hypersurfaces $\Sigma$
with normals $u^i$. Next, from the relation
$R_{abcd}u^d = (\nabla_a \nabla_b - \nabla_b \nabla_a) u_c$,
we obtain the identity
\begin{eqnarray}
R_{bd} u^b u^d &=& g^{ac} R_{abcd} u^b u^d
 = u^b \nabla_a \nabla_b u^a - u^b \nabla_b \nabla_a u^a \nonumber \\
&=& \nabla_a(u^b \nabla_b u^a) - (\nabla_au^b)(\nabla_b u^a)
   -\nabla_b(u^b \nabla_a u^a) + (\nabla_b u^b)^2 \nonumber \\
&=& \nabla_i(Ku^i +a^i) - K_{ab}K^{ab} + K_a^a K^b_b
\label{rabcd}
\end{eqnarray}
where  $K_{ij} = K_{ji} = -\nabla_i u_j - u_i a_j$, is the extrinsic curvature with
 $K \equiv K^i_i = -\nabla_i u^i$ and
$K_{ij} K^{ij} = (\nabla_iu^j) (\nabla_ju^i)$).
Further using
\begin{equation}
R = -R~g_{ab} u^a u^b = 2 (G_{ab}-R_{ab}) u^a u^b ~,
\label{rinhone}
\end{equation}
and the identity
\begin{equation}
2~G_{ab} u^a u^b = K_a^a K^b_b   - K_{ab}K^{ab} + {}^{(3)}\mathcal{R}~,
\label{ginr}
\end{equation}
where ${}^{(3)}\mathcal{R}$ is the scalar curvature of the 3-dimensional space,
we can write the scalar curvature as
\begin{equation}
R = {}^{(3)}\mathcal{R} +K_{ab}K^{ab} - K_a^a K^b_b - 2 \nabla_i (Ku^i + a^i)
  \equiv L_{\rm ADM} -2 \nabla_i (Ku^i + a^i) ~,
\label{rinh}
\end{equation}
where $L_{\rm ADM}$ is the ADM Lagrangian.

Let us now integrate Eq.~(\ref{rinh}) over a four volume $\mathcal{V}$
bounded by two space-like hypersurfaces $\Sigma_1$ and $\Sigma_2$
and a time-like hypersurface $\mathcal{S}$.
The space-like hypersurfaces are constant time slices with normals $u^i$,
and the time-like hypersurface has normal $n^i$ orthogonal to $u^i$.
The induced metric on the space-like hypersurface $\Sigma$ is
$h_{ab} = g_{ab} + u_a u_b$, while the induced metric on the time-like
hypersurface $\mathcal{S}$ is $\gamma_{ab} = g_{ab} - n_a n_b$. The
$\Sigma$ and $\mathcal{S}$ intersect along a 2-dimensional surface $\mathcal{Q}$,
with the induced metric
$\sigma_{ab} = h_{ab} - n_a n_b = g_{ab} + u_a u_b - n_a n_b$. 
With $g_{00}=-N^2$, we get
\begin{eqnarray}
A_{\rm EH} = \frac{1 }{16\pi} \int_\mathcal{V} d^4x~\sqrt{-g}~R 
  &=& \frac{1 }{16\pi} \int_\mathcal{V} d^4x~\sqrt{-g}~L_{\rm ADM}
    -\frac{1 }{ 8\pi} \int_{\Sigma_1}^{\Sigma_2} d^3x~\sqrt{h}~K\nonumber\\
   && \quad - \frac{1 }{8\pi} \int_\mathcal{S}
      dt~d^2x~N~\sqrt{\sigma}(n_ia^i) ~.
\label{bigeqn}
\end{eqnarray}
 
Let the hypersurfaces $\Sigma, \mathcal{S}$ as well as their intersection
2-surface $\mathcal{Q}$ have the corresponding extrinsic curvatures
$K_{ab}, \Theta_{ab}$ and $q_{ab}$. 
To express  the Einstein-Hilbert action in the form in Eq.(\ref{gibbhawk}),
as a term having only the first derivatives, plus an integral of the trace
of the extrinsic curvature over the bounding surfaces, we use the
 foliation condition $n_i u^i=0$
between the surfaces, and note that
\begin{equation}
n_i a^i = n_i u^j \nabla_j u^i = -u^j u^i \nabla_j n_i
= (g^{ij} - h^{ij}) \nabla_jn_i =q- \Theta 
\label{thetaink}
\end{equation} 
where $\Theta\equiv\Theta^a_a$ and $q\equiv q^a_a$ are the traces of the
extrinsic curvature of the surfaces, when treated as embedded in the
4-dimensional or 3-dimensional enveloping manifolds.
Using Eq.~(\ref{thetaink}) to replace $(n_i a^i)$ in the last term of
Eq.~(\ref{bigeqn}), we get the result   
\begin{eqnarray}
&&A_{\rm EH} + \frac{1 }{ 8\pi} \int_{\Sigma_1}^{\Sigma_2} d^3x\sqrt{h}K
 - \frac{1}{ 8\pi} \int_\mathcal{S} dtd^2xN\sqrt{\sigma}\Theta \nonumber\\
&&\qquad\quad = \frac{1 }{ 16\pi} \int_\mathcal{V} d^4x\sqrt{-g}L_{\rm ADM}
 - \frac{1 }{ 8\pi} \int_\mathcal{S} dtd^2xN\sqrt{\sigma}q
\label{EHtoADM}
\end{eqnarray}
The left hand side is in the form we want as the sum of $A_{\rm EH}$ and the traces of extrinsic curvatures on the bounding surfaces. In the right hand side,
 the first term,
$L_{\rm ADM}$ is {\it not} purely quadratic in the first derivatives of the metric tensor, since it contains ${}^{(3)}\mathcal{R}$,
which in turn contains second derivatives of the metric tensor. We can now use a formula,
 analogous to Eq.~(\ref{eqnone}), to separate the second derivatives from ${}^{(3)}\mathcal{R}$. The relation is
\begin{equation}
{}^{(3)}\mathcal{R}\sqrt{h }={}^{(3)}L_{\rm quad}\sqrt{h }+\partial_\mu Q^\mu,
\end{equation}
 where $h $ is the determinant of the spatial metric, ${}^{(3)}L_{\rm quad}$ is made from the spatial metric and its spatial derivatives and $Q^\mu$ is same as $P^i$ but built from spatial metric. The  sign reflects
 the fact that $g$ is negative definite while $h $ is positive definite.  What we need in Eq.(\ref{EHtoADM}) 
is $\sqrt{-g}{}^{(3)}\mathcal{R}=
N\sqrt{h }{}^{(3)}\mathcal{R}$ which becomes:
\begin{eqnarray}
\label{threeR}
\hskip-0.8cm\sqrt{-g}{}^{(3)}\mathcal{R}&=&{}^{(3)}L_{\rm quad}\sqrt{-g}+N\partial_\mu Q^\mu\\
&=&{}^{(3)}L_{\rm quad}\sqrt{-g}-\sqrt{-g}\left(\frac{\partial_\mu N}{N}\right)\frac{\partial_\nu(h h^{\mu\nu})}{h}
+\partial_\mu (NQ^\mu)\nonumber
\end{eqnarray}
On integration, the last term becomes a surface integral and using the result analogous to Eq.~(\ref{pkrel}),
we find that
\begin{equation}
\int dt d^3x\partial_\mu (NQ^\mu)=\int dt d^2x N Q^\mu n_\mu=\int dt d^2x N \sqrt{\sigma} q
\end{equation}
When we substitute Eq.(\ref{threeR}) into the $L_{ADM}$ in Eq.(\ref{EHtoADM}), the terms with $q$ 
cancel and we get the final result:
\begin{eqnarray}
A_{\rm EH} + \sum \frac{1}{ 8\pi} \int_{\Sigma_1}^{\Sigma_2} d^3x~\sqrt{h}~K
&=& \frac{1 }{16\pi} \int_\mathcal{V} d^4x~\sqrt{-g}[(K_{ab}K^{ab} - K_a^a K^b_b)\nonumber\\
&&+{}^{(3)}L_{\rm quad}+\frac{\partial_\mu N}{Nh}\partial_\nu(h h^{\mu\nu})]
\label{final}
\end{eqnarray}
which is precisely $A_{\rm quad}$. The terms with $K_{ab}$ are quadratic in time derivatives of spatial metric,
the ${}^{(3)}L_{\rm quad}$ has quadratic terms of  spatial derivatives of spatial metric and the last term gives a (quadratic) cross term between spatial derivatives of spatial metric and $g_{00,\mu}$.
This is the standard result often used,
which---unfortunately---misses the importance of the $(n_i a^i)$ term
in the action by splitting it as in Eq.~(\ref{thetaink}).

Let us now get back to some  features of Eq.~(\ref{eqnone}) which are not adequately emphasised in the literature. 
  The first interesting result that can be obtained from Eq.~(\ref{eqnone})  is a direct relation between 
  $P^j$ and $L_{\rm quad}$. Differentiation of $L_{\rm quad}$ followed by
  contraction with $g_{ab} $ gives
  \begin{equation}
  g_{ab} \frac{\partial L_{\rm quad}}{\partial (g_{ab,c})} =g_{ij,k} \left[ g^{ij} g^{ck} - g^{ik} g^{cj}
  \right] = V^c = \frac{1}{\sqrt{-g}} P^c
  \end{equation}
  This remarkable result shows that the scalar curvature can be written in the form
  \begin{equation}
  R = L_{\rm quad} - \frac{1}{\sqrt{-g}}\partial_c \left[ \sqrt{-g} g_{ab}
  \frac{\partial L_{\rm quad}}{\partial (g_{ab,c})}
  \right]
  \label{rlquadrel}
  \end{equation}
  Comparing this result with Eq.~(\ref{pkrel}), we get a more dynamical interpretation of $K$. We have
  \begin{equation}
  2K=n_c g_{ab} \frac{\partial L_{\rm quad}}{\partial (g_{ab,c})}\equiv n_cg_{ab}\pi^{abc}
  \end{equation}
The quantity  $\Pi^{ab}=n_c\pi^{abc}$ is the energy-momentum conjugate to $g_{ab}$ with respect to the
surface defined by the normal $n_c$. 

If we take the Lagrangian to be $L(q_A, \partial_i q_A)$
  which depends on a set of dynamical variables $q_A$ where $A$ could denote a collection
  of indices (in the case of gravity $q_A \to g_{ab}$ with $A$ denoting a pair of indices),
then one can obtain a second Lagrangian by
\begin{equation}
L_\pi=L-\partial_i\left[ q_A \frac{\partial L}{\partial (\partial_i q_A)}
  \right]=L-\partial_i(q_Ap^{Ai})
\label{legendre}
\end{equation}
Both will lead to the same equations of motion provided $q_A$ is fixed while varying $L$ and $p^{Ai}$ is fixed while varying
$L_\pi$. [See discussion leading to Eq.~(\ref{lbtp}).]  In the case of gravity, $L$ corresponds to the quadratic Lagrangian while $L_\pi$ corresponds to the Einstein-Hilbert Lagrangian and Eq.(\ref{legendre}) corresponds to Eq.(\ref{rlquadrel}). 
 
It is possible to understand  Eq.(\ref{rlquadrel})  from the fact that $L_{\rm quad}$ has
  certain degrees of homogeneity in terms of $g_{ab}$ and $g_{ab,c}$. 
  The argument proceeds as follows: Consider any Lagrangian $L(q_A, \partial_i q_A)$
  which depends on a set of dynamical variables $q_A$ where $A$ could denote a collection
  of indices as before.
  Let the Euler-Lagrange function resulting from $L$ be: 
  \begin{equation}
  F^A \equiv \frac{\partial L}{\partial q_A} - \partial_i \left[\frac{\partial L}{\partial(\partial_i q_A)} \right]
  \end{equation}
  Taking the contraction $q_A F^A$ and manipulating the terms we get
  \begin{equation}
  q_A F^A = q_A\frac{\partial L}{\partial q_A} - \partial_i \left[ q_A \frac{\partial L}{\partial (\partial_i q_A)}
  \right] +  (\partial_i q_A) \frac{\partial L}{\partial (\partial_i q_A)}
  \end{equation}
  If $L$ is a homogeneous function of degree $\mu$ in $q_A$ and a homogeneous function
  of degree $\lambda$ in $\partial_i q_A$, then the first term on the right hand side
  is $\mu L$ and the third term is $\lambda L$ because of Euler's theorem.
  Hence 
  \begin{equation}
  q_A F^A = (\lambda + \mu) L  - \partial_i \left[ q_A \frac{\partial L}{\partial (\partial_i q_A)}
  \right] 
  \label{qafa}
  \end{equation}
  In the case of gravity, $F^A =- (R^{ab} - (1/2) g^{ab} R) \sqrt{-g}$  with the minus sign
  arising from the fact that $F^A$ corresponds to contravariant indices. So 
  \begin{equation}
  q_A F^A
  = g_{ab}[- (R^{ab} - \frac{1}{2} g^{ab} R) \sqrt{-g} = R\sqrt{-g}
  \end{equation}
  Further, if we change $g_{ab} \to f g_{ab}$ then  $g^{ab} \to f^{-1} g^{ab}, \sqrt{-g} \to f^2 \sqrt{-g}$.
  If the first derivatives $g_{ab,c}$ are held fixed, the above changes will change 
  $\sqrt{-g}L_{\rm quad}  $ in Eq.~(\ref{eqnone}) by the factor $f^2 f^{-3} = f^{-1}$ showing 
  that $\sqrt{-g} L_{\rm quad}$ is of degree $\mu = -1$ in $g_{ab}$. When $g_{ab}$ is held fixed
  and $g_{ab,c}$ is changed by a factor $f, \sqrt{-g} L_{\rm quad}$ changes by factor
  $f^2$; so $\sqrt{-g} L_{\rm quad}$ is of degree $\lambda = +2$ in the derivatives.
  Using $q_AF^A = R \sqrt{-g}$ and $\mu + \lambda =1$ in Eq.~(\ref{qafa})
  we get the result which is identical to Eq.~(\ref{rlquadrel}).
  
  From the  relation Eq.~(\ref{legendre}), it is possible to derive
the variations of $A_{\rm EH}$ and $A_{\rm quad}$ for arbitrary variations of $\delta g_{ab}$. We get:
\begin{eqnarray}
\delta (16\pi A_{\rm EH})&=&\int_\mathcal{V}d^4x\sqrt{-g}G_{ab}\delta g^{ab}+\int_{\partial\mathcal{V}}d^3xh_{ab}\delta[\sqrt{h}(K^{ab}-h^{ab}K)]\nonumber\\
&=&
\int_{\partial\mathcal{V}}d^3xh_{ab}\delta\Pi^{ab}
\end{eqnarray}
where 
$\Pi^{ab} = \sqrt{h} (K^{ab} - h^{ab}K)$ and
the last equality holds  when equation of motion $(G_{ab}=0)$ are satisfied (``on-shell'').
 Similarly,
\begin{eqnarray}
\delta (16\pi A_{\rm quad})&=&\int_\mathcal{V}d^4x\sqrt{-g}G_{ab}\delta g^{ab}
-\int_{\partial\mathcal{V}}d^3x[\sqrt{h}(K^{ab}-h^{ab}K)]\delta h_{ab}
\nonumber\\
&=&
-\int_{\partial\mathcal{V}}d^3x\Pi^{ab}\delta h_{ab}
\end{eqnarray}
with the last equality holding on shell. Subtracting one from the other, we have
\begin{eqnarray}
 16\pi \delta(A_{\rm quad}- A_{\rm EH})&=&-\int_{\partial\mathcal{V}}d^3x(\Pi^{ab}\delta h_{ab}+h_{ab}\delta\Pi^{ab})\nonumber\\
&=&-\int_{\partial\mathcal{V}}d^3x\delta (h_{ab}\Pi^{ab})=2\delta\int_{\partial\mathcal{V}}d^3x \sqrt{h} K
\end{eqnarray}
irrespective of the equations of motion (``off-shell'')
which is precisely what is needed for consistency. Thus Einstein-Hilbert Lagrangian describes gravity in the momentum space
and leads to the field equations when the momenta $\Pi^{ab}$ are fixed at the boundaries while the quadratic Lagrangian
describes gravity in the coordinate space with the metric $h_{ab}$ fixed on the boundary.

  Finally, we shall provide a direct derivation of the ADM form of the action starting from Eq.(\ref{eqnone}) and 
   separating out the space and time components.
  To do this, we shall assume a metric of the form $g_{00} = -N^2, g_{0\alpha} =0$
  and $g_{\alpha\beta}$ arbitrary. In evaluating the kinetic energy term of the form
  $(1/4) M \partial g \partial g$ in Eq.(\ref{eqnone}), one can separate out the terms made of (i) the time derivatives
  of $g_{\alpha\beta}$, (ii) time derivatives of $g_{00}$, (iii) spatial derivatives of $g_{\alpha\beta}$,
  (iv) spatial derivatives of $g_{00}$, (v) mixed terms involving one spatial derivative of $g_{00}$ 
  and one spatial derivative of $g_{\alpha\beta}$. Of these, it is easy to verify that (ii) and (iv)
  vanishes identically since the corresponding component of $M$ is zero. The remaining 
  three terms give in $L_{\rm quad}$: 
  \begin{equation}
 \label{lquadsplit}
  L_{\rm quad} = \frac{1}{4N^2} \dot g_{\alpha\beta} \dot g_{\mu\nu}
  \left[
   g^{\alpha\beta} g^{\mu\nu}- g^{\alpha\mu} g^{\beta\nu}
  \right]
 + \left( \frac{\partial_\mu N}{N}\right) \partial_\nu g_{\alpha\beta}
  \left[
    g^{\alpha\beta} g^{\mu\nu}-g^{\alpha\nu} g^{\beta\mu}
  \right]
   + (\cdots ) 
  \end{equation}
where $(\cdots)$ denote purely spatial terms.
  The first three terms in $L_{\rm quad}$ correspond to (i), (v) and (iii) respectively. The last term
  made entirely out of spatial derivatives of spatial metric is not explicitly written down. Next
  consider the terms that arise from $(-g)^{-1/2} \partial_c P^c$ which can be classified as 
  follows: 
  (a) The time derivative term arises from $c=0$. 
  (b) Spatial derivatives  involving $\partial_\alpha g_{00}$.
  (c) In calculating the spatial derivative
  terms, one should note that $\sqrt{-g} = N \sqrt{h}$. This will give terms involving
  product of spatial derivatives of $N$ and $g_{\alpha\beta}$. 
  (d) Spatial derivatives of purely spatial metric. Working out the terms, we get
  \begin{eqnarray}
 \label{dcpcsplit}
  && \frac{1}{\sqrt{-g}} \partial_c \left( \sqrt{-g} V^c \right) =
   \frac{1}{\sqrt{-g}} \partial_0 \left( \sqrt{-g} g^{00} g^{\alpha\beta} \dot g_{\alpha\beta} \right)
   + \frac{2}{\sqrt{-g}} \partial_\alpha \left( \sqrt{-g} g^{\alpha\beta} \frac{\partial_\beta N}{N} \right)\nonumber\\
   &&\hskip2cm+\frac{\partial_\mu N}{N} \partial_\nu g_{\alpha\beta} \left[ g^{\alpha\beta} g^{\mu\nu} 
     - g^{\alpha \nu} g^{\beta \mu} \right]
     + (\cdots)
    \end{eqnarray}
     When Eq.~(\ref{lquadsplit}) and Eq.~(\ref{dcpcsplit}) are added, the cross term involving $\partial_\mu N 
     \partial_\nu g_{\alpha\beta}$ cancels out precisely. All the spatial terms combine together
     to give ${}^{(3)}\mathcal{R}$. This leads to the result
     \begin{eqnarray}
     R&=&  \frac{1}{4N^2} \dot g_{\alpha\beta} \dot g_{\mu\nu}
  \left[
  g^{\alpha\beta} g^{\mu\nu}- g^{\alpha\mu} g^{\beta\nu}
  \right] + {}^3R  \nonumber\\
  &&\qquad - \frac{1}{\sqrt{-g}} \partial_0 \left[ \sqrt{-g} g^{00} g^{\alpha\beta} \dot g_{\alpha\beta} \right]
   -\frac{2}{\sqrt{-g}} \partial_\alpha \left[ \sqrt{-g} g^{\alpha\beta} \frac{\partial_\beta N}{N} \right]
  \end{eqnarray}
  The terms in the first line give what is conventionally called the ADM Lagrangian $L_{\rm ADM}$.
  The time derivative term (in the second line) leads to the integral of twice  the trace of the extrinsic curvature $K$
  on the $t=$constant surfaces. The spatial derivative term leads to the integral of twice 
  the normal component
  of the acceleration on the timelike boundaries. 
  Incidentally, note that the last two terms can be expressed more symmetrically in the form
  \begin{equation}
  - \frac{1}{\sqrt{-g}}\left[ \partial_0 \left( \sqrt{-g} g^{00} g^{\alpha\beta} \partial_0 g_{\alpha\beta} \right)
  - \partial_\alpha \left( \sqrt{-g}  g^{\alpha\beta}g^{00} \partial_\beta g_{00} \right)\right]
  \end{equation}
It is clear that the structure of Einstein-Hilbert Lagrangian is very special.

\bibliography{physrepthermo}

\begin{thebibliography}{100}
\expandafter\ifx\csname url\endcsname\relax
  \def\url#1{\texttt{#1}}\fi
\expandafter\ifx\csname urlprefix\endcsname\relax\def\urlprefix{URL }\fi

\bibitem{Eddington:1920}
Eddington, Space, Time and Gravitation, Cambridge University Press, UK, 1920.

\bibitem{Dirac:1962}
P.~A.~M. Dirac, Proc.Roy.Soc (London) A 270 (1962) 354.

\bibitem{Bardeen:1973gs}
J.~M. Bardeen, B.~Carter, S.~W. Hawking, The four laws of black hole mechanics,
  Commun. Math. Phys. 31 (1973) 161--170.

\bibitem{BHLS}
C.~DeWitt, B.~DeWitt (Eds.), Black holes, University of Grenobe, 1972.

\bibitem{membrane}
K.~S. Thorne, R.~H. Price, D.~A. Macdonald (Eds.), Black Holes: The Membrane
  Paradigm, Yale University Press, London, 1986.

\bibitem{Bekenstein:1972tm}
J.~D. Bekenstein, Black holes and the second law, Nuovo Cim. Lett. 4 (1972)
  737--740.

\bibitem{Bekenstein:1973ur}
J.~D. Bekenstein, Black holes and entropy, Phys. Rev. D7 (1973) 2333--2346.

\bibitem{Bekenstein:1974ax}
J.~D. Bekenstein, Generalized second law of thermodynamics in black hole
  physics, Phys. Rev. D9 (1974) 3292--3300.

\bibitem{Hawking:1975sw}
S.~W. Hawking, Particle creation by black holes, Commun. Math. Phys. 43 (1975)
  199--220.

\bibitem{Fulling:1973md}
S.~A. Fulling, Nonuniqueness of canonical field quantization in riemannian
  space-time, Phys. Rev. D7 (1973) 2850--2862.

\bibitem{Davies:1975th}
P.~C.~W. Davies, Scalar particle production in schwarzschild and rindler
  metrics, J. Phys. A8 (1975) 609--616.

\bibitem{Gerlachbh}
U.~H. Gerlach, The mechanism of blackbody radiation from an incipient black
  hole, Phys. Rev. D14 (1976) 1479--1508.

\bibitem{mtw}
C.~Misner, K.~Thorne, J.~Wheeler, Gravitation, Freeman and Co., 1973.

\bibitem{Birrel:bkqft}
N.~D. Birrel, P.~C.~W. Davies, Quantum Field Theory in Curved Space-Time,
  Cambridge University Press, Cambridge, 1982.

\bibitem{Fulling:bkqft}
S.~A. Fulling, Aspects of Quantum Field Theory in Curved Space-Time, Cambridge
  University Press, Cambridge, 1989.

\bibitem{Wald:bkqft}
R.~M. Wald, Quantum Field Theory in Curved Spacetime and Black Hole
  Thermodynamics, The University of Chicago Press, Chicago, 1994.

\bibitem{Dewitt:1975ys}
B.~S. Dewitt, Quantum field theory in curved space-time, Phys. Rept. 19 (1975)
  295--357.

\bibitem{Takagi:1986kn}
S.~Takagi, Vacuum noise and stress induced by uniform accelerator:
  Hawking-unruh effect in rindler manifold of arbitrary dimensions, Prog.
  Theor. Phys. Suppl. 88 (1986) 1--142.

\bibitem{Sriramkumar:1999nw}
L.~Sriramkumar, T.~Padmanabhan, Probes of the vacuum structure of quantum
  fields in classical backgrounds, Int. J. Mod. Phys. D11 (2002) 1--34.

\bibitem{Bousso:2002ju}
R.~Bousso, The holographic principle, Rev. Mod. Phys. 74 (2002) 825--874.

\bibitem{Brout:1995rd}
R.~Brout, S.~Massar, R.~Parentani, P.~Spindel, A primer for black hole quantum
  physics, Phys. Rept. 260 (1995) 329--454.

\bibitem{Wald:1999vt}
R.~M. Wald, The thermodynamics of black holes, Living Rev. Rel. 4 (2001) 6.

\bibitem{Sciama:1981hr}
D.~W. Sciama, P.~Candelas, D.~Deutsch, Quantum field theory, horizons and
  thermodynamics, Adv. Phys. 30 (1981) 327--366.

\bibitem{Hawking:1973el}
S.~W. Hawking, G.~F.~R. Ellis, The large scale structure of space-time,
  Cambridge University Press, Cambridge, 1973.

\bibitem{Letaw:1981ik}
J.~R. Letaw, J.~D. Pfautsch, The quantized scalar field in the stationary
  coordinate systems of flat space-time, Phys. Rev. D24 (1981) 1491.

\bibitem{Letaw:1981yv}
J.~R. Letaw, Vacuum excitation of noninertial detectors on stationary world
  lines, Phys. Rev. D23 (1981) 1709.

\bibitem{Padmanabhan:1982apsci}
T.~Padmanabhan, General covariance, accelerated frames and the particle
  concept, Astroph. Sp. Sci 83 (1982) 247.

\bibitem{lltwo}
L.~Landau, E.~M. Lifshitz, Classical Theory of Fields: Volume II, Pergamon
  Press, New York, 1975.

\bibitem{Ashtekar:2003hk}
A.~Ashtekar, B.~Krishnan, Dynamical horizons and their properties, Phys. Rev.
  D68 (2003) 104030.

\bibitem{Date:2001xj}
G.~Date, Isolated horizon, killing horizon and event horizon, Class. Quant.
  Grav. 18 (2001) 5219--5226.

\bibitem{Grove:1986fz}
P.~G. Grove, On an inertial observer's interpretation of the detection of
  radiation by linearly accelerated particle detectors, Class. Quant. Grav. 3
  (1986) 801--809.

\bibitem{Hu:1996br}
B.~L. Hu, A.~Raval, Thermal radiance from black hole and cosmological space-
  times: A unified view, Mod. Phys. Lett. A11 (1996) 2625--2638.

\bibitem{Koks:1997rk}
D.~Koks, B.~L. Hu, A.~Matacz, A.~Raval, Thermal particle creation in
  cosmological spacetimes: A stochastic approach, Phys. Rev. D56 (1997)
  4905--4915.

\bibitem{Hu:1996vu}
B.~L. Hu, Hawking-unruh thermal radiance as relativistic exponential scaling of
  quantum noise, [gr-qc/9606073] (1996).

\bibitem{Raval:1997vt}
A.~Raval, B.~L. Hu, D.~Koks, Near-thermal radiation in detectors, mirrors and
  black holes: A stochastic approach, Phys. Rev. D55 (1997) 4795--4812.

\bibitem{Visser:2001kq}
M.~Visser, Essential and inessential features of hawking radiation, Int. J.
  Mod. Phys. D12 (2003) 649--661.

\bibitem{tplsksrinia}
K.~Srinivasan, L.~Sriramkumar, T.~Padmanabhan, Plane waves viewed from an
  accelerated frame: Quantum physics in classical setting, Phys. Rev. D 56
  (1997) 6692.

\bibitem{tplsksrinib}
K.~Srinivasan, L.~Sriramkumar, T.~Padmanabhan, Possible quantum interpretation
  of certain power spectra in classical field theory, Int. J. Mod. Phys. D 6
  (1997) 607--623.

\bibitem{Padmanabhan:2002ha}
T.~Padmanabhan, Thermodynamics and / of horizons: A comparison of
  schwarzschild, rindler and de sitter spacetimes, Mod. Phys. Lett. A17 (2002)
  923--942.

\bibitem{Bekenstein:1998aw}
J.~D. Bekenstein, Black holes: Classical properties, thermodynamics, and
  heuristic quantization, in: M.~Novello (Ed.), Cosmology and Gravitation,
  Atlantisciences, France, 2000, pp. 1--85, [gr-qc/9808028].

\bibitem{Srinivasan:1998ty}
K.~Srinivasan, T.~Padmanabhan, Particle production and complex path analysis,
  Phys. Rev. D60 (1999) 024007.

\bibitem{Carlip:1999cy}
S.~Carlip, Entropy from conformal field theory at killing horizons, Class.
  Quant. Grav. 16 (1999) 3327--3348.

\bibitem{Park:1999tj}
M.-I. Park, J.~Ho, Comments on 'black hole entropy from conformal field theory
  in any dimension', Phys. Rev. Lett. 83 (1999) 5595.

\bibitem{Park:2001zn}
M.-I. Park, Hamiltonian dynamics of bounded spacetime and black hole entropy:
  Canonical method, Nucl. Phys. B634 (2002) 339--369.

\bibitem{Kim:2002uz}
Y.-b. Kim, C.~Y. Oh, N.~Park, Classical geometry of de sitter spacetime: An
  introductory review, [hep-th/0212326] (2002).

\bibitem{Gibbons:1977mu}
G.~Gibbons, S.~Hawking, Cosmological event horizons, thermodynamics, and
  particle creation, Phys. Rev. D15 (1977) 2738--2751.

\bibitem{Feynman:1988}
R.~Feynman, The reason for anti-particle, in: R.~Feynman, S.~Weinberg (Eds.),
  Elementary Particles and the Laws of Physics: The 1986 Dirac Memorial
  Lectures, Cambridge University Press, Cambridge, 1988, pp. 1--59.

\bibitem{Parker:1968mv}
L.~Parker, Particle creation in expanding universes, Phys. Rev. Lett. 21 (1968)
  562--564.

\bibitem{Parker:1969au}
L.~Parker, Quantized fields and particle creation in expanding universes. 1,
  Phys. Rev. 183 (1969) 1057--1068.

\bibitem{Damour:76}
T.~Damour, R.~Ruffini, Black-hole evaporation in the
  klein-sauter-heisenberg-euler formalism, Phys. Rev. D 14 (1976) 332.

\bibitem{Hartle:1976tp}
J.~B. Hartle, S.~W. Hawking, Path integral derivation of black hole radiance,
  Phys. Rev. D13 (1976) 2188--2203.

\bibitem{Gibbons:1977ue}
G.~W. Gibbons, S.~W. Hawking, Action integrals and partition functions in
  quantum gravity, Phys. Rev. D15 (1977) 2752--2756.

\bibitem{Parikh:1999mf}
M.~K. Parikh, F.~Wilczek, Hawking radiation as tunneling, Phys. Rev. Lett. 85
  (2000) 5042--5045.

\bibitem{Shankaranarayanan:2000gb}
S.~Shankaranarayanan, K.~Srinivasan, T.~Padmanabhan, Method of complex paths
  and general covariance of hawking radiation, Mod. Phys. Lett. A16 (2001)
  571--578.

\bibitem{Shankaranarayanan:2000qv}
S.~Shankaranarayanan, T.~Padmanabhan, K.~Srinivasan, Hawking radiation in
  different coordinate settings: Complex paths approach, Class. Quant. Grav. 19
  (2002) 2671--2688.

\bibitem{Schutzhold:2000ju}
R.~Schutzhold, On the hawking effect, Phys. Rev. D64 (2001) 024029.

\bibitem{Christensen:1978tw}
S.~M. Christensen, M.~J. Duff, Flat space as a gravitational instanton, Nucl.
  Phys. B146 (1978) 11.

\bibitem{Troost:1978yk}
W.~Troost, H.~van Dam, Thermal propagators and accelerated frames of reference,
  Nucl. Phys. B152 (1979) 442.

\bibitem{Padmanabhan:2003dc}
T.~Padmanabhan, Topological interpretation of the horizon temperature, Mod.
  Phys. Letts.A 18 (2003) 2903.

\bibitem{Unruh:1976db}
W.~G. Unruh, Notes on black hole evaporation, Phys. Rev. D14 (1976) 870.

\bibitem{Padmanabhan:2004tz}
T.~Padmanabhan, Entropy of horizons, complex paths and quantum tunneling,
  Mod.Phys.Letts. A 19 (2004) 2637--2643.

\bibitem{Keski-Vakkuri:1997xp}
E.~Keski-Vakkuri, P.~Kraus, Microcanonical d-branes and back reaction, Nucl.
  Phys. B491 (1997) 249--262.

\bibitem{Medved:2002zj}
A.~J.~M. Medved, Radiation via tunneling from a de sitter cosmological horizon,
  Phys. Rev. D66 (2002) 124009.

\bibitem{llthree}
L.~Landau, E.~M. Lifshitz, Course of Theoretical Physics: Volume III, Quantum
  Mechanics, Pergamon Press, New York, 1977.

\bibitem{lee}
T.~D. Lee, Are black holes blackbodies?, Nucl. Phys. B 264 (1986) 437.

\bibitem{Unruh:1983ac}
W.~G. Unruh, N.~Weiss, Acceleration radiation in interacting field theories,
  Phys. Rev. D29 (1984) 1656.

\bibitem{Gerlach:1989rz}
U.~H. Gerlach, Quantum states of a field partitioned by an accelerated frame,
  Phys. Rev. D40 (1989) 1037--1047.

\bibitem{Dewitt:1979}
B.~S. DeWitt, Quantum gravity: The new synthesis, in: S.~Hawking, W.~Israel
  (Eds.), General Relativity: An Einstein Centenary Survey, Cambridge
  University Press, Cambridge, 1979, pp. 680--745.

\bibitem{Grove:1983rp}
P.~G. Grove, A.~C. Ottewill, Notes on 'particle detectors.', J. Phys. A16
  (1983) 3905--3920.

\bibitem{Padmanabhan:1985}
T.~Padmanabhan, Why does an accelerated detector click?, Class. Quan. Grav. 2
  (1985) 117.

\bibitem{Padmanabhan:2002ji}
T.~Padmanabhan, Cosmological constant: The weight of the vacuum, Phys. Rept.
  380 (2003) 235--320.

\bibitem{Padmanabhan:2002trctp}
T.~Padmanabhan, T.~R. Choudhury, Can the clustered dark matter and the smooth
  dark energy arise from the same scalar field?, Phys. Rev. D66 (2002) 081301.

\bibitem{Choudhury:2003tj}
T.~R. Choudhury, T.~Padmanabhan, Cosmological parameters from supernova
  observations: A critical comparison of three data sets, Astron. Ap.,(in
  press), [astro-ph/0311622].

\bibitem{Boulware:1975dm}
D.~G. Boulware, Quantum field theory in schwarzschild and rindler spaces, Phys.
  Rev. D11 (1975) 1404.

\bibitem{Padmanabhan:2002sh}
T.~Padmanabhan, Classical and quantum thermodynamics of horizons in spherically
  symmetric spacetimes, Class. Quant. Grav. 19 (2002) 5387--5408.

\bibitem{Davis:2003ye}
T.~M. Davis, P.~C.~W. Davies, C.~H. Lineweaver, Black hole versus cosmological
  horizon entropy, Class. Quant. Grav. 20 (2003) 2753--2764.

\bibitem{Davies:2003me}
P.~C.~W. Davies, T.~M. Davis, How far can the generalized second law be
  generalized?, [astro-ph/0310522] (2003).

\bibitem{Deser:1997ri}
S.~Deser, O.~Levin, Accelerated detectors and temperature in (anti) de sitter
  spaces, Class. Quant. Grav. 14 (1997) L163--L168.

\bibitem{Deser:1998xb}
S.~Deser, O.~Levin, Mapping hawking into unruh thermal properties, Phys. Rev.
  D59 (1999) 064004.

\bibitem{Choudhury:2004ph}
T.~R. Choudhury, T.~Padmanabhan, Concept of temperature in multi-horizon
  spacetimes: Analysis of schwarzschild-de sitter metric, , [gr-qc/0404091]
  (2004).

\bibitem{Markovic:1991ua}
D.~Markovic, W.~G. Unruh, Vacuum for a massless scalar field outside a
  collapsing body in de sitter space-time, Phys. Rev. D43 (1991) 332--339.

\bibitem{Tadaki:1990cg}
S.~Tadaki, S.~Takagi, Quantum field theory in two-dimensional schwarzschild-de
  sitter space-time. 2: Space with a collapsing star, Prog. Theor. Phys. 83
  (1990) 1126--1139.

\bibitem{Tadaki:1990aa}
S.-I. Tadaki, S.~Takagi, Quantum field theory in two-dimensional
  schwarzschild-de sitter space-time 1. empty space, Prog. Theor. Phys. 83
  (1990) 941--952.

\bibitem{Unruh:1982ic}
W.~G. Unruh, R.~M. Wald, Acceleration radiation and generalized second law of
  thermodynamics, Phys. Rev. D25 (1982) 942--958.

\bibitem{Frolov:1993fy}
V.~P. Frolov, D.~N. Page, Proof of the generalized second law for
  quasistationary semiclassical black holes, Phys. Rev. Lett. 71 (1993)
  3902--3905.

\bibitem{Padmanabhan:1990gm}
T.~Padmanabhan, Statistical mechanics of gravitating systems, Phys. Rept. 188
  (1990) 285.

\bibitem{Gerlach:1976hbh}
U.~H. Gerlach, Why is a black hole hot ?, Phys. Rev. D14 (1976) 3290--3293.

\bibitem{Hawking:1976de}
S.~W. Hawking, Black holes and thermodynamics, Phys. Rev. D13 (1976) 191--197.

\bibitem{Bekenstein:1975tw}
J.~D. Bekenstein, Statistical black hole thermodynamics, Phys. Rev. D12 (1975)
  3077--3085.

\bibitem{Wheelerbk:1990}
J.~A. Wheeler, A Journey into Gravity and Spacetime, Freeman and Co., N.Y.,
  1990.

\bibitem{'tHooft:1990fr}
G.~'t~Hooft, The black hole interpretation of string theory, Nucl. Phys. B335
  (1990) 138--154.

\bibitem{Susskind:1993if}
L.~Susskind, L.~Thorlacius, J.~Uglum, The stretched horizon and black hole
  complementarity, Phys. Rev. D48 (1993) 3743--3761.

\bibitem{Wald:1993nt}
R.~M. Wald, Black hole entropy in noether charge, Phys. Rev. D48 (1993)
  3427--3431.

\bibitem{Jacobson:1994vj}
T.~Jacobson, G.~Kang, R.~C. Myers, On black hole entropy, Phys. Rev. D49 (1994)
  6587--6598.

\bibitem{Visser:1993nu}
M.~Visser, Dirty black holes: Entropy as a surface term, Phys. Rev. D48 (1993)
  5697--5705.

\bibitem{Banados:1994qp}
M.~Banados, C.~Teitelboim, J.~Zanelli, Black hole entropy and the dimensional
  continuation of the gauss-bonnet theorem, Phys. Rev. Lett. 72 (1994)
  957--960.

\bibitem{Susskind:1993ws}
L.~Susskind, Some speculations about black hole entropy in string theory, in:
  C.~Teitelboim (Ed.), The black hole, 25 years after, World Scientific, 1998,
  pp. 118--131, [hep-th/9309145].

\bibitem{Dowker:1994fi}
J.~S. Dowker, Remarks on geometric entropy, Class. Quant. Grav. 11 (1994)
  L55--L60.

\bibitem{Israel:1976ur}
W.~Israel, Thermo field dynamics of black holes, Phys. Lett. A57 (1976)
  107--110.

\bibitem{Callan:1994py}
J.~Callan, Curtis~G., F.~Wilczek, On geometric entropy, Phys. Lett. B333 (1994)
  55--61.

\bibitem{Frolov:1998vs}
V.~P. Frolov, D.~V. Fursaev, Thermal fields, entropy, and black holes, Class.
  Quant. Grav. 15 (1998) 2041--2074.

\bibitem{Frolov:1997up}
V.~P. Frolov, D.~V. Fursaev, Mechanism of generation of black hole entropy in
  sakharov's induced gravity, Phys. Rev. D56 (1997) 2212--2225.

\bibitem{Frolov:1997aj}
V.~P. Frolov, D.~V. Fursaev, A.~I. Zelnikov, Statistical origin of black hole
  entropy in induced gravity, Nucl. Phys. B486 (1997) 339--352.

\bibitem{Unruh:1995je}
W.~G. Unruh, Sonic analog of black holes and the effects of high frequencies on
  black hole evaporation, Phys. Rev. D51 (1995) 2827--2838.

\bibitem{Jacobson:1993hn}
T.~Jacobson, Black hole radiation in the presence of a short distance cutoff,
  Phys. Rev. D48 (1993) 728--741.

\bibitem{Brout:1995wp}
R.~Brout, S.~Massar, R.~Parentani, P.~Spindel, Hawking radiation without
  transplanckian frequencies, Phys. Rev. D52 (1995) 4559--4568.

\bibitem{'tHooft:1985re}
G.~'t~Hooft, On the quantum structure of a black hole, Nucl. Phys. B256 (1985)
  727.

\bibitem{Padmanabhan:1986rs}
T.~Padmanabhan, On the quantum structure of horizons, Phys. Lett. B173 (1986)
  43--45.

\bibitem{Frolov:1993ym}
V.~Frolov, I.~Novikov, Dynamical origin of the entropy of a black hole, Phys.
  Rev. D48 (1993) 4545--4551.

\bibitem{Zurek:1985gd}
W.~H. Zurek, K.~S. Thorne, Statistical mechanical origin of the entropy of a
  rotating, cha rged black hole, Phys. Rev. Lett. 54 (1985) 2171.

\bibitem{Bombelli:1986rw}
L.~Bombelli, R.~K. Koul, J.-H. Lee, R.~D. Sorkin, A quantum source of entropy
  for black holes, Phys. Rev. D34 (1986) 373.

\bibitem{Srednicki:1993im}
M.~Srednicki, Entropy and area, Phys. Rev. Lett. 71 (1993) 666--669.

\bibitem{Padmanabhan:1998jp}
T.~Padmanabhan, Quantum structure of spacetime and black hole entropy, Phys.
  Rev. Lett. 81 (1998) 4297--4300.

\bibitem{Padmanabhan:1998vr}
T.~Padmanabhan, Event horizon: Magnifying glass for planck length physics,
  Phys. Rev. D59 (1999) 124012.

\bibitem{Unruh:1994zw}
W.~G. Unruh, Dumb holes and the effects of high frequencies on black hole
  evaporation, [gr-qc/9409008] (1994).

\bibitem{Corley:1996ar}
S.~Corley, T.~Jacobson, Hawking spectrum and high frequency dispersion, Phys.
  Rev. D54 (1996) 1568--1586.

\bibitem{Myers:1997qi}
R.~C. Myers, Pure states don't wear black, Gen. Rel. Grav. 29 (1997)
  1217--1222.

\bibitem{Rovelli:1998yv}
C.~Rovelli, Loop quantum gravity, Living Rev. Rel. 1 (1998) 1.

\bibitem{Das:2000su}
S.~R. Das, S.~D. Mathur, The quantum physics of black holes: Results from
  string theory, Ann. Rev. Nucl. Part. Sci. 50 (2000) 153--206.

\bibitem{Peet:2000hn}
A.~W. Peet, Tasi lectures on black holes in string theory, in: Boulder 1999,
  Strings, branes and gravity, 1999, pp. 353--433, [hep-th/0008241].

\bibitem{Strominger:1996sh}
A.~Strominger, C.~Vafa, Microscopic origin of the bekenstein-hawking entropy,
  Phys. Lett. B379 (1996) 99--104.

\bibitem{Breckenridge:1996sn}
J.~C. Breckenridge, et~al., Macroscopic and microscopic entropy of
  near-extremal spinning black holes, Phys. Lett. B381 (1996) 423--426.

\bibitem{Sen:1994yi}
A.~Sen, Dyon - monopole bound states, selfdual harmonic forms on the multi -
  monopole moduli space, and sl(2,z) invariance in string theory, Phys. Lett.
  B329 (1994) 217--221.

\bibitem{Vafa:1994tf}
C.~Vafa, E.~Witten, A strong coupling test of s duality, Nucl. Phys. B431
  (1994) 3--77.

\bibitem{Sen:1999mg}
A.~Sen, Non-bps states and branes in string theor, in: Cargese 1999, Progress
  in string theory and M-theory, 3rd APCTP Winter School on Duality in Fields
  and Strings, 1999, pp. 187--234, [hep-th/9904207].

\bibitem{Rovelli:1998gg}
C.~Rovelli, P.~Upadhya, Loop quantum gravity and quanta of space: A primer,
  [gr-qc/9806079] (1998).

\bibitem{Thiemann:2001yy}
T.~Thiemann, Introduction to modern canonical quantum general relativity,
  submitted to Living Rev.Rel., [gr-qc/0110034] (2001).

\bibitem{Krasnov:1998wc}
K.~V. Krasnov, On statistical mechanics of gravitational systems, Gen. Rel.
  Grav. 30 (1998) 53--68.

\bibitem{Rovelli:1996dv}
C.~Rovelli, Black hole entropy from loop quantum gravity, Phys. Rev. Lett. 77
  (1996) 3288--3291.

\bibitem{Ashtekar:2000eq}
A.~Ashtekar, J.~C. Baez, K.~Krasnov, Quantum geometry of isolated horizons and
  black hole entropy, Adv. Theor. Math. Phys. 4 (2000) 1--94.

\bibitem{Ashtekar:1998yu}
A.~Ashtekar, J.~Baez, A.~Corichi, K.~Krasnov, Quantum geometry and black hole
  entropy, Phys. Rev. Lett. 80 (1998) 904--907.

\bibitem{Corichi:2002ty}
A.~Corichi, On quasinormal modes, black hole entropy, and quantum geometry,
  Phys. Rev. D67 (2003) 087502.

\bibitem{Kokkotas:1999bd}
K.~D. Kokkotas, B.~G. Schmidt, Quasi-normal modes of stars and black holes,
  Living Rev. Rel. 2 (1999) 2.

\bibitem{Berti:2003jh}
E.~Berti, V.~Cardoso, K.~D. Kokkotas, H.~Onozawa, Highly damped quasinormal
  modes of kerr black holes, Phys. Rev. D68 (2003) 124018.

\bibitem{Cardoso:2003vt}
V.~Cardoso, J.~P.~S. Lemos, S.~Yoshida, Quasinormal modes of schwarzschild
  black holes in four and higher dimensions, Phys. Rev. D69 (2004) 044004.

\bibitem{Padmanabhan:2003fx}
T.~Padmanabhan, Quasi normal modes: A simple derivation of the level spacing of
  the frequencies, Class. Quant. Grav. 21 (2004) L1.

\bibitem{Choudhury:2003wd}
T.~R. Choudhury, T.~Padmanabhan, Quasi normal modes in schwarzschild-desitter
  spacetime: A simple derivation of the level spacing of the frequencies, Phys.
  Rev. D69 (2004) 064033.

\bibitem{Motl:2003cd}
L.~Motl, A.~Neitzke, Asymptotic black hole quasinormal frequencies, Adv. Theor.
  Math. Phys. 7 (2003) 307--330.

\bibitem{Motl:2002hd}
L.~Motl, An analytical computation of asymptotic schwarzschild quasinormal
  frequencies, Adv. Theor. Math. Phys. 6 (2003) 1135--1162.

\bibitem{Dreyer:2002vy}
O.~Dreyer, Quasinormal modes, the area spectrum, and black hole entropy, Phys.
  Rev. Lett. 90 (2003) 081301.

\bibitem{Hod:1998vk}
S.~Hod, Bohr's correspondence principle and the area spectrum of quantum black
  holes, Phys. Rev. Lett. 81 (1998) 4293.

\bibitem{Padmanabhan:2003ub}
T.~Padmanabhan, A.~Patel, Role of horizons in semiclassical gravity: Entropy
  and the area spectrum, [gr-qc/0309053] (2003).

\bibitem{Padmanabhan:2003grg}
T.~Padmanabhan, Why gravity has no choice: Bulk spacetime dynamics is dictated
  by information entanglement across horizons, Gen.Rel.Grav. 35 (2003)
  2097--2103, fifth prize essay, Gravity Research Foundation Essay Contest,
  2003.

\bibitem{York:1986it}
J.~York, James~W., Black hole thermodynamics and the euclidean einstein action,
  Phys. Rev. D33 (1986) 2092--2099.

\bibitem{Braden:1990hw}
H.~W. Braden, J.~D. Brown, B.~F. Whiting, J.~York, James~W., Charged black hole
  in a grand canonical ensemble, Phys. Rev. D42 (1990) 3376--3385.

\bibitem{Brown:1991fk}
J.~D. Brown, E.~A. Martinez, J.~York, James~W., Complex kerr-newman geometry
  and black hole thermodynamics, Phys. Rev. Lett. 66 (1991) 2281--2284.

\bibitem{Martinez:1989hn}
E.~A. Martinez, J.~York, James~W., Additivity of the entropies of black holes
  and matter in equilibrium, Phys. Rev. D40 (1989) 2124--2127.

\bibitem{Hayward:1990zm}
G.~Hayward, Euclidean action and the thermodynamics of manifolds without
  boundary, Phys. Rev. D41 (1990) 3248--3251.

\bibitem{Hayward:1991}
G.~Hayward, General first law and thermodynamics of horizon/matter systems,
  Phys. Rev. D43 (1991) 3861--3872.

\bibitem{Jacobson:1995ab}
T.~Jacobson, Thermodynamics of space-time: The einstein equation of state,
  Phys. Rev. Lett. 75 (1995) 1260--1263.

\bibitem{Volovik:2000ua}
G.~E. Volovik, Superfluid analogies of cosmological phenomena, Phys. Rept. 351
  (2001) 195--348.

\bibitem{Volovik:2003fe}
G.~E. Volovik, The universe in a helium droplet, Oxford University Press, UK,
  2003.

\bibitem{Brown:1995su}
J.~D. Brown, Black hole entropy and the hamiltonian formulation of
  diffeomorphism invariant theories, Phys. Rev. D52 (1995) 7011--7026.

\bibitem{Padmanabhan:2002xm}
T.~Padmanabhan, Is gravity an intrinsically quantum phenomenon? dynamics of
  gravity from the entropy of spacetime and the principle of equivalence, Mod.
  Phys. Lett. A17 (2002) 1147--1158.

\bibitem{Padmanabhan:2002ma}
T.~Padmanabhan, Gravity from spacetime thermodynamics, Astroph. Sp. Sci. 285
  (2003) 407.

\bibitem{Padmanabhan:2004xk}
T.~Padmanabhan, From gravitons to gravity: Myths and reality, [gr-qc/0409089]
  (2004).

\bibitem{Arnowitt1962}
R.~e. Arnowitt, Gravitation: An Introduction to Current Research, Wiley, New
  York, 1962.

\bibitem{York1988}
J.~York, in: W.~H. Zurek, et~al. (Eds.), Between Quantum and Cosmos, Princeton
  University Press, Princeton, 1988, p. 246.

\bibitem{Sakharov:1968pk}
A.~D. Sakharov, Vacuum quantum fluctuations in curved space and the theory of
  gravitation, Sov. Phys. Dokl. 12 (1968) 1040--1041.

\bibitem{Padmanabhan:2004kf}
T.~Padmanabhan, Gravity as elasticity of spacetime: A paradigm to understand
  horizon thermodynamics and cosmological constant, Int. Jour. Mod. Phys. D (in
  press)[gr-qc/0408051].

\bibitem{Padmanabhan:2003pk}
T.~Padmanabhan, Gravitational entropy of static spacetimes and microscopic
  density of states, Class. Quant. Grav. 21 (2004) 4485--4494.

\bibitem{tolman}
R.~C. Tolman, On the use of the energy-momentum principle in general
  relativity, Phys. Rev. 35 (1930) 875--895.

\bibitem{Padmanabhan:2002jr}
T.~Padmanabhan, The holography of gravity encoded in a relation between
  entropy, horizon area and action for gravity, Gen. Rel. Grav. 34 (2002)
  2029--2035, second prize essay, Gravity Research Foundation Essay Contest,
  2002.

\bibitem{Padmanabhan:1987au}
T.~Padmanabhan, Limitations on the operational definition of space-time events
  and quantum gravity, Class. Quant. Grav. 4 (1987) L107--L113.

\bibitem{Amelino-Camelia:1994vs}
G.~Amelino-Camelia, Limits on the measurability of space-time distances in the
  semiclassical approximation of quantum gravity, Mod. Phys. Lett. A9 (1994)
  3415--3422.

\bibitem{Amelino-Camelia:2003uc}
G.~Amelino-Camelia, The three perspectives on the quantum-gravity problem and
  their implications for the fate of lorentz symmetry, [gr-qc/0309054] (2003).

\bibitem{Bekenstein:1974jk}
J.~D. Bekenstein, The quantum mass spectrum of the kerr black hole, Lett. Nuovo
  Cim. 11 (1974) 467.

\bibitem{Gour:1999yu}
G.~Gour, Quantum mechanics of a black hole, Phys. Rev. D61 (2000) 124007.

\bibitem{Louko:1996md}
J.~Louko, J.~Makela, Area spectrum of the schwarzschild black hole, Phys. Rev.
  D54 (1996) 4982--4996.

\bibitem{Mukhanov:1986me}
V.~F. Mukhanov, Are black holes quantized?, JETP Lett. 44 (1986) 63--66.

\bibitem{Kogan:1986yd}
Y.~I. Kogan, Quantization of the mass of a black hole in string theory, JETP
  Lett. 44 (1986) 267--270.

\bibitem{Mazur:1987jf}
P.~O. Mazur, Are there topological black hole solitons in string theory?, Gen.
  Rel. Grav. 19 (1987) 1173--1180.

\bibitem{Lousto:1995jd}
C.~O. Lousto, The emergence of an effective two-dimensional quantum description
  from the study of critical phenomena in black holes, Phys. Rev. D51 (1995)
  1733--1740.

\bibitem{Peleg:1995gg}
Y.~Peleg, The spectrum of quantum dust black holes, Phys. Lett. B356 (1995)
  462--465.

\bibitem{Das:2002xb}
S.~Das, P.~Ramadevi, U.~A. Yajnik, Black hole area quantization, Mod. Phys.
  Lett. A17 (2002) 993--1000.

\bibitem{Danielsson:1993um}
U.~H. Danielsson, M.~Schiffer, Quantum mechanics, common sense and the black
  hole information paradox, Phys. Rev. D48 (1993) 4779--4784.

\bibitem{Maggiore:1994ww}
M.~Maggiore, Black holes as quantum membranes, Nucl. Phys. B429 (1994)
  205--228.

\bibitem{Bekenstein:1995ju}
J.~D. Bekenstein, V.~F. Mukhanov, Spectroscopy of the quantum black hole, Phys.
  Lett. B360 (1995) 7--12.

\bibitem{Kastrup:1996pu}
H.~A. Kastrup, On the quantum levels of isolated spherically symmetric
  gravitational systems, Phys. Lett. B385 (1996) 75--80.

\end{thebibliography}

\end{document}